\documentclass[10pt,aps,pra,twocolumn,showpacs,superscriptaddress,nobalancelastpage,longbibliography,nofootinbib,floatfix]{revtex4-2}

\usepackage{graphicx}
\usepackage{bm}
\usepackage{color}
\usepackage{MnSymbol}
\usepackage{enumerate}
\usepackage{bbold,soul}
\usepackage{lipsum}
\usepackage{array}
\usepackage{hyperref}
\usepackage{textgreek}
\usepackage[dvipsnames]{xcolor}
\usepackage{tabularx}
\usepackage{graphics,epsfig,subfigure}
\usepackage{wrapfig}
\usepackage{mathrsfs}
\usepackage{cleveref}

\usepackage{multirow}

\DeclareMathAlphabet\mathbfcal{OMS}{cmsy}{b}{n}

\def\be{\begin{equation}}
\def\ee{\end{equation}}

\newcommand{\beq}{\begin{equation}}
\newcommand{\eeq}{\end{equation}}
\newcommand{\nn}{\nonumber}

\newcommand{\erf}[1]{Eq.~(\ref{#1})}

\newcommand{\erfand}[2]{Eqs.~(\ref{#1}) and (\ref{#2})}

\newcommand{\smallfrac}[2]{\mbox{$\frac{#1}{#2}$}}
\newcommand{\smallsum}[2]{\mbox{$\sum_{#1}^{#2}$}}

\newcommand{\half}{\smallfrac{1}{2}}
\newcommand{\bra}[1]{\langle{#1}|}
\newcommand{\ket}[1]{|{#1}\rangle}

\newcommand{\sign}{{\rm sign}}

\newcommand{\sq}[1]{\left[ {#1} \right]}
\newcommand{\cu}[1]{\left\{ {#1} \right\}}
\newcommand{\ro}[1]{\left( {#1} \right)}

\newcommand{\tp}{^{\top}}

\newcommand{\s}[1]{\hat{\sigma}_{#1}}
\renewcommand{\ul}[1]{\underline{{#1}}}

\newcommand{\dd}{{\rm d}}

\newcommand{\kd}{\kappa}
\newcommand{\ks}{K}
\newcommand{\phis}{\Phi}
\newcommand{\phid}{\phi}
\newcommand{\cd}{c}
\newcommand{\xd}{X}
\newcommand{\dt}{{\rm d}t}
\newcommand{\ddt}{\tau} 
\newcommand{\noise}{\Xi}
\newcommand{\info}{Y}
\newcommand{\mc}{\lambda}
\newcommand{\nc}{\eta}
\newcommand{\gu}{\gamma_\uparrow}
\newcommand{\gd}{\gamma_\downarrow}
\newcommand{\gud}{\gamma_{\uparrow,\downarrow}}
\newcommand{\bg}{\breve\gamma}
\newcommand{\mg}{\bar\gamma}
\newcommand{\rtp}{z}

\newcommand{\coh}{\mathcal{C}}
\newcommand{\cohc}{\coh^{\rm c}}
\newcommand{\cohnc}{\coh^{\rm nc}}
\newcommand{\ccc}{\mathcal{A}}

\newcommand{\ar}{>}
\newcommand{\al}{<}

\newcommand{\GreedyAi}{\underline{A}^{\rm (i)}}
\newcommand{\GreedyAii}{\underline{A}^{\rm (ii)}}
\newcommand{\Greedythi}{\theta_{\rm (i)}}
\newcommand{\Greedythii}{\theta_{\rm (ii)}}
\newcommand{\meas}{{\mu}}

\newcommand{\opt}{^\star}

\newcommand{\hypoT}{{\mathfrak T}}

\DeclareMathOperator*{\argmax}{arg\,max}

\definecolor{nblue}{rgb}{0.06,0.3,0.73}
\definecolor{nblack}{rgb}{0,0,0}
\definecolor{nred}{rgb}{0.9,0.1,0.1}
\definecolor{nmagenta}{rgb}{0.7,0.0,0.3}
\definecolor{neditcolor}{rgb}{0.3,0.3,0.9}

\newcommand{\ie}{\emph{i.e.}}
\newcommand{\eg}{\emph{e.g.}}

\newcommand{\ea}{\emph{et al.}}

\begin{document}

\title{Greedy versus Map-based Optimized Adaptive Algorithms
\texorpdfstring{\\}{}
for random-telegraph-noise mitigation by spectator qubits}
\author{Behnam Tonekaboni}
\email{behnam.tfn@gmail.com}
\affiliation{Centre for Quantum Dynamics, Griffith University, Yuggera Country, Brisbane, Queensland 4111, Australia}
\author{Areeya Chantasri}
\email{areeya.chn@mahidol.ac.th}
\affiliation{Optical and Quantum Physics Laboratory, Department of Physics, Faculty of Science, Mahidol University, Bangkok, 10400, Thailand}%
\affiliation{Centre for Quantum Computation and Communication Technology (Australian Research Council), \texorpdfstring{\\}{} Centre for Quantum Dynamics, Griffith University, Yuggera Country, Brisbane, Queensland 4111, Australia}
\author{Hongting Song}
\email{shtfc@163.com}
\affiliation{Centre for Quantum Computation and Communication Technology (Australian Research Council), \texorpdfstring{\\}{} Centre for Quantum Dynamics, Griffith University, Yuggera Country, Brisbane, Queensland 4111, Australia}
\affiliation{Qian Xuesen Laboratory of Space Technology, China Academy of Space Technology,  Beijing 100094, China}%
\author{Yanan Liu}
\email{yanan.liu@griffith.edu.au}
\affiliation{Centre for Quantum Computation and Communication Technology (Australian Research Council), \texorpdfstring{\\}{} Centre for Quantum Dynamics, Griffith University, Yuggera Country, Brisbane, Queensland 4111, Australia}

\author{Howard M. Wiseman}
\email{h.wiseman@griffith.edu.au}
\affiliation{Centre for Quantum Computation and Communication Technology (Australian Research Council), \texorpdfstring{\\}{} Centre for Quantum Dynamics, Griffith University, Yuggera Country, Brisbane, Queensland 4111, Australia}

\date{\today}

\begin{abstract}
In a scenario where data-storage qubits are kept in isolation as far as possible, with minimal  measurements and controls, noise mitigation can still be done using additional noise probes, with corrections applied only when needed. Motivated by the case of solid-state qubits, we consider dephasing noise arising from a two-state fluctuator, described by random telegraph process, and a noise probe which is also a qubit, a so-called spectator qubit (SQ). We construct the  theoretical model assuming projective measurements on the SQ, and derive the performance of different measurement and control strategies in the regime where the noise mitigation works well. 
We start with the Greedy algorithm; that is, the strategy that always maximizes the data qubit coherence in the immediate future. We show numerically that this algorithm works very well, and find that its adaptive strategy can be well approximated by a simpler algorithm with just a few parameters. Based on this, and an analytical construction using Bayesian maps, we design a one-parameter ($\Theta$) family of algorithms. In the asymptotic regime of high noise-sensitivity of the SQ, we show analytically that this $\Theta$-family of algorithms reduces the data qubit decoherence rate by a divisor scaling as the square of this sensitivity. Setting $\Theta$ equal to its optimal value, $\Theta\opt$, yields the Map-based Optimized Adaptive Algorithm for Asymptotic Regime (MOAAAR). 
We show, analytically and numerically, that MOAAAR outperforms the Greedy algorithm, especially in the regime of high noise sensitivity of SQ. 
\end{abstract}
\maketitle

\section{Introduction}
Qubit decoherence from environmental noises is one of the major obstacles to building useful large-scale quantum computers~\cite{TemBra2017,Pre2018,EndBen2018,KanTem2019}. In the past years, a wide range of noise-mitigation techniques have been proposed to prolong the coherence time of computational qubits. Some of the techniques, such as the dynamical decoupling (DD), aim at removing average effects of noise continuously in time, via applying carefully designed controls onto qubits~\cite{Vio99,viola2003robust, biercuk2011dynamical, NgLid2011,SouAlv2011, MedCyw2012, PazLid2013, ZhaSou2014}. Other techniques, such as those that fit in the category of quantum error correction (QEC), are designed to monitor effects of noise and correct errors via information-redundancy encoding~\cite{Shor1995,Steane1996,Terhal2015}. 

However, both DD and QEC approaches require direct controls or measurements on the qubits of interest, which could in turn create more sources of noise. Therefore, in the particular case where it is possible to keep the computing (\emph{data}) qubits well-isolated from their environment, new techniques, that can both correct noise effects while minimizing direct contact to the data qubits, are required.
Such techniques would demand an additional entity, \eg, a sensor or a probe, in proximity of the data qubits, that can sense the problematic noise and any correction to the data qubits can be applied only when needed. 

In this paper, we are interested in the recently proposed idea of spectator qubits (SQs)~\cite{GupEdm2020, MajAnd2020,SinBra2022}, which are qubit-type probes that are assumed to be much more sensitive to target noise than data qubits and can be easily measured. This could be applicable in solid-state quantum computing, where different types of qubits with very different properties can co-exist. For example, the data qubits could be the spin qubits with low sensitivity to noise~\cite{HanKou2007, PlaTan2012, MorPla2020} and SQs could be quantum dots nearby with an ability of high-efficiency readout~\cite{MorPla2010, KeiHou2019, BluPan2022}. Assuming a well-isolated spin-donor qubit in silicon substrate, there will still be lingering effect from low-frequency noise, such as charge noise, which can affect qubits at different locations simultaneously~\cite{CulZim2013,BerKei2014} and is considered difficult to remove~\cite{CulHu2009}. 
Therefore, we are interested in constructing a theoretical model using the data-spectator setup to investigate a plausible optimal regime for the noise mitigation. It turns out that the solution is not trivial, even with a simple regime with single data and spectator qubits.

We have presented our most important findings on this problem in a Companion Letter (CL)~\cite{PRL}.  Specifically, we showed that by performing projective measurements on the SQ at certain times and in certain bases, that are chosen adaptively in an optimized manner, the decoherence of the data qubit can be suppressed by an amount which is quadratic in the SQ sensitivity. This result holds under ideal conditions, and the optimal suppression is quite non-trivial, as it is better than that can be achieved by simply maximizing the data qubit coherence in each, arbitrarily small, time step (Greedy algorithm). In this paper, we give the detailed derivations of the results in the CL~\cite{PRL}, and explore in much more depth the comparison between the Greedy algorithm and the optimized algorithm. But to explain more precisely the contents of this paper, it is necessary first to give some more theoretical background.

\begin{figure}[t!]
\includegraphics[width=0.95\columnwidth]{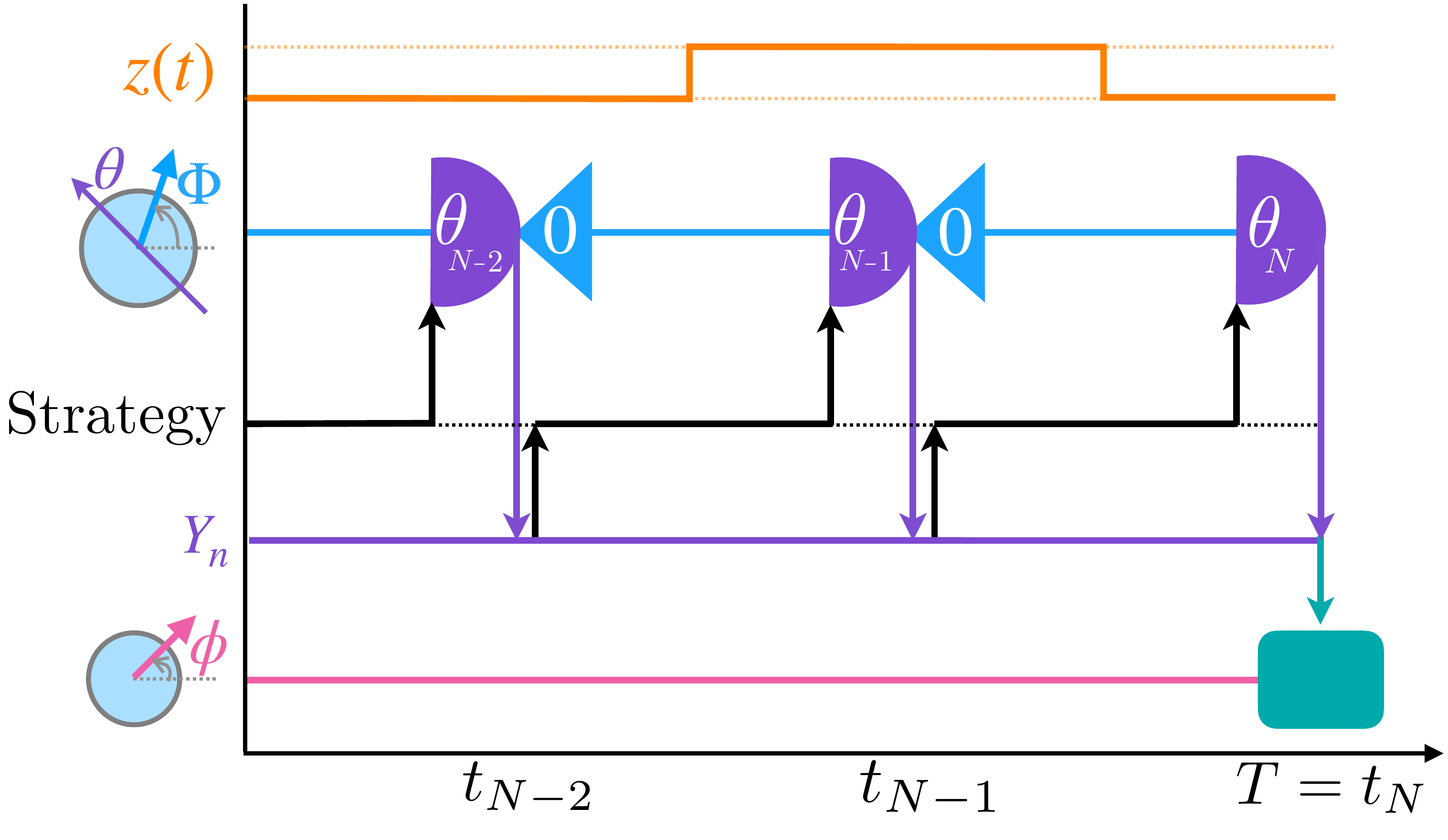}
\caption{Schema for the adaptive measurement strategy on SQ. The orange line shows the dynamics of RTP which jumps between two values indicated with pale dashed orange lines. The blue line represents the SQ which is reset (triangle) after each measurement (purple semicircle). Storing measurement results are represented by purple arrows toward the measurement record line $Y_n$ (purple). These measurement records are used via the adaptive strategy (black arrows) to decide on the next measurement time $t_n$ and next measurement angle $\theta_n$. Lastly, at final time $T$, the data qubit (pink) is corrected using information $Y_N$, (Teal arrow downward).
\label{fig1}}
\end{figure}

We model the decoherence of the data qubit as being caused by low-frequency phase noise arising from the random motion of a trapped charge caused by impurities, as is common in the experimental solid-state qubit setting~\cite{ZorAhl1996,PalFao2002,fujisawa2000charge, CulHu2009, PalGal2014}. Specifically, we model the noise $z(t)$ as a two-state fluctuator~\cite{ItaTok2003, BerGal2006, BerGal2009, MuhDeh2014}, with values $\pm1$ as shown in Fig.~\ref{fig1}. This can be mathematically described by the random telegraph process (RTP), with two flip rates: $\gu$ and $\gd$~\cite{Gar85, jacobs2010stochastic}. To monitor the noise, a SQ is placed nearby, experiencing the same phase noise, but with the higher sensitivity.  This is described by a total Hamiltonian of both qubits,
\begin{equation}\label{eq:totalHamiltonian}
    \hat{H}_{\rm tot} = \frac{\kd}{2} \hat{\sigma}_z^{\rm d} \, z(t) + \frac{\ks}{2} \hat{\sigma}_z^{\rm s}\, z(t).
\end{equation}
Here $\kd$ and $\ks$ are the data qubits' and SQ's noise sensitivities and  $\hat\sigma_z^{\rm d,s}$ are their respective $z$-Pauli matrices. Also, we are working in a rotating frame for both qubits (their frequencies need not be the same or even similar) and with units where $\hbar=1$.

We assume that, in order to obtain information about the noise $z(t)$, the SQ can be measured after arbitrary waiting times $\{ \tau_1, \tau_2, ... \}$, with arbitrary angles $\{\theta_1, \theta_2, ... \}$,  and re-prepared in an arbitrary state $\ket{\theta}^{\rm s}$. Here $\theta$ always  indicates an angle on the equator of the Bloch sphere, defining either a Pauli observable, $\hat\sigma_{\theta}^{\rm s} \equiv \cos\theta\,  \hat \sigma_x^{\rm s} + \sin\theta\, \hat\sigma_y^{\rm s}$, or a state $\ket{\theta}^{\rm s}$ which is the $+1$ eigenstate of $\hat\sigma_{\theta}^{\rm s}$.   The resulting string of measurement readouts can then be used in estimating the data qubit phase correction at some final time $T$,  which is when the data qubit is required. The goal is to maximize the resultant coherence of the data qubit, averaging over all possible realizations of the noise. 

Given this model and goal, the question is how one should choose the $\tau_n$'s, the $\theta_n$'s, and the final phase correction. In this work, we apply the {\em separation principle} from control theory, and the structure of the reward function (which defines the goal), to greatly simplify the problem. We show that the information necessary for all future controls can be stored in a complex 2-vector and propagated from one measurement to the next by a \emph{Bayesian map}, a complex $2\times 2$ matrix.  We then use the maps to numerically find the locally optimal---also known as \emph{Greedy}---algorithm.

Locally optimal algorithms, in general, are not always globally optimal, and we find that to be the case for the problem here. We are interested in the regime where the SQ approach is most useful, to wit, 
\begin{equation}\label{eq:regime}
T^{-1}, ~ \kd \ll \gud \ll \ks.
\end{equation} 
We analyze how the Greedy algorithm works in this regime, and this suggests a simple, single parameter ($\Theta$) family of models where $\tau_n = \Theta/K$ and $\theta_n =\pm \Theta$.  Here the sign of $\theta_n$ is determined \emph{adaptively}, and  reflects the algorithm's best estimate of $z(t_n)$, the current state of RTP. In the asymptotic regime (\ref{eq:regime}), we can solve analytically for the performance of this family of algorithms. Choosing the optimal parameter value $\Theta\opt$ yields a measurement and control strategy that is plausibly optimal, and which we call MOAAAR (Map-based Optimized Adaptive Algorithm in the Asymptotic Regime). This can reduce the decoherence rate of the data qubit, relative to the no control rate, by the multiplicative factor
\begin{align} \label{multfac}
    H\opt \,(\mg/\ks)^2, \text{ where } H\opt \approx 1.254,
\end{align}
where $\mg := (\gu + \gd)/2$. That is, as mentioned above, the data qubit lifetime enhancement offered by the MOAAAR algorithm scales like the square of the sensitivity of the SQ. While the same quadratic scaling is observed for the Greedy algorithm, we find that the prefactor $H\opt$ for the decoherence rate with MOAAAR is smaller than that from Greedy.

The structure of this paper is as follows.  First, in Sec. \ref{sec:RTP}, we briefly discuss the physics and mathematical description of the RTP. We then, in Sec. \ref{sec:Dephasing}, calculate the dephasing of the data qubit in the absence of any control, and formulate it as a matrix product, which will be useful later. We end this section by looking at how, in general, additional information can be used to improve the coherence in the data qubit. In Sec.~\ref{sec:SQ}, we construct the setup and measurement on the SQ and discuss the notion of adaptive measurement strategy in the context of the SQ. Then, in Sec.~\ref{sec:baymap}, we use Bayesian statistics and formulate the expected coherence using the gathered information via the SQ. We write the Bayesian map as a matrix product which updates a \emph{coherence vector} after each measurement on the SQ. We also introduce a phase space formalism which includes all the sufficient information. We then optimize our measurement strategy using the locally optimization algorithm (Greedy) in Sec.~\ref{sec:Greedy} and the global optimization approach (MOAAAR) in Sec.~\ref{sec:MOAAAR}. In Sec.~\ref{sec:CF}, we compare these two optimization strategies by constructing a closed-form formula for each that applies even outside the asymptotic regime. We conclude our work with a discussion of future research directions in Sec.~\ref{sec:conclusion}.

\section{Charge noise and random telegraph process (RTP)} \label{sec:RTP}
Charge noise is one of the causes of decoherence in a wide range of physical systems including superconducting qubits~\cite{christensen2019anomalous}, quantum dots~\cite{freeman2016comparison}, Nitrogen vacancy centers~\cite{kim2015decoherence}, trapped ions~\cite{brownnutt2015ion}, and semiconductors~\cite{ItaTok2003, fujisawa2000charge, GalAlt2006, kuhlmann2013charge, CulHu2009, BerGal2009}. The noise is often described by stochastic motion of electrons or holes trapped at impurities in the devices, which can be mathematically modeled as a random telegraph process (RTP), whose value jumps between two values~\cite{Gar85, jacobs2010stochastic}. Consider the RTP noise parameter $z_t := \rtp(t)$, which can be either $+1$ or $-1$. It makes a jump from $\rtp=-1$ to $\rtp = +1$ with rate $\gu$, and $\rtp=+1 \rightarrow -1$ with rate $\gd$. Define the probability vector 
\begin{equation} \label{defPt}
    \underline{P}_t :=
    \left(\begin{matrix}
    \wp(z_t=+1) \\
    \wp(z_t=-1)
    \end{matrix}\right),
\end{equation}
where $\wp(z_t=\pm 1)$ is the probability of $z_t =\pm 1$ at time $t$. Then, with the jump rates defined above, one may write a (classical) master equation for vector $\ul{P}_t$ as
\begin{equation} \label{eq:RTPmaster}
    \underline{\dot{P}}_t = \mathbf{J} \underline{P}_t,
\end{equation}
where $\mathbf{J}$ is the jump rate matrix,
\begin{equation} \label{eq:J_matrix}
    \mathbf{J}=\left( 
    \begin{matrix}
        -\gd  & +\gu  \\ +\gd & -\gu
    \end{matrix} 
    \right).  
\end{equation}
If the RTP evolves for a duration of $\tau$, between times $t'$ and $t'' = t' +\tau$, the master equation~\eqref{eq:RTPmaster} has a solution of the form $\underline{P}_{t''}= e^{\mathbf{J} \tau} \underline{P}_{t'}$. One may  rearrange the matrix exponential and write the solution as
\begin{equation}\label{eq:Pt_RTP}
    \underline{P}_{t''}= \left( \mathbf{I}+ \frac{1-e^{-2\mg \tau}}{2\mg}\mathbf{J} \right) \underline{P}_{t'}.
\end{equation}
where $\mg=(\gu+\gd)/2$ as before, and $\mathbf{I}$ is the identity matrix.
It is worth mentioning that when $\tau \rightarrow \infty$, the probability vector $\underline{P}_{t''}$ converges to its steady state given by
\begin{equation}\label{eq:Pss}
\ul{P}_{\rm ss} = \frac{1}{2 \mg} \left( \begin{matrix} \gu\\ \gd \end{matrix} \right).
\end{equation}
This will be used throughout the paper as the state of RTP when no information about the noise is available, such as at the initial time. 

We model the effect of the charge noise on the data and spectator qubits by the Hamiltonian in Eq.~\eqref{eq:totalHamiltonian}.  From this, it is apparent that, under free evolution in some time interval $(t',t'')$, the qubits' accumulated phase (angle of rotation around the $\s{z}$ axis) will be proportional to the integral of the noise in that interval, 
\begin{equation} \label{IRTP}
	x := \int_{t'}^{t''}\! \!\rtp(s){\rm d}s.
\end{equation} 

To determine the dephasing caused by this random accumulated phase  (more on this in the next section) it is useful to calculate the  average, $\int \wp(x)e^{ikx}\dd x$, of the exponential factor, $e^{ikx}$, for arbitrary $k$. Actually, it is even more useful to calculate the integral $\int e^{i k x} \wp(x,z_{t''}|z_{t'}) \dd x$ for arbitrary $k$, $z_{t''}$ and $z_{t'}$. This is, of course, just the Fourier transform of $\wp(x,z_{t''}|z_{t'})$, and can be calculated using \erfand{eq:Pt_RTP}{IRTP} as  
\begin{align}\label{eq:FR-pofX}
\int &e^{i k x} \wp(x,z_{t''}|z_{t'}) \dd x = \exp(-\bar{\gamma}\tau)  \nonumber \\
& \times 
    \begin{cases}
        \cosh\left(\dfrac{\mc}{2} \tau \right)- s \dfrac{\nc}{\mc}\sinh\left(\dfrac{\mc }{2}\tau\right),    & \text{for } z_{t'}=z_{t''} = s\\
        \dfrac{2 \gu}{\mc} \sinh\left(\dfrac{\mc}{2}\tau\right),  & \text{for } z_{t''} = - z_{t'} = +1\\
        \dfrac{2 \gd}{\mc} \sinh\left(\dfrac{\mc}{2}\tau\right),  & \text{for } z_{t''} = - z_{t'} = -1 
    \end{cases}
\end{align}
where $s \in \{ -1, +1\}$ and
\begin{subequations}
    \begin{eqnarray}
        \mc(k)&=&\sqrt{(\gd+\gu)^2-4i k (\gd-\gu) -4 k^2},\\
        \nc(k)&=&(\gd-\gu) -2i k,
    \end{eqnarray}
\end{subequations}
are functions of the Fourier variable $k$ (See appendix~\ref{app:H} and Refs.~\cite{BerGal2009,DanBen2018} for the details of its derivation). In the following sections, we will use these results to derive the statistics of data and spectator phases, conditioned on measurement results and/or qubit controls. 

\section{Data qubit dephasing} \label{sec:Dephasing}
In this section, we define the data qubit's coherence and analyze its dephasing due to the RTP noise, both for the cases with no control and with the control phase correction. For the no-control case, we show that the qubit's coherence can be written as a matrix equation involving a complex two-by-two  matrix, ${\bf H}$. The matrix representation for the no-control coherence will be extended to the version including measurement strategies and readouts of the spectator qubit in Section~\ref{sec:baymap}.

\subsection{Formulation of data qubit coherence} \label{ssec:formdqc}

From the total Hamiltonian in Eq.~\eqref{eq:totalHamiltonian}, if we include the control (phase correction) applied at the final time, $t=T$, we can write the Hamiltonian of the data qubit as
\begin{equation}\label{eq:H_DataNoise}
    \hat{H}_{\rm d}(t) = \frac{\kd}{2}\,  \hat{\sigma}_z^{\rm d}\,  z(t) + \hat H_{\rm ctrl}\delta(t-T).
\end{equation}
The first term of the Hamiltonian describes a stochastic phase rotation around the $z$-axis of the data qubit's Bloch sphere. To evaluate the data qubit's decoherence due to this term, we consider a qubit's density matrix in the $z$-basis affected by a random noise,
\begin{align}\label{eq:dqubitmatrix}
\rho^{\rm d}(\noise) = \, & \rho_{++}
\ket{+1}_z^{\rm d}\bra{+1} + \rho_{+-} e^{-i \phi(\noise)}\ket{+1}_z^{\rm d}\bra{-1}  \nonumber \\
& + \rho_{-+}e^{+i\phi(\noise)}\ket{-1}_z^{\rm d}\bra{+1} + \rho_{--}\ket{-1}_z^{\rm d}\bra{-1},
\end{align}
where $\ket{+1}_z^{\rm d}$ and $\ket{-1}_z^{\rm d}$ are the two eigenstates of $\hat\sigma_z^{\rm d}$, \ie, $\hat \sigma_z^{\rm d} \ket{\pm1}_z^{\rm d} = \pm \ket{\pm1}_z^{\rm d}$. The terms $\rho_{++},\rho_{--}, \rho_{+-},\rho_{-+}$ are density matrix elements of the data qubit's state at the initial time ($t=0$), before affected by the noise. The Hamiltonian Eq.~\eqref{eq:H_DataNoise} thus introduces an additional phase, denoted by $\phi = \phi(\noise)$, which is a function of a set of variables, $\noise$, which can include both the noise that affects the data qubit's phase, and the control applied to it. 

Since the value of $\noise$ varies from run-to-run, we are interested in the average final state of the qubit, i.e., $\langle \rho^{\rm d}(\noise)\rangle_{\noise}$,
%
where the subscript $\noise$ indicates that this is the random variable with respect to which the average is taken.   The data qubit's final purity is completely determined by the magnitude of the off-diagonal elements of Eq.~\eqref{eq:dqubitmatrix}, which gives $|\rho_{+-}\left\langle e^{i\phid(\noise)}\right\rangle_\noise|$. Since the element $\rho_{+-}$ is not affected by the noise, 
we can thus define the quantity to be maximized, which we call the \emph{coherence}, as  
\begin{equation} \label{eq:Coh}
\coh := \left|\left\langle e^{i\phid(\noise)}\right\rangle_\noise\right|.
\end{equation}
The expected average of random phases typically leads to the value of coherence being less than one (its maximum value) and can be as low as zero for the case of  uniformly distributed phases in $[0,2\pi)$. The reduction of coherence from random phases is called qubit dephasing. It is also convenient later to define a \emph{complex coherence},
\begin{equation}
\ccc := \left\langle e^{i\phid(\noise)}\right\rangle_\noise,
\end{equation}
without the absolute value, where $\coh = |\ccc|$. The information of the argument of $\ccc$ will be useful later, in estimating the phase information of the data qubit.

Let us define capital-letter variables for a total accumulated noise, from time $t=0$ to any time $t$,
\begin{equation}\label{eq:X}
	X := \int_{0}^{t}\! \!\rtp(s){\rm d}s,
\end{equation}
and for a total information, 
\begin{equation}
    Y := \text{all acquired information about } X,
\end{equation}
which is from any measurements up to the time $t$. In the following subsections, we simplify the representation of a data qubit's state in Eq.~\eqref{eq:dqubitmatrix} to showing only its phase. Let us define a state of the data's qubit phase as 
\begin{equation}\label{eq:equatorial_state}
    \ket{\phi}^{\rm d} := \frac{1}{\sqrt{2}} \left(\ket{+1}_z^{\rm d} + e^{i \phi} \ket{-1}_z^{\rm d} \right).
\end{equation}
Then, the initial phase of the data qubit is $\ket{\phi = 0}^{\rm d}$ (a zero noisy phase) and we consider two cases: the no-control case and the case with the noise correction. For the former, $\hat H_{\rm ctrl} = 0$, the total data noisy phase will be a function of noise, \ie, $\phi(\noise) = \phi(X)$. For the latter, we introduce a phase-correction control, $\hat H_{\rm ctrl} = - i c(Y) \hat{\sigma}_z^{\rm d}/2$, which describes an instantaneous rotation of the data qubit's state around the $z$-axis with an angle $-c(Y)$ that depends on the information, $Y$. In this case, the data qubit's phase will be a function of both the noise and the control, \ie, $\phi(\noise) = \phi(X, Y)$. We will show how to compute the data qubit's coherence for both cases in the following subsections.

\subsection{Decay of no-control (nc) coherence}\label{sec:noctrl}
For the `no-control' (nc) case, $\hat{H}_{\rm ctrl} = 0$, the data qubit only evolves stochastically due to the RTP noise. Given the Hamiltonian, Eq.~\eqref{eq:H_DataNoise}, the data qubit's state (represented by its noisy phase) at any time $t$, becomes
\begin{equation}
\ket{\phi(\noise)}^{\rm d} = \exp\left(- i \tfrac{\kappa}{2}\hat \sigma_z^{\rm d} \! \int_0^t \! \dd s\,  z(s) \right) \ket{\phi = 0}^{\rm d} = \ket{ \kappa X}^{\rm d},
\end{equation}
which means that the data noisy phase, $\phi(\noise) = \phi(X) = \kappa X$, is simply proportional to the total accumulated noise. Following Eq.~\eqref{eq:Coh}, replacing $\noise$ with $X$, we can calculate the no-control coherence as
\begin{align}
    \cohnc &:= \left| \left\langle e^{i \phi(X)} \right\rangle_X \right| = \left| \int e^{i\kappa X} \wp(X) \dd X \right|, \label{eq:noctrldef}
\end{align}
and its no-control complex coherence
\begin{align}\label{eq:noctrlAcal}
    \ccc^{\rm nc} &:= \int e^{i\kappa X} \wp(X) \dd X,
\end{align}
where $\wp(X)$ is the probability density function of the accumulated noise. 

Since the variable $X$ is an accumulated RTP noise, it is convenient to write the probability function $\wp(X)$ as marginalized over the initial $z_0 = z(t = 0)$ and the end-point $z_t = z(t)$ of RTP, \ie,
\begin{equation}
    \wp(X) = \sum_{z_t} \sum_{z_0} \wp(X,z_t|z_0) \wp(z_0).
\end{equation}
Substituting the above equation into Eq.~\eqref{eq:noctrldef}, we find that the no-control coherence is
\begin{align}\label{eq:noCtrlIntegralform}
    \cohnc
    &= \left|\sum_{z_t} \sum_{z_0} \left[\int e^{i\kappa X} \wp(X,z_t|z_0) \dd X \right] \wp(z_0) \right|,
\end{align}
where the term inside the square brackets is the Fourier form as we defined in Eq.~\eqref{eq:FR-pofX}. Now, we see that Eq.~\eqref{eq:noCtrlIntegralform} can be written as a map-based equation by defining a vector 
\begin{equation}\label{eq:initialA}
\underline{A}_0 := \underline{P}_{t=0} = \left( \begin{matrix} \wp(z_0 = +1) \\ \wp(z_0 = -1) \end{matrix}\right),
\end{equation}
using the definition in Eq.~\eqref{defPt} and defining matrix elements,
\begin{equation}\label{eq:H_element_def}
H_{z_0}^{z_t}(t,\kappa) := \int e^{i\kappa X} \wp(X,z_t|z_0) \dd X.
\end{equation}
Since $z_0, z_t \in \{ -1 , +1\}$, this defines a two-by-two matrix,
\begin{equation}\label{eq:Hmatrix}
{\bf H}(t,\kappa):= \left( \begin{matrix}H^{+1}_{+1}(t,\kappa) &  H^{+1}_{-1}(t,\kappa) \\[5pt] H^{-1}_{+1}(t,\kappa) &  H^{-1}_{-1}(t,\kappa) \end{matrix} \right).
\end{equation}
Thus, the coherence in Eq.~\eqref{eq:noCtrlIntegralform} is then written as
\begin{equation}\label{eq:Coh_nc_A}
    \cohnc = \left|\underline{I}^\top 
    {\bf H}(t,\kappa) \underline{A}_0 \right|,
\end{equation}
where $\ul{I}^\top = (1,1)$. We can calculate the coherence Eq.~\eqref{eq:Coh_nc_A} for the no-control case analytically (see details in the Appendix~\ref{app:H}). As we will see later, reformulating the coherence as matrix equations simplifies  analytical and numerical calculations for optimal controls, particularly when measurements on the spectator qubit are involved.

To simplify the discussion later, let us also define a complex 2-dimensional coherence vector for the no-control case,
\begin{equation}\label{interpAnc}
\underline{A}^{\rm nc} =  {\bf H}(t,\kappa) \underline{A}_0 =  \left(\begin{matrix}
    \wp(z_t=+1) \ccc_{|z_t=+1} \\
    \\
    \wp(z_t=-1) \ccc_{|z_t=-1}
    \end{matrix}\right).
\end{equation}
Here we have introduced a complex {\em conditional} coherence,  
\begin{equation}
\ccc_{|z_t} := \int e^{i\kappa X} \wp(X|z_t ) dX = \langle e^{i\kappa X} \rangle_{X|z_t},
\end{equation}
of the data qubit. This quantity is conditioned on (hypothetically) finding out the value of the RTP $z_t$ at time $t$;  later we will condition on other data, that may be hypothetical or actual or a combination of both. We can get back the no-control complex coherence in Eq.~\eqref{eq:noctrlAcal} by
\begin{equation}\label{eq:nc-ccc}
\ccc^{\rm nc} = \ul{I}^\top \ul{A}^{\rm nc} =  \ul{I}^\top {\bf H}(t,\kappa) \ul{A}_0 =  \sum_{z_t} \wp(z_t) \ccc_{|z_t},
\end{equation}
where the sum is over $z_t \in \{ -1,1\}$, showing the connection among matrix equations and the complex coherence. The absolute value of this complex coherence is  the no-control coherence as shown in Eq.~\eqref{eq:Coh_nc_A}.

\subsection{Decay of coherence in asymptotic regime}
Here, we show that, in the long-time limit, $t\rightarrow \infty$, and the asymptotic regime, where $\kd$ is much smaller than the RTP jump rate, \ie, $\kd \ll \mg$, the no-control coherence $\cohnc$ decays exponentially. We start with the no-control complex coherence, $\mathcal{A}^{\rm nc}$ in Eq.~\eqref{eq:noctrlAcal}, and calculate the ratio of the change over an infinitesimal time and its value. That is, the ratio is defined as
\begin{equation}\label{eq:Lambdat}
     \Lambda_t := \frac{\dd \mathcal{A}^{\rm nc}\big/ \dd t}{\mathcal{A}^{\rm nc}},
\end{equation}
which in general can be a time-dependent function denoted by the subscript $t$. We use the RTP steady state as the initial vector $\ul{A}_0  = \ul{P}_{t=0} = \ul{P}_{\rm ss}$ in~\eqref{eq:Pss}, and expand the right-hand side of Eq.~\eqref{eq:Lambdat} to second order in $\kappa$ and find
\begin{align}
     \Lambda_t =  i\left(\gu-\gd\right)\frac{\kd}{2\mg} +  \left(e^{-2t\mg} - 1 \right) \frac{\kd^2 \bg}{2 \mg^2} +  \mathcal{O}(\kd^3),
\end{align}  
where $\bg \equiv 2\gu\gd/(\gu+\gd)$ is the harmonic mean of $\gu$ and $\gd$. Taking the long-time limit, \ie, $t\rightarrow \infty$, of above the expansion, we find that the factor $\Lambda_t$ approaches a constant complex value,
\begin{align} \label{asymcomdec}
    \Lambda = \lim_{t\rightarrow\infty} \Lambda_t 
   &= -\Gamma^{\rm nc} + i \Omega^{\rm nc},
\end{align}
where, in the last line, we have defined the average no-control frequency (phase drift rate) and decay rate (phase diffusion rate) as 
\begin{equation}
    \Gamma^{\rm nc} := \frac{\kd^2 \bg}{2 \mg^2} \quad \text{and} \quad \Omega^{\rm nc}:=\frac{\kd (\gu-\gd)}{2\mg}.
\end{equation}

As a result, we substitute $\Lambda_t$ in Eq.~\eqref{eq:Lambdat} with the constant $\Lambda$ and solve the differential equation for the no-control complex coherence to get
\begin{equation}
    \mathcal{A}^{\rm nc}(t) = \mathcal{A}^{\rm nc}(0) \exp\left(-\Gamma^{\rm nc}t +i \Omega^{\rm nc}\, t \right),
\end{equation}
which has the initial condition $\mathcal{A}^{\rm nc}(0)=\ul{I}^\top \ul{P}_{\rm ss} = 1$. We note that the complex coherence has its imaginary contribution only when the RTP has asymmetric jump rates. That is, for the symmetric case, $\gu = \gd$, we have $\Omega^{\rm nc}=0$. Finally, since $\cohnc = |\ccc^{\rm nc}|$, we obtain the no-control coherence in the asymptotic regime,
\begin{equation}
\label{uncond_exp}
\cohnc(t)  = \exp(- \Gamma^{\rm nc} t).
\end{equation}
In the CL~\cite{PRL} (Figure 1), we show the comparison between   \erf{uncond_exp} and the exact coherence in \eqref{eq:Coh_nc_A}. They differ significantly at short times, this is expected since \erf{asymcomdec} is only valid in the limit $t \gg 1/\mg$. Moreover, even by $t=1/\mg$ one can see that their decay is almost identical.

\subsection{Noise correction and optimal choice of control}\label{sec:wthctrl}

The data qubit's coherence can be improved, from the no-control case, by utilising additional information gathered by a SQ. Before discussing how to obtain this information (in the following section), we will show how additional information, in general, helps increase the coherence of the data qubit. 

Let us assume that we have access to an additional information $\info$ about the unknown accumulated noise $X$. We can use the information to compute an appropriate phase correction, $-c(\info)$, and apply it to the data qubit as described in Sec.~\ref{ssec:formdqc}, so that $\phi(\noise) = \phi(X,\info) = \kd X -c(\info)$. Given the general form of coherence in Eq.~\eqref{eq:Coh}, and a control function $c(\bullet)$, we write a phase-corrected coherence as
\begin{equation} \label{eq:CtrlCoh}
    \coh_{c(\bullet)} := \Big|\big\langle e^{i [\kd X - c(\info)]}\big\rangle_{X,Y} \Big|,
\end{equation}
which is similar to Eq.~\eqref{eq:noctrldef}, but now the expected average is over all possible values of $X$ and $Y$. 

If $X$ were known, we could trivially find that $c(\info) = \kappa X$ is a perfect correction that completely removes the phase error from the noise and the data qubit's coherence is maximized at one. However, the noise is actually unknown and we only have its indirect information from $Y$. Therefore, the main task is to find the control function, $c(\bullet)$, that maximizes the coherence, $\coh_{c(\bullet)}$ in Eq.~\eqref{eq:CtrlCoh}. To accomplish this, we first write the expected average in Eq.~\eqref{eq:CtrlCoh} as an explicit summation over values for $Y$ (assumed discrete) and an integral over the continuous value of $X$ with appropriate probability weights as
\begin{align} \label{eq:CtrlCoh_expand}
    \coh_{c(\bullet)} &= \bigg| \sum_{Y} \int e^{i \kd X}  e^{-i c(Y)} \wp(Y) \wp(X|Y) \dd X \bigg| \notag \\
    &= \bigg| \sum_{Y} e^{-i c(Y)} \wp(Y) \big\langle e^{i \kd X} \big\rangle_{X|\info} \bigg| \notag \\
    & \le\, \sum_{Y}  \wp(Y) \left|\big\langle e^{i \kd X} \big\rangle_{X|\info}\right|,
\end{align}
where, in the second line, we have rearranged terms into a conditional expected average, $\big\langle e^{i \kd X} \big\rangle_{X|\info}$. In the last line, 
the inequality is,  in general, 
a strict inequality. However, it becomes an equality when 
\be\label{eq:optctrl}
\cd(\bullet) =\arg \big\langle e^{i \kd X} \big\rangle_{X|\bullet},
\ee
which thus maximizes the phase-corrected coherence $\coh_{c(\bullet)}$. Given the formula of the coherence in Eq.~\eqref{eq:CtrlCoh}, we can regard 
\erf{eq:optctrl} as defining the {\em optimal} 
estimator of the real phase, $\phi = \kappa X$, for the purpose of the final phase correction. Indeed, it is just the mean  under \emph{circular statistics}~\cite{fisher1995statistical}. We indicate the maximized \emph{control coherence} defined as  
\begin{equation} \label{eq:maxcoh}
    \cohc := \coh_{c(\bullet) = \arg \ccc_{|\bullet}}  = \sum_\info \wp(\info) \left|\ccc_{|\info} \right|.
\end{equation}
Here we are  again considering the  conditional complex coherence, 
\begin{equation}\label{eq:calAcondY}
    \ccc_{|\info} := \langle e^{i \kd X} \rangle_{X|\info} = \int e^{i \kd X} \wp(X|Y) \dd X,
\end{equation}
this time conditioned on the information $Y$.

\section{Spectator qubit (SQ)\texorpdfstring{\\}{} for noise sensing}
\label{sec:SQ}
In our model, the information $Y$ used in the phase correction can be obtained by measuring the SQ which senses the same RTP noise affecting the data qubit. The Hamiltonian of the SQ, \begin{equation}\label{eq:SQHamiltonian}
    \hat H_{\rm s}(t) = \frac{\ks}{2} \hat \sigma_z^{\rm s} \, z(t),
\end{equation}
is identical to that of the data qubit, but with $\ks$ as the SQ's noise sensitivity in place of the much smaller $\kd$ for the data qubit. Thus the SQ's Bloch vector also  rotates around the $z$-axis. We assume that we have complete control over the SQ. To maximize its usefulness as a noise probe, we take it be reset to an equatorial state after each measurement. Thus its state can always be written as
\begin{equation}\label{eq:specqbstate}
    \ket{\phis}^{\rm s} = \frac{1}{\sqrt{2}} \left(\ket{+1}_z^{\rm s} + e^{i \phis} \ket{-1}_z^{\rm s} \right),
\end{equation}
given the eigenstates $\ket{\pm1}_z^{\rm s}$ of the observable $\hat \sigma_z^{\rm s}$, \ie, $\hat \sigma_z^{\rm s} \ket{\pm1}_z^{\rm s} = \pm \ket{\pm1}_z^{\rm s}$. The definition in Eq.~\eqref{eq:specqbstate} should be compared to Eq.~\eqref{eq:equatorial_state} for the data qubit.

\subsection{SQ's measurement and likelihood function}
To acquire information about the RTP noise, we probe the SQ at various times throughout the process. Let us denote the measurement times by 
\begin{equation}\label{eq:measurementtime}
t_1, t_2, \dots, t_n, \dots , t_{N-1}, t_N = T,    
\end{equation}
where $T$ is the final time that the phase correction is applied on the data qubit. Given the SQ's Hamiltonian~\eqref{eq:SQHamiltonian}, the integrated noise over time intervals of  
duration 
\begin{equation}
\tau_n :=t_n - t_{n-1}
\end{equation}
can be independently probed. Without loss of generality, we initialize the SQ at the zero-phase state, $\ket{\phis=0}^{\rm s}$, at the initial time, $t_0 = 0$, and after the $n$th measurement, for $n \in \{1,N-1\}$. 

Just before the $n$th measurement, at  time $t_n$, the SQ's state is given by $\ket{\phis(t_n)}^{\rm s}$, where
\begin{equation}\label{eq:SQphase}
    \phis(t_n) = K \!\! \int_{t_{n-1}}^{t_n} \!\dt ~ z(t) = K (X_n- X_{n-1}) := K x_n.
\end{equation}
Here we have defined the  integrated noise up to time $t_n$,
\begin{equation}
X_n := \int_0^{t_n} \!\! z(s)\dd s,
\end{equation}
similar to Eq.~\eqref{eq:X}, and defined $x_n:= X_n- X_{n-1}$, similar to Eq.~\eqref{IRTP}. We also define $z_n:=z(t_n)$ for the RTP at the measurement time $t_n$.

We assume that the measurement of the SQ at time $t_n$ is projective. Given the parameter of interest, $\Phi(t_n)$, the optimal  observable to measure will be of the form  
\begin{align}\label{eq:thetaop}
    \hat{\theta}_n := { \mathbb{1}}- \ket{\theta_n}^{\rm s}\bra{\theta_n}.
\end{align}
Here  $\theta_n$ is the chosen measurement angle and   $\ket{\theta_n}^{\rm s} = (1/\sqrt{2}) \left(\ket{+1}_z^{\rm s} + e^{i \theta_n} \ket{-1}_z^{\rm s} \right)$, and ${ \mathbb{1}}= \ket{+1}_z^{\rm s}\bra{+1} + \ket{-1}_z^{\rm s}\bra{-1}$ is the identity operator. The outcomes of the projective measurement, 
\begin{equation}
y_n\in\{0,1\},
\end{equation}
are associated with projecting the SQ's state to the two eigenstates, $\{\ket{\theta_n}^{\rm s},\ket{\theta_n+\pi}^{\rm s}\}$, respectively. 
We define $\hat{\theta}_n$, Eq.~\eqref{eq:thetaop}, in this way so that
the outcomes have a significance as null and non-null outcomes, respectively, for a natural choice of $\theta_n$, as we will see in Sec.~\ref{subsec:Greedy_result}. 

Given the SQ's state, $\ket{\phis(t_n)}^{\rm s} = \ket{K x_n}^{\rm s}$,  Born's rule gives the probability of outcome $y_n$ as 
\begin{align} \label{eq:forwardP}
      \wp(y_n|\theta_n, x_n) &:= \big| \, ^{\rm s}\langle \theta_n+\pi \, y_n | \ks x_n \rangle^{\rm s}\, \big|^2 \notag \\
        &=  y_n + (-1)^{y_n}\cos^2\sq{\half (\theta_n - \ks x_n ) }  .
\end{align}
In the context of Bayesian estimation (see next section), this is the \emph{likelihood function},   expressed slightly differently as  Eq.~(9) of the CL~\cite{PRL}.

\subsection{Adaptive strategies and the separation principle} \label{sec:ASASP}

Say one has just measured and reprepared the SQ at time $t_{n-1}$. 
There are two real parameters that define the next measurement on it: the measurement angle, $\theta_{n}$, and the waiting time, $\tau_{n}$. Let us combine them and define---for any $n$---a \emph{measurement setting}
\begin{equation}
    \meas_n:=\{ \theta_n, \tau_n \},
\end{equation}
and define a \emph{measurement strategy} as
\begin{equation} \label{deffrakS}
    {\mathfrak S} = \{ \meas_{n} : n \in \{1,\cdots,N\}\}.
\end{equation}
The aim of our work is to search for measurement strategies, ${\mathfrak S}$, that maximizes the data qubit's coherence. 

In particular, to be general, we must include adaptive strategies, where, at a local measurement time $t_{n-1} \in \{ t_1, ..., t_{N-1}\}$, one can use the information 
\begin{equation}
    Y_{n-1} := \{ y_1, y_2,..., y_{n-1}\},
\end{equation}
obtained in the past up to $t_{n-1}$, to find the optimal choice for the next measurement at $t_n$. That is, we can explicitly write a measurement setting, $\meas_{n}(Y_{n-1})$,  as a function, 
\begin{equation} \label{defmun}
    \meas_{n} : \{0,1\}^{(n-1)} \mapsto U(1)\times \mathbb{R}_+,
\end{equation}
that maps measurement results, $Y_{n-1} \in \{0,1\}^{(n-1)}$ (in a set of $(n-1)$-string of binary numbers), to a measurement angle, $\theta_{n} \in U(1)$ (in a set of the 1D-circular group), and a waiting time, $\tau_{n} \in \mathbb{R}_+$ (in a set of positive real numbers).  

From the above, it is apparent that the space of possible strategies $\cu{{\mathfrak S}}$ is very large for $N$ large. However, we can simplify the problem of trying to find an optimal strategy by applying the {\em separation principle}~\cite{JosTou61,Astrom1970}, as mentioned in the introduction.  First, let us formally define the \emph{expected reward function}: 
\begin{equation} \label{eq:exprew}
    \cohc(T) = \sum_{\info_N} \wp(\info_N) \left|\ccc_{|\info_N} \right|,
\end{equation}
which is the control coherence, Eq.~\eqref{eq:maxcoh}, for the information, $Y = Y_N$, at the final time $T=t_N$. It is implicit that the complex conditional coherence $\ccc_{|Y_N}$, as in \erf{eq:calAcondY}, is defined with $X=X_N$, \ie, at the same time ($T=t_N$) as $Y_N$. We also note that the summation, $\sum_{Y_N}$, is over $Y_N \in \{ 0,1 \}^{(N)}$ and that $\wp(Y_N) = \wp(y_1, y_2, ..., y_N)$ is a joint probability of the string of readouts. Now, the separation principle says that if the reward function is additive in time---which ours, \erf{eq:exprew}, is trivially since it is evaluated at the final time $T$ only---then the optimization problem separates into two parts. 

The first part is optimal estimation of the system. This means finding, at any time $t_n$, the Bayesian probability for the relevant parameters conditioned on the results obtained so far. That is, one can construct a conditional distribution $\wp_n \equiv \wp(X_n,z_n|Y_n)$ for the RTP noise ($z_n$) and its accumulated one ($X_n$) at time $t_n$ conditioned on the current information $Y_n$. The second part is determining the optimal control based on this information. That is, instead of considering $\mu_{n+1}(Y_n)$, we can consider $\mu_{n+1}(\wp_n)$. Moreover, the goal of the control in the future of $t_n$ (\eg, at the final time $T$) is to maximize the expected reward function {\em given} $Y_n$, that is 
\begin{equation}\label{eq:cond_coh0}
    \cohc_{|Y_n}(T) := \sum_{Y_N|Y_n} \wp(y_{n+1},...,y_N|Y_n)  \left|{\cal A}_{|Y_N}\right|,
\end{equation}
where the summation is defined as over possible \emph{future} measurement outcomes, \ie, $\info_N|\info_n :=(y_{n+1}, \dots, y_N)$. 

Now, given that $\wp_n$ is a {\em function} of a real variable $X_n$ and a binary variable $z_n$, it is not immediately obvious that replacing $\mu_{n+1}(Y_n)$ by $\mu_{n+1}(\wp_n)$ is progress. However, as we will see in the next section (after considerable calculation), for the case of our particular conditional expected reward function Eq.~\eqref{eq:cond_coh0}, only a couple of statistics derived from the distribution $\wp(X_n,z_n|Y_n)$ are relevant (see Section~\ref{subsec:suff_stat}), which greatly simplifies the problem. 

Before turning to the details of that calculation, we here also introduce a more general definition of the conditional control coherence:
\begin{equation}\label{eq:cond_coh1}
    \cohc_{|D_n}(t_m) := \sum_{Y_m} \wp(Y_m|D_n)  \left|\langle e^{i \kd X_m} \rangle_{X_m|Y_m}\right|,
\end{equation}
which will be of use in later sections. 
Here the generalizations beyond (\ref{eq:cond_coh1}) are two-fold. First, we allow conditioning on an arbitrary set of data, $D_n$, which is available at measurement time $t_n$, not necessarily equal to the full record $Y_n$ up to that time. Second, the time at which we imagine implementing the control on the data qubit is any measurement time $t_m$, with $m\geq n$, rather than necessarily the final time $t_N = T$ as in Eq.~\eqref{eq:cond_coh0}. 


\section{Bayesian maps\texorpdfstring{\\}{} for phase estimation}\label{sec:baymap}

In this section, we formulate a Bayesian map, which is central to our results, especially the MOAAAR algorithm of Sec.~\ref{sec:MOAAAR}. We start with the expected reward function, Eq.~\eqref{eq:exprew}, and show that it can be written as a matrix equation, similar to that of the no-control case, with the map ${\bf H}$ in Eq.~\eqref{eq:Coh_nc_A}. In contrast to the no-control case, where there is only one map for the entire process, the coherence must be updated whenever new information from the SQ measurements at $t \in \{ t_1, t_2,..., t_N\}$ are obtained. We will see that the new map, denoted by ${\bf F}(\mu_n = \{ \theta_n, \tau_n\} , y_n)$, is an explicit function of the measurement setting, $\meas_n$, and the readout, $y_n$, and can be derived analytically using Bayes' theorem.

\subsection{Bayes' theorem for the expected reward function}\label{subsec:BayesMap}
The expected reward function, defined in Eq.~\eqref{eq:exprew}, can be explicitly written as
\begin{align}\label{eq:CohCtrlYn} 
    \cohc(T) &= \sum_{\info_N} \wp(\info_N) \left|\ccc_{|\info_N} \right|, \notag \\
    &= \sum_{\info_N} \wp(\info_N) \left|\int \dd X_N ~\wp(X_N|Y_N) e^{i \kd X_N} \right|,\notag\\
    & = \sum_{\info_N}  \left|\int \dd X_N ~\wp(Y_N,X_N) e^{i \kd X_N} \right|,
\end{align}
where $X_N$ is the final accumulated noise, 
\begin{equation}\label{eq:accum_noise}
    X_N = x_1 + x_2 + ... + x_N,
\end{equation}
and $Y_N$ is a string of measurement results from the SQ,
\begin{align}
    Y_N = ( y_1, y_2, ..., y_N),
\end{align}
up until the final time.

The integral in Eq.~\eqref{eq:CohCtrlYn} is not trivial since we do not know an explicit form of the joint probability function $\wp(Y_N,X_N)$. However, we have the likelihood functions $\wp(y_n|\theta_n, x_n)$ and $\wp(x_n,z_n|z_{n'})$, via Eqs.~\eqref{eq:forwardP} and~\eqref{eq:FR-pofX} respectively. Thus, we write Eq.~\eqref{eq:CohCtrlYn} in terms of these likelihood functions, by first defining a noise vector, $\vec{x}:=\{x_1,\dots,x_N\}$, and its  1-norm $||\vec{x}||_1=\sum_n x_n$.  Using Eq.~\eqref{eq:accum_noise} allows us to write 
\begin{align}
    \wp(Y_N,X_N)                &=\int \!\!\! \wp(Y_N, \vec{x}) \delta\big(X_N - ||\vec{x}||_1 \big) \dd^{ N} \vec{x}, \notag \\
                &=\int \!\!\! \wp(Y_N| \vec{x}) \wp(\vec{x}) \delta\Big(X_N - ||\vec{x}||_1 \Big) \dd^{N} \vec{x}.
\end{align} 
The integral measure is $\dd^{ N} \vec{x} \equiv \dd x_1 \cdots \dd x_N$.
Substituting the above joint distribution back into Eq.~\eqref{eq:CohCtrlYn} and integrating over $X_N$, we obtain
\begin{eqnarray}\label{eq:CohCvecx} 
\cohc(T) = \sum_{Y_N} \left|\int \wp(Y_N|\vec{x}) \wp(\vec{x}) e^{i \kd ||\vec{x}||_1} \dd^{N} \vec{x} \right|.
\end{eqnarray}
For the joint distribution, $\wp(Y_N | \vec{x})$, since we reset the SQ after every measurement, each result $y_n$ depends only on the accumulated noise $x_n$ during the waiting time $\tau_n$. Thus, we can write
\begin{align}\label{eq:baycon1}
    \wp(Y_N | \vec{x}) &\equiv \wp(y_1,...,y_N| x_1,...,x_N), \notag \\
    &= \prod_{n=1}^N \wp_{\meas_n}(y_n | x_n),
\end{align}
where we have added the measurement setting $\meas_n = \{ \theta_n, \tau_n\}$ as a subscript of the likelihood function~\eqref{eq:forwardP}.
Furthermore, we construct a joint probability of the accumulated noises, $x_1,..., x_N$, with the help of RTP variables and Eq.~\eqref{eq:FR-pofX}, and write,
\begin{align}\label{eq:baycon2} 
\wp(\vec{x}) &\equiv \wp(x_1,..., x_N), \notag \\
             &= \sum_{z_0,..., z_N} \left( \prod_{n=1}^N  \wp(x_n,z_n|z_{n-1})  \right)\wp(z_0),
\end{align}
as a marginalized probability function over the RTP noises, $z_0, ..., z_N$, at all times. Here the sum $\sum_{z_0,...,z_N}$ is over all possible values of the RTP. We then substitute Eqs.~\eqref{eq:baycon1} and \eqref{eq:baycon2} into the reward function in Eq.~\eqref{eq:CohCvecx} and get
\begin{widetext}
\begin{align}
        \cohc(T) 
        = \sum_{\info_N} \wp(\info_N) \left|\ccc_{|\info_N} \right| \notag &= \,\, \sum_{Y_N}
        \left| \sum_{z_0,...,z_N}  \left( \prod_{n=1}^N \int dx_n ~\wp_{\meas_n}(y_n|x_n) \wp(x_n,z_n|z_{n-1}) e^{i \kappa x_n}\right) \wp (z_0) \right|, \nonumber \\
        &=\,\, \sum_{Y_N}
        \left| \sum_{z_0,...,z_N}   \left( \prod_{n=1}^N F_{z_{n-1}}^{z_n}\big(\meas_n, y_n\big) \right) \wp (z_0) \right| , \label{eq:prematrix}
\end{align}
\end{widetext}
where in the second line we have defined 
\begin{align}
\label{eq:F_elements_def}
    F_{z_{n-1}}^{z_n}\!\left(\mu_n, y_n\right) :=\!\!
    \int\!\! dx_n \,\wp_{\meas_n}(y_n|x_n)\,\wp(x_n,z_n|z_{n-1})e^{i \kd x_n}.
\end{align}  
These are the linear maps that are the core elements of our Bayesian-map formalism. In the next subsection, we will derive an analytic expression for $F_{z_{n-1}}^{z_n}\!\left(\mu_n, y_n\right)$, and rewrite it explicitly as a matrix.

\subsection{Analytical expression for \texorpdfstring{$F$}{}-map}\label{subsec:Fmap}
In the similar manner as in the construction of the matrix $\mathbf{H}$ in Eq.~\eqref{eq:Hmatrix}, we see  that the sums and products in \erf{eq:prematrix} can be rewritten as matrix multiplication. Specifically, defining the two-by-two matrix 
\begin{equation}\label{eq:Fmatrix}
{\bf F}\left(\mu_n, y_n\right):= 
\left( 
\begin{matrix}
F^{+1}_{+1}\left(\mu_n, y_n\right) &  F^{+1}_{-1}\left(\mu_n, y_n\right) \\[5pt] 
F^{-1}_{+1}\left(\mu_n, y_n\right) & F^{-1}_{-1}\left(\mu_n, y_n\right) 
\end{matrix} 
\right),
\end{equation}
we can write the expected reward function Eq.~\eqref{eq:prematrix} as 
\begin{align} \label{eq:CohCtrlAn}
        \cohc(T) 
        &= \,\, \sum_{Y_N}\big |  \underline{I}^\top\,  {\bf F}(\meas_N, y_N)\,\cdots {\bf F}(\meas_{2},y_{2})  {\bf F}(\meas_1,y_1) \underline{A}_0 \big| \notag \\
        &= \sum_{Y_N}\big | \underline{I}^\top\,  \underline{A}_N \big| ,
\end{align}
where $\underline{A}_0$ and $\underline{I}$ are defined in Eqs.~\eqref{eq:initialA} and \eqref{eq:Coh_nc_A}. Also, in the second line of Eq.~\eqref{eq:CohCtrlAn}, we defined a new coherence vector
\begin{equation}\label{eq:AN}
\ul{A}_N \equiv {\bf F}(\meas_N, y_N)\,\cdots {\bf F}(\meas_{2},y_{2})  {\bf F}(\meas_1,y_1) \underline{A}_0.
\end{equation}
Moreover, we can use convolutions and Fourier transform (see details in the Appendix~\ref{app:F_Derivation}) to find that the matrices, $\mathbf{F}$ and $\mathbf{H}$, are related via
\begin{align}
\label{eq:FH}
\mathbf{F}\big(\mu_n{=}\{\theta_n,\ddt_n\},y_n\big)  = & \frac{1}{4}\Big[2 \, \mathbf{H}(\ddt_n,\kd) \nonumber \\
 & +(-1)^{y_n}e^{-i\theta_n} \mathbf{H}(\ddt_n,\kd+\ks) \nonumber \\
 & +(-1)^{y_n}e^{+i\theta_n}\mathbf{H}(\ddt_n,\kd-\ks)\Big].
\end{align}
One can easily confirm from Eq.~\eqref{eq:FH} that 
\begin{align}
    \sum_{y_n} \mathbf{F}(\meas_n,y_n) = \mathbf{H}(\tau_n,\kd),
\end{align}
which can be understood that the no-control map $\mathbf{H}$ (no measurement information) is equivalent to ignoring the gained information from the conditional map $\mathbf{F}(\meas_n,y_n)$.

\subsection{Interpretation of complex 2-vector \texorpdfstring{$\underline{A}_n$}{}}
Following the definition of $\ul{A}_N$ in Eq.~\eqref{eq:AN}, we  define a coherence vector $\underline{A}_n =(A_n^{z=+1},A_n^{z=-1})^\top$ for an intermediate time $t_n$. This can be computed from the vector at the previous time step using the measurement result $y_{n}$ via
\begin{equation} \label{eq:An}
\underline{A}_{n} = {\bf F}(\meas_{n}, y_{n})\underline{A}_{n-1}.
\end{equation}
We can interpret the two elements of $\ul{A}_n$, similar to \erf{interpAnc}, as
\begin{equation} \label{interpAn}
\underline{A}_{n} = \left(\begin{matrix}
    \ul{A}_n^{+1} \\
    \ul{A}_n^{-1}
    \end{matrix}\right)= \left(\begin{matrix}
    \wp(Y_n,z_n=+1) {\cal A}_{|Y_n,z_n=+1} \\
    \wp(Y_n,z_n=-1) {\cal A}_{|Y_n,z_n=-1}
    \end{matrix}\right).
\end{equation}
That is, the coherence vector $\ul{A}_n$ contains information about the measurement results $Y_n = (y_1,...,y_n)$ and the complex conditional coherence defined as
\begin{equation}\label{eq:calAYnzn}
{\cal A}_{|Y_n,z_n} := \langle e^{i\kappa X_n} \rangle_{X_n|Y_n,z_n},
\end{equation}
conditioned on both the  results and the hypothetically knowable $z_n$, at time $t_n$.  
With this, we can see a relationship between two types of complex conditional coherences, in Eq.~\eqref{eq:calAcondY} and Eq.~\eqref{eq:calAYnzn}, via
\be\label{eq:relvecAcalA}
\wp(Y_n)\ccc_{|Y_n} = \ul{I}^\top \ul{A}_n = \sum_{z_n} \wp(Y_n,z_n) \ccc_{|Y_n,z_n},
\ee
where the summation is over the RTP state $z_n \in \{ -1 ,+1\}$. 

The vector $\ul{A}_n$ does not let one calculate exactly
the probability of the measurement results, $\wp(Y_n)$. However, 
if one considers $\kd$ (the sensitivity of the data qubit to noise) to be a variable rather than a fixed number, then one can take the limit $\kappa \to 0$ in Eq.~\eqref{eq:calAYnzn}. This gives $\lim_{\kappa \rightarrow 0} {\cal A}_{|Y_n,z_n}=1$ and, from Eq.~\eqref{interpAn}, we see that 
\begin{equation}\label{eq:probPvec}
    \wp(Y_n) = \lim_{\kappa \rightarrow 0} \ul{I}^\top \ul{A}_n.
\end{equation}
This motivates  defining a probability vector, 
\begin{equation}\label{eq:Pn}
    \ul{\check A}_n = \lim_{\kappa \rightarrow 0}  \ul{A}_n = \left(
    \begin{matrix}
    \wp(Y_n,z_n=+1) \\
    \wp(Y_n,z_n=-1)
    \end{matrix}
    \right), 
\end{equation}
which should not be confused with the unconditioned RTP probability vector $\ul{P}_t$.

Similar to the coherence vector $\ul{A}_n$ in Eq.~\eqref{eq:An}, the probability vector can be updated after every SQ measurement via
\begin{equation} \label{eq:Pn+1}
\ul{\check A}_{n} = \check{\bf F}(\meas_{n}, y_{n})\ul{\check A}_{n-1},
\end{equation}
where we have defined a new probability map with $\kappa \rightarrow 0$,
\begin{equation}\label{eq:Fcheck}
\check{\bf F}(\meas_n,y_n):=\lim_{\kd\to0}{\bf F}(\meas_n,y_n).
\end{equation}
This $\check{\bf F}$ matrix contains only real numbers and thus can map between two real-number vectors containing only  probability functions. Interestingly, the initial state of the probability vector is the same as $\ul{A}_0$ and the RTP initial state, $\ul{P}_0$, \ie,
\begin{align}\label{eq:intstateA}
\ul{\check A}_0 = \ul{A}_0 = \ul{P}_0 = \ul{P}_{\rm ss}.
\end{align}
Here we have also set them to be equal to the RTP steady state as per  Eq.~\eqref{eq:Pss}. We will see later that $\ul{\check A}_n$ and $\wp(Y_n)$ are useful to both understanding dynamics of the SQ and simulating the data qubit's coherence numerically.

\subsection{Two real sufficient statistics: \texorpdfstring{$(\alpha,\zeta)$}{}} \label{subsec:suff_stat}

As mentioned in Section~\ref{sec:ASASP}, the separation principle allows us to separate the problem at any time $t_n$ into the estimation part,  using the information $Y_n$ obtained up to that time, and the control part,  which uses that estimate to maximize the expected future cost function, conditioned on the present $Y_n$. Here we show that, for our cost function, the estimation problem reduces to calculating the complex 2-vector $\ul{A}_n$. Furthermore, except at the final time $T$, just two real numbers derived from $\ul{A}_n$ are sufficient for the control problem. 

We start with the expected reward function conditioned on $Y_n$ given in  Eq.~\eqref{eq:cond_coh0}, which we rewrite here for convenience: 
\begin{equation}\label{eq:cond_coh}
    \cohc_{|Y_n}(T) := \sum_{Y_N|Y_n} \wp(y_{n+1},...,y_N|Y_n)  \left|{\cal A}_{|Y_N}\right|.
\end{equation}
We then use Bayes' rule to write $\wp(y_{n+1},...,y_N|Y_n) = \wp(Y_N)/\wp(Y_n)$ and use the relationship 
$\wp(Y_N) |{\cal A}_{|Y_N}| = | \underline{I}^\top\,\underline{A}_N|$ from Eq.~\eqref{eq:relvecAcalA} to obtain
\begin{subequations}
\begin{eqnarray}\label{eq:cond_coh_A}
\cohc_{|Y_n}(T)
&=& [\wp(Y_n)]^{-1}\sum_{Y_N|Y_n} \big | \underline{I}^\top\,\underline{A}_N \big|, \\
&=&[\wp(Y_n)]^{-1} \sum_{Y_N|Y_n}\big |  \underline{I}^\top\,  {\bf F}(\meas_N, y_N)\,\cdots \label{eq:cohalze}\\ 
&& \hspace{1cm} {\bf F}(\meas_{n+2},y_{n+2})  {\bf F}(\meas_{n+1},y_{n+1}) \underline{A}_n \big|, \nn
\end{eqnarray}
\end{subequations}
where in the second line we used Eq~\eqref{eq:AN}, only expanding out the $\bf F$ maps for the future outcomes $y_{n+1},...,y_N$. 

Now, to maximize this reward function, the only things we have the ability to control, in the future of $t_n$, in \erf{eq:cohalze} is the SQ measurement strategy, $\meas_{n+1},..., \meas_{N}$. What current information is relevant to this strategy? First, since $Y_n$ is already known, $\wp(Y_n)$ is a fixed scalar multiple, and so is not relevant. That leaves only the vector $\ul{A}_n$. In other words, if we know $\ul{A}_n$, we know {\em everything} relevant for choosing the SQ measurement strategy to maximize the final objective $\cohc_{|Y_n}(T)$.  From \erf{eq:maxcoh}, it also contains all the relevant present information for calculating, at the final time, the optimal control to apply to arrive at \erf{eq:cohalze}.  Here we are concerned with the question `what parts of $\ul{A}_n$ are relevant for the SQ measurement strategy?'   The complex-2 vector $\ul{A}_n$ defines four real parameters. It turns out that only two of them are relevant as we now show.  

Since the two (complex) elements of $\ul{A}_n$ are associated with two plausible values of the unknown RTP, \ie, $z_n \in \{-1, +1 \}$ as in Eq.~\eqref{interpAn}, 
one might guess that these elements should tell 
us something relevant about the data qubit's phase and coherence if one could, hypothetically, find out which state $z_n$ the RTP is actually in. One such quantity is the scaled ratio of the two elements of $\ul{A}_n$, 
\begin{subequations}
\begin{align} \label{defalpha}
\alpha_n :&=  \frac{K}{\kappa}\arg\frac{A_n^{z=+1}}{A_n^{z=-1}}, \\ 
          &=\frac{K}{\kappa}\ro{\arg \ccc_{|Y_n,+1}-\arg \ccc_{|Y_n,-1} }, \label{alreadyseen}
\end{align}
\end{subequations}
using the definition in Eq.~\eqref{interpAn}. This quantity $\alpha_n$ is the first of the two sufficient statistics and can be interpreted as the \emph{maximum difference of an optimal control phase} from finding out if the RTP state was $z_n = +1$ or $z_n = -1$, scaled so as to be of order unity.

To understand Eq.~\eqref{defalpha}, we recall that the optimal control (phase correction) is given by $\cd(Y) =\arg \ccc_{|Y}$ from Eq.~\eqref{eq:optctrl}. If, at a current time $t_n$, a phase correction, $\cd(Y_n)$, were to be applied to the data qubit, by further assuming the hypothetical $z_n$, one could consider refining the control to 
\begin{equation} \label{eq:refined_ctrl}
\cd'(Y_n,\rtp_n) = \arg \ccc_{|Y_n,z_n} = \arg \left\langle e^{i\kd \xd_n}\right\rangle_{{\xd_n}|\info_n,z_n},
\end{equation}
where $z_n$ is included in the condition of the expected average. The bracketed term in Eq.~\eqref{alreadyseen} is therefore the difference of the control phases for the two hypothetical $z_n$. To understand the scaling factor that makes this phase difference of order unity, $O(1)$, it is necessary to consider the dynamics of the RTP and the measurements, as we do in the following paragraph. 

The value of $z_n$ is relevant  to the data qubit phase  only for the recent period of evolution, of order $1/\bg$.  For any time earlier than that, the RTP state is pretty much uncorrelated with its current state. Thus the maximum difference that finding out $z_n$ could make on the data qubit's phase estimate is the difference in the cumulative phase, $\kappa X_n$, over that time scale, for $z_n=+1$ versus $z_n=-1$. That is, of order $\kappa/\bg$, which is small. In actuality, when we are gathering information about $z$ through a good adaptive protocol (see later sections) the difference is even smaller, because from the preceding measurement result we have a very good idea of the value of $z$. Hence if $z$ is unknown that is substantially because it may have  changed its value since the last measurement. In other words, the preceding time scale for which finding out the value of $z$ is relevant is of order $\tau$, not of order $1/\bg$. As we will see later, the optimal value for $\tau$ is of order $1/K$.  Thus, the difference that finding out $z_n$ could make for the phase estimate is of order $\kappa/K$. Hence, to scale the phase difference to the order unity, we divide it by the factor $\kappa/K$ and get  $\alpha_n$ as in Eq.~\eqref{defalpha}. 

The second of the two sufficient statistics is related to the modulus of the two elements of $\ul{A}_n$, defined as
\begin{equation} \label{defzeta}
    \zeta_n :=\frac{|A_n^{\rtp=+1}|-|A_n^{\rtp=-1}|}
    {|A_n^{\rtp=+1}|+|A_n^{\rtp=-1}|}.
\end{equation}
This can be interpreted as an \emph{approximated mean of $z_n$ conditioned on $Y_n$}. To see this, we consider the conditional complex coherence, $\ccc_{|Y_n,\pm1}$. As we have already seen above, their arguments are very close, differing by only $O(\kappa/K)$ as per \erf{alreadyseen}. But this implies that their moduli must also be very close. In fact, the relative difference in their moduli is  quadratically smaller than the difference in their arguments.  That is because the relative change in the data qubit coherence $|\ccc|$ over some short time interval is quadratic in the uncertainty in the data qubit's phase. (One can see this by Taylor expanding the decoherence to the second order of the phase.) But the {\em difference} in the uncertainty in the data qubit's phase, for the different values of $z_n$, will not be larger than the corresponding difference in their mean, because the difference between the two RTP values is the maximum uncertainty the RTP value can have. 

Therefore, in the relevant time for the refined best-estimate phase shift, $\tau = O(1/K)$, the difference in the relative decoherence resulting from refining one's knowledge to $z=+1$ versus $z=-1$ would be at most of order $O((\kappa/K)^2)$. That is, for a good measurement protocol, we can be confident that  $|\ccc_{|Y_n,+1}|/|\ccc_{|Y_n,-1}| = 1 + O\big((\kappa/K)^2\big)$. With relative errors of the same magnitude, we can thus replace $|\ccc_{|Y_n, z_n}|$ by $|\ccc_{|Y_n}|$ in \erf{interpAn}, to get
\begin{equation}
\label{eq:AInterpretation}
|A_n^{z_n}|\approx|{\cal A}_{|Y_n}|\, \wp(Y_n,z_n). 
\end{equation}
Hence, $\zeta_n$ in Eq.~\eqref{defzeta} can be approximated as 
\begin{equation}\label{eq:zeta_z}
\zeta_n \approx \frac{\sum_{z_n} z_n \, \wp(Y_n, z_n)}{\sum_{z_n} \wp(Y_n, z_n)},
\end{equation}
which is exactly the mean of $\rtp_n$ conditioned on the current measurement record, $Y_n$.

There are two more (third and fourth) parameters from the vector $\ul{A}_n$ that are not relevant to finding the optimal SQ control. The third parameter is the argument of the sum of elements, \ie, 
\begin{equation}\label{defvarphi}
    \varphi_n:= \arg \ccc_{|Y_n} =  \arg\left(A_n^{z=+1}+ A_n^{z=-1} \right).
\end{equation}
This would be relevant as the final control on the {\em data} qubit, $\cd(Y_n)$, if $t_n=T$ were the final time. However, because $t_n$ is not the end of the protocol, the quantity simply expresses how far the mean phase has drifted so far. Since the random phase driven by the RTP is a true $U(1)$ process, having an estimate of the phase at any particular time is not relevant to future decoherence. One can see this from Eq.~\eqref{eq:cohalze} that a global phase change to the vector $\ul{A}_n$, \ie, changing the value of $\varphi_n$, does not affect the coherence on the left of the equation.

Fourthly, we can define the final real parameter from the sum of moduli of the elements as
\begin{equation} \label{defr}
    r_n: = |A_n^{\rtp=+1}|+|A_n^{\rtp=-1}| \approx 
    \wp(Y_n) |\ccc_{|Y_n}|,
\end{equation}
where we have used Eq.~\eqref{eq:AInterpretation} with the same degree of approximation as  discussed around that equation. As already argued, $\wp(Y_n)$ is irrelevant to optimizing the future control, and so is any loss of coherence already suffered such that $|\ccc_{|Y_n}| < 1$, by a similar argument used for $\varphi_n$ above.

\begin{table}[t!]
  \begin{center}
    \begin{tabular}{|p{1.9cm}|m{6.3cm}|}
        \hline
        Parameter &  \hspace{2cm} Description\\
        \hline
         $r_n$, Eq.~\eqref{defr} & $\approx$  $Y_n$-probability $\times$ conditional coherence.\\ 
         \hline
        $\varphi_n$, Eq.~\eqref{defvarphi} & $=$ optimal control phase, not knowing $z_n$.   \\
                \hline
        $\zeta_n$, Eq.~\eqref{defzeta} & $\approx$ mean of $z_n$ conditioned on $Y_n$. \\
         \hline
          $\alpha_n$, Eq.~\eqref{defalpha} & $=$ maximum difference that finding out $z_n$ could make to optimal control phase, scaled. \\
 \hline
    \end{tabular}
  \end{center}
  \caption{We present four (real) statistics parameters along with their physical explanations in this table. The vector $\ul{A}_n$ is described by these parameters.}
  \label{tab:parameters}
\end{table}

We summarize the meaning of the four real parameters that make up the vector $\ul{A}_n$ in table~\ref{tab:parameters}. Another way to summarize them is to note that in the regime of interest it is possible to approximate $\ul{A}_n$ by a simple expression using the above four parameters,
\begin{equation} \label{eq:Aapprox}
{A}_n^{z=\pm 1} \approx r_n \exp(i \varphi_n)  \times
\frac{1\pm \zeta_n}{2}\exp\left( \pm i \frac{1\mp\zeta_n}{2}\frac{\alpha_n \kappa}{K} \right).
\end{equation}
Here the approximation sign allows for a {\em relative} error in $r_n$ of $O \left( (\kappa/K)^2 \right)$ and an {\em absolute} error in $\varphi_n$ of $O \left( (\kappa/K)^3 \right)$. This expression is for interest only; we do not use it in the remainder of the paper. 

Returning to the two  sufficient statistics, $\alpha_n\in \mathbb{R}$ in Eq.~\eqref{defalpha} and $\zeta_n \in [-1,1]$ in Eq.~\eqref{defzeta}, these can thus be regarded as the only two moments of $\wp(X_n,z_n|Y_n)$ (the solution to the estimation part of the problem) that are relevant to the SQ measurement part of the problem. Thus, as mooted in Sec.~\ref{sec:ASASP}, we can vastly reduce in size the set of control protocols $\mathfrak{S}$ Eq.~\eqref{deffrakS} that we consider by simplifying the functional form of $\mu_{n+1}$ from that in \erf{defmun} to 
\begin{equation} \label{defnumun}
    \meas_{n+1} : \mathbb{R}\times[-1,1] \mapsto  U(1)\times \mathbb{R}_+.
\end{equation} 
That is, $\mu_{n+1}$ is now a function of $(\alpha_n,\zeta_n)$ rather than of $Y_n$ directly. Finding the exactly optimal control strategy  $\cu{\mu_n:n\in\{1,\cdots,N\}}$ is still a difficult task, especially as, in general, $N$ will not, in fact, be fixed for a fixed $T$. However, if we seek a close-to-optimal, rather than exactly optimal,  control strategy ${\mathfrak S}$, then this difficulty can be avoided. Because we are interested in the long time limit, we will have $N\gg 1$, and for the vast majority of the control choices we can ignore the \emph{edge effects} (where $n$ is small or $n$ is close to $N$), and choose the $\mu_{n+1}$ to all be the {\em same} function of their two arguments. That is, we have only to optimize a single function, 
\begin{equation} \label{defmun_ig_ed}
    \meas : \mathbb{R}\times[-1,1]  \mapsto U(1)\times \mathbb{R}_+.
\end{equation}
In the future sections, we will make further assumptions to restrict the function $\mu$, and we will see how the stochastic behaviour of $\alpha_n$ and $\zeta_n$ is greatly useful in developing the intuition to do this.

\section{Local optimisation algorithm (Greedy)} \label{sec:Greedy}

The first strategy we explore is a `greedy' one~\cite{cormen2009AlgorithmGreedy}. That is, it based on the natural idea of maximizing the expected reward function $\cohc$ locally in time. This is easier than maximizing the actual reward function, which is $\cohc$ at the final time $T$. Greedy algorithms are not optimal for typical hard problems, but even if they are not, they may be relatively close to optimal. That appears to be the situation for our problem, as we will find.  We call this strategy the Greedy algorithm, or simply `Greedy' for short.

We conceptualize the Greedy algorithm as follows. Say  that 
we are at some time $t$ such that $t_n < t \le t_{n+1}$. By this we mean that exactly $n$ measurements on the SQ have already happened and their measurement results, $\info_n$, are known, and it is still possible to decide to make the $(n+1)$th measurement at time $t$. Imagine now that the final time is the immediate future, 
\be\hypoT = t+\dt.\ee
Then the Greedy algorithm aims to maximize the conditional expected reward function $\cohc_{|Y_n}(\hypoT)$ as in \erf{eq:cond_coh_A} where we have replaced $T$ with $\hypoT$. 

We emphasize that any---even random---measurement on the SQ provides some information that can be used to increase the coherence of the data qubit. Thus to maximize $\cohc_{|Y_n}(\hypoT)$ at least one measurement on the SQ at time $\hypoT=t+\dt$ should be performed. Greedy maximizes the expected reward function at the immediate future time, by deciding whether or not to measure at time $t$ and optimizing the measurement angles at both times. 
In the following subsections, we present the details of Greedy in Sec.~\ref{subsec:Greedy_prot} and then demonstrate its numerical implementation in Sec.~\ref{subsec:Greedy_num}. Studying the patterns in the numerically found Greedy algorithm in Sec.~\ref{subsec:Greedy_result} then motivate us to propose and analyze various `experimentally' simpler greedy algorithms in  Secs.~\ref{subsec:Greedy_4} and~\ref{subsec:Greedy_2}, where in Sec.\ref{subsec:Greedy_phase} we analyze the phase space dynamics of the simpler greedy algorithms.

\subsection{The Greedy algorithm to maximize \texorpdfstring{$\cohc_{|Y_n}(\hypoT)$}{}}
\label{subsec:Greedy_prot}
Let the current time be denoted  $t=t_n+\tau$, with $\tau>0$. The goal is to determine the next measurement setting, $\meas_{n+1}$, based on the current information, $Y_n$. To do that, the Greedy algorithm compares the reward $\cohc_{|\info_n}(\hypoT)$ from two prospective scenarios:
\begin{enumerate}
\item[(i)] The SQ is measured only at time $\hypoT = (t_n +\tau) + \dt$,
\item[(ii)] The SQ is measured now, at $t = t_n + \tau$, and again at $\hypoT = (t_n + \tau) + \dt$.
\end{enumerate}
We then calculate the coherence vectors associated with the two scenarios at time $\hypoT$ using equation \eqref{eq:An}:
\begin{subequations}
\begin{align}
\GreedyAi &:= \mathbf{F}(\meas_{(\rm i)}, y_{n+1}) \underline{A}_n,	\\
\GreedyAii &:= \mathbf{F}(\meas_{(\rm ii)}',y_{n+2}) \mathbf{F}(\meas_{(\rm ii)},y_{n+1}) \underline{A}_n,
\end{align}
\end{subequations}
where $y_{n+1}$ and $y_{n+2}$ are measurement results at time $t$ and $\hypoT=t+\dt$ respectively. Scenario~(i) consists of only one measurement with a setting $\meas_{(\rm i)} := \{\Greedythi, \tau + \dt \}$ whereas scenario~(ii) consists of two consecutive measurements with settings $\meas_{(\rm ii)} := \{\Greedythii, \tau \}$ and $\meas_{(\rm ii)}' := \{\Greedythii', \dt \}$ for the first and second measurement respectively. Note that these angles should be thought of not as numbers but as {\em functions} of $\underline{A}_n$ and, for $\Greedythii'$, also of $y_{n+1}$.

Despite the fact that scenario~(ii) has one more measurement than scenario~(i), the coherence vectors are both calculated at the hypothetical final time $\hypoT=t+\dt$. Thus, the reward functions---as it is  defined in Eq.~\eqref{eq:cond_coh_A}--- for the two cases at time $\hypoT$ are
\begin{subequations}\label{eq:GreedyCoh}
\begin{align}
\cohc_{\rm(i)}(\Greedythi) 
:&= \cohc_{|Y_n}(\hypoT)\Big|_\text{(i)}  \notag \\
&= [\wp(Y_n)]^{-1} \sum_{y_{n+1}} \left| \underline{I}\tp \GreedyAi \right|, \\
\cohc_{\rm(ii)}(\Greedythii,\Greedythii') 
:&= \cohc_{|Y_n}(\hypoT)\Big|_\text{(ii)}  \notag \\
&= [\wp(Y_n)]^{-1}\sum_{y_{n+2},y_{n+1}} \left| \underline{I}\tp \GreedyAii \right|,
\end{align}
\end{subequations}
where on the left hand sides of the above equations, we defined $\cohc_{\rm(i)}$ and $\cohc_{\rm(ii)}$ as functions of the prospective angles. We then find the optimal angles by maximizing both $\cohc_{\rm(i)}$ and $\cohc_{\rm(ii)}$ over all possible angles. Formally, the maximized reward functions are given by
\begin{subequations}
\label{eq:GreedyOpt}
\begin{align}
    \coh^{\rm op}_{\rm (i)}  &:= \max_{\Greedythi} ~\cohc_{\rm (i)} (\Greedythi), \\
    \coh^{\rm op}_{\rm (ii)} &:= \max_{\Greedythii,\Greedythii'} \cohc_{\rm (ii)} (\Greedythii,\Greedythii').
\end{align}
\end{subequations}
We compare the optimum values in Eqs.~\eqref{eq:GreedyOpt} to decide whether or not to measure at time $t$. The decision-making procedure is as follows: if $\coh^{\rm op}_{\rm (i)} > \coh^{\rm op}_{\rm (ii)}$, it means that no measurement at time $t$ is required to maximize the reward function at $\hypoT$. Thus, Greedy decides to wait and moves to the next infinitesimal time to repeat the procedure again. If, on the other hand, $\coh^{\rm op}_{\rm (i)} \le \coh^{\rm op}_{\rm (ii)}$, Greedy chooses to measure the SQ at $t$, so that $\tau$ is the actual waiting time $\tau_{n+1}$ from $t_n$ to $t_{n+1}=t$ and the measurement angle $\Greedythii$ is the actual $\theta_{n+1}$. Thus, we write
\begin{align}
    \meas_{n+1} = \{\Greedythii, \tau\},
\end{align}
as the next measurement setting chosen by the Greedy. We note that, the pre-factor $[\wp(Y_n)]^{-1}$ in Eq.~\eqref{eq:GreedyCoh} does not influence the Greedy decision-making procedure and knowing only $\ul{A}_n$ suffices. Indeed, following the argument of Sec.~\ref{subsec:suff_stat}, knowing only the two real parameters $\alpha_n$ and $\zeta_n$ suffices. 

Now that we have identified $\meas_{n+1}$, we repeat the procedure, starting at time $t_{n+1}$, to find the measurement settings for the next step, \ie, $\meas_{n+2}$, using  $\ul{A}_{n+1}$. Repeating in this fashion till $t_N=T$ (where we always measure), equips us with a single realization of $Y_N$ that corresponds to a specific sequence of settings, $\{\meas_1, \dots, \meas_N\}$ and a trajectory of the reward function, $\cohc_{|Y_n}(t_n)$ for $n \in \{1, \dots, N\}$. Then, to calculate the expected reward function, similar to what we defined in Eq.~\eqref{eq:CohCtrlYn}, we average over the information $Y_n$ as
\begin{align}\label{eq:GreedyCohAve}
    \cohc(t_n) &= \left\langle \cohc_{|Y_n}(t_n) \right\rangle_{Y_n} = \sum_{Y_n} \, \wp(Y_n)\,  \cohc_{|Y_n}(t_n).
\end{align}
The hope, then, is that by locally maximizing $\cohc$ in time, that the final $\cohc(T)$ will be much larger than the unconditioned coherence, $\cohnc$, and perhaps even close to optimal. 

In the following subsection, we will implement the ``full-Greedy" algorithm as presented here, and evaluate \erf{eq:GreedyCohAve} numerically. However, instead of numerically generating all possible trajectories associated with all possible measurement results $Y_N$, we will perform a stochastic simulation where the measurement output at each step, \ie, $y_n$, is generated randomly according to its probability. Then, by generating enough stochastic trajectories, we can evaluate the expected reward, $\cohc(t_n)$ in Eq.~\eqref{eq:GreedyCohAve} as an average. But in later subsections we will see how it can be possible to simplify the algorithm so as to perform the exact sum.

\begin{figure}[t!]
	\includegraphics[width=\columnwidth]{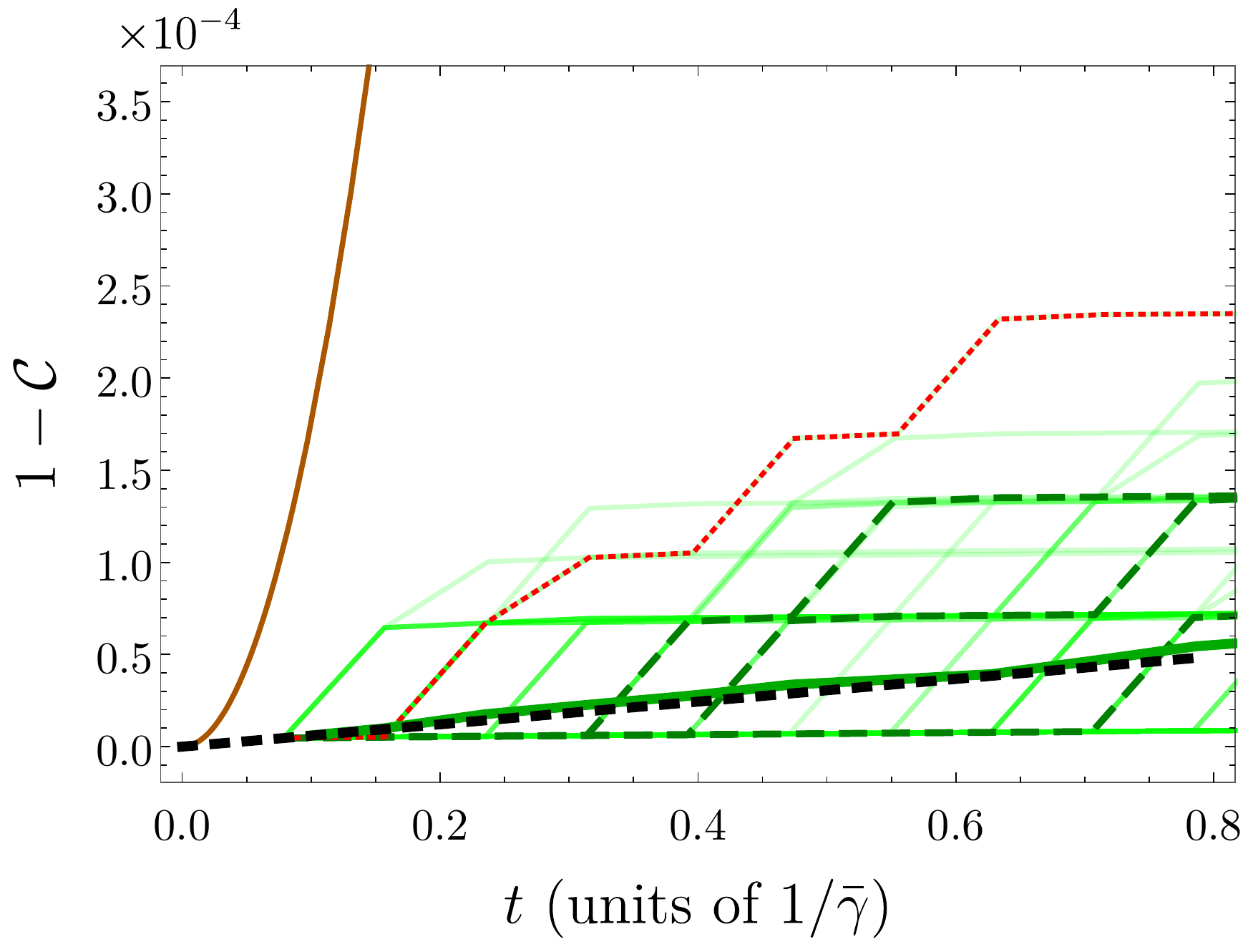}
\caption{Decoherence in data qubit for $\kd=0.2$ and $K=20$. The thick brown  curve represents no-control decoherence $(1-\cohnc)$ in asymptotic regime as in Eq.~\eqref{uncond_exp}, whereas the green lines represent $100$ trajectories from numerical simulations of the Greedy algorithm. We plotted the trajectories in light green so that the saturation in the colour shows the relative probability of a particular decoherence value at a particular time. Note that  Greedy selects a different waiting time at each step, but the difference between them in this case is of order $10^{-3}/\mg$, which is too small to see in the figure, giving the appearance that all branches occur at the same time. Three typical trajectories are bolded with thick dashed lines and one atypical trajectory, which happened only once, is shown in dotted red. The thick dark green line is the average of the $100$ trajectories, while the black dashed line represents the exact mean of Greedy$_4$, introduced in Sec.~\ref{subsec:Greedy_4}.}
\label{fig:GreedyDecoherence}
\end{figure}

\subsection{Numerical implementation of Greedy}
\label{subsec:Greedy_num}
For our numerical calculations, we choose $\gu=\gd= \mg = 1$ for simplicity and the other dynamical parameters ($\kd$ and $\ks$) are in the units of $\mg$. Equating RTP up and down jump rates yields symmetric dynamics that are easier to analyze. Also, we assume that, at $t=0$, the RTP is in the steady state, \ie, $\underline{A}_0 = \underline{P}_{\rm ss}$, where $\underline{P}_{\rm ss}$ is defined in Eq.~\eqref{eq:Pss}. We choose $\kd=0.2$  and a range of $\ks$ values from $2$ to $100$. At the higher end of ranges for $\ks$, condition~\eqref{eq:regime} will be satisfied. For the (theoretically infinitesimal) 
time step $\dt$, it is most critical that it is small compared to $1/\ks$. We choose $\dt =0.001/\ks$.

The Greedy algorithm requires optimization over the three measurement angles as in Eqs.~\eqref{eq:GreedyOpt}. Fortunately, our locally optimal strategy has the same structure as that introduced in Ref.~\cite{BerWis00}. This means we can use analytical solutions described in \cite{BerWisBre01} to reduce the range of each possible angle from a real interval $[-\pi/2,\pi/2]$ to only three discrete choices. This is described in Appendix~ \ref{app:BerryWiseman}.

As we discussed above, after finding an optimal $\meas_{n+1}$, we need to simulate the measurement outcome $y_{n+1}$ in order to calculate $\ul{A}_{n+1}$. For our stochastic simulation we randomly generate $y_{n+1}$ using a probability function  $\wp(y_{n+1}|Y_n)$, where $Y_n$ is the measurement record assumed known up to that time. This function is encoded in vector $\ul{\check A}_{n+1}$, following Eq.~\eqref{eq:probPvec}, where we write
\begin{equation}
    \wp(y_{n+1}|Y_n) = \frac{\wp(Y_{n+1})}{\wp(Y_n)} =  \frac{\ul{I}^\top \ul{\check A}_{n+1}}{\ul{I}^\top \ul{\check A}_{n}}.
\end{equation}
To be able to use the above equation, we update the vector $\ul{\check A}_n$ after each measurement using Eq.~\eqref{eq:Pn+1}, with the initial probability vector $\ul{\check A}_0 = \ul{P}_{\rm ss}$, same as $\ul{A}_0$ as in Eq.~\eqref{eq:intstateA}.

We show the conditional coherence $\cohc_{|\info_n}(t_n)$ in Fig.~\ref{fig:GreedyDecoherence} from numerical implementation of the Greedy algorithm using aforementioned parameters and $K=20$. The light green lines are $100$ trajectories of $\cohc_{|Y_n}(t_n)$ where three of them are exaggerated (bold-dashed lines) as examples. To calculate the expected reward function $\cohc(t_n)$, we average over these $100$ trajectories. It should be noted that the waiting times, $\tau_n$, in different trajectories do not match, so we could not simply calculate the average decoherence at each $t_n$. Instead, for each trajectory we interpolated a line between consecutive $(t_n,\cohc_{|Y_n}(t_n))$ points, giving rise to the broken lines shown in Fig.~\ref{fig:GreedyDecoherence}. Then we took the average of the broken lines, which is shown in dark green. Comparing the average with the no-control coherence $\cohnc(t)$, we see that the Greedy algorithm can massively improve the data qubit's coherence.

\begin{figure}[t!]
	\includegraphics[width=\columnwidth]{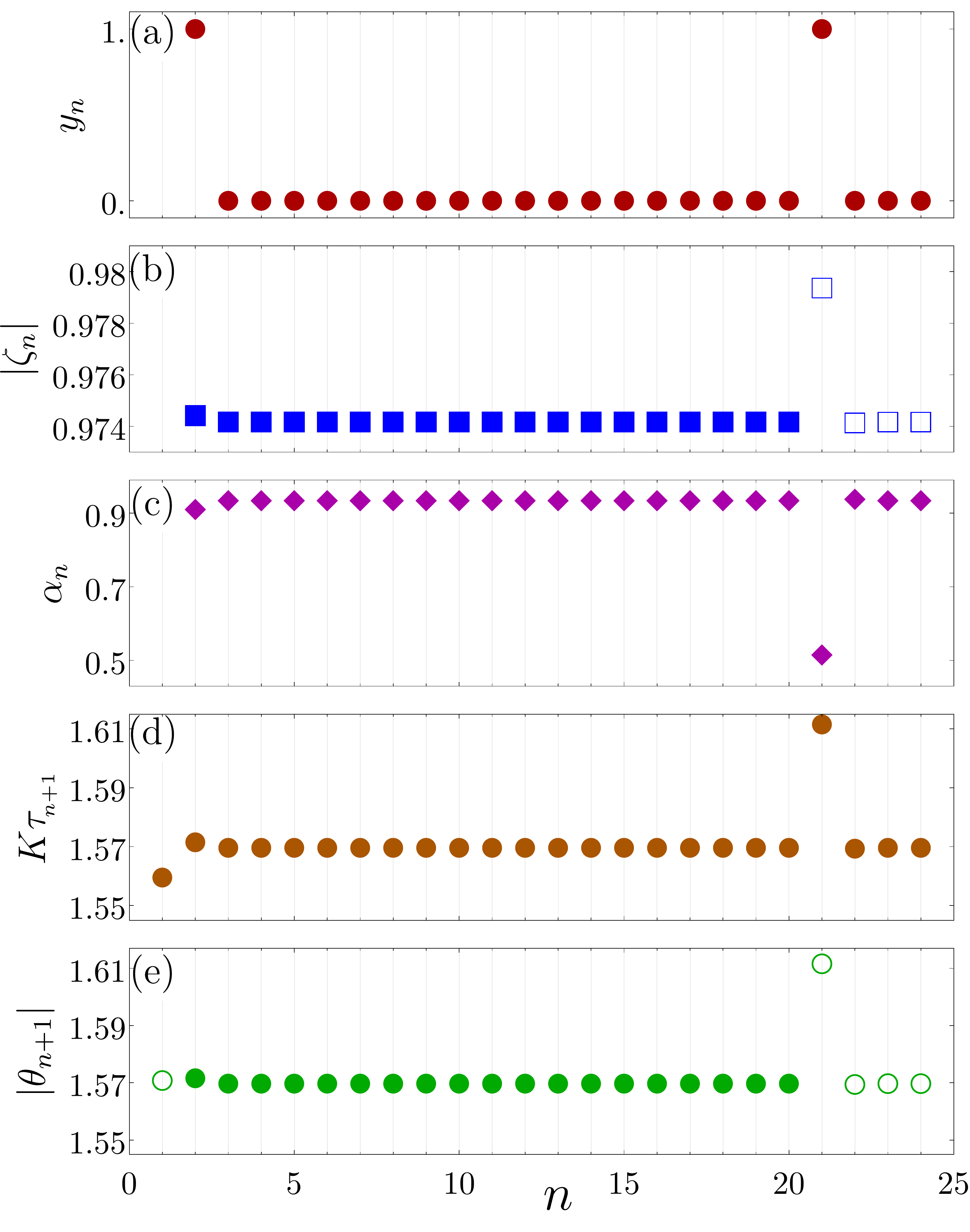}
	\caption{Greedy chosen values of next measurement parameters, $\{\theta_{n+1},\tau_{n+1} \}$, together with the information gained via $n$th measurement, $\zeta_n$, $\alpha_n$ and $y_n$, in a single trajectory of the Greedy algorithm for $\kd=0.2$ and $\ks=20$. For $\theta_{n+1}$ and $\zeta_n$, their absolute values are shown in the vertical axes, while their signs are depicted via filled and hollow markers for positive and negative signs, respectively.}
	\label{fig:greedychosen}
\end{figure}

\subsection{Greedy choices of measurement settings} \label{subsec:Greedy_result}
We would like to understand how the greedy algorithm chooses the measurement settings $\meas_{n+1} = \{ \theta_{n+1}, \tau_{n+1} \}$, based on the measurement result $y_n$ and relevant statistics $\alpha_n$ and $\zeta_n$. To do that, we look at individual trajectories of the Greedy chosen settings. A typical one is presented in Fig.~\ref{fig:greedychosen}. For $\zeta_n$ and $\theta_{n+1}$, their values can change sign, so we plotted the absolute values of them in order to see the pattern better, while the sign of them are depicted by filled and hollow markers for positive and negative respectively.

We observe in Fig.~\ref{fig:greedychosen} that, after an initial transient, the statistical parameters, $\alpha_n$ and $|\zeta_n|$, jump between only two values each, and these jumps are correlated with the measurement result $y_n$. We also observe that whenever a non-null measurement $(y_n=1)$ occurs, the sign of $\zeta_n$ changes, that is depicted by the filled markers switched to hollow markers. Moreover, from Fig.~\ref{fig:greedychosen}(d), we see that the Greedy choices for the waiting time $\tau_{n+1}$ and the absolute value of measurement angle $|\theta_{n+1}|$ are, apart from transients, again confined to two values and their jumps are correlated with jumps in $\{\alpha_n, \zeta_n\}$. Finally, the sign of $\theta_{n+1}$ is the same as the sign of $\zeta_n$. These observations lead us to proposing a new approach to implement a simpler version of Greedy algorithm that we explain next.

\subsection{``Greedy\texorpdfstring{$_4$}{}'' with four parameters}\label{subsec:Greedy_4}
We can capture the Greedy choices of measurement parameters as just explored by setting 
\begin{subequations}\label{eq:Gthetatau}
\begin{align}
     \theta_{n+1} &:= s_n  \Theta_{\rm G} (\alpha_n,|\zeta_n|),\\
     \tau_{n+1} &:= \frac{\Delta_{\rm G}(\alpha_n,|\zeta_n|) }{\ks}. 
\end{align}
\end{subequations}
Here the function $\Theta_{\rm G}$ sets the absolute value of the Greedy choice of measurement angle, while 
\begin{equation}
s_n:=\sign(\zeta_n),
\end{equation}
determines its sign, and the function $\Delta_{\rm G}$ sets the  waiting time $\tau_{n+1}$. This $\Delta_{\rm G}$ is dimensionless and of order unity, since, as we have discussed, the waiting time scales with $1/\ks$.

Although, in general,  $\Delta_{\rm G}$ and $\Theta_{\rm G}$ may be functions of both $\alpha_n$ and $\zeta_n$, we see from Fig.~\ref{fig:greedychosen} that knowing $\alpha_n$ is sufficient.  In fact, since (after some transient steps) $\alpha_n$ jumps between two values, denoted by $\{\alpha^{\ar},\alpha^{\al}\}$, it is sufficient to know the binary variable $a_n \in \{>,<\}$, defined as 
\begin{equation}
    a_n =  
    \begin{cases} 
        > & \text{if } \alpha_n > \bar{\alpha} \ , \\
        < & \text{if } \alpha_n < \bar{\alpha} \ ,
    \end{cases}
\end{equation}
where $\bar{\alpha}:=(\alpha^{\ar}+\alpha^{\al})/2$. Note that, most of the time, $a_n = \ >$. Thus, we simplify $\Theta_{\rm G}$ and $\Delta_{\rm G}$ as binary functions, each with two values as 
\begin{subequations} \label{eq:GreedyBinary}
\begin{equation}\label{eq:GreedyBinary1}
    \Theta_{\rm G}(a_n) =  
    \begin{cases} 
        \Theta_{\rm G}^> & \text{if } 
        a_n = \ > \ , \\
        \Theta_{\rm G}^< & \text{if }
        a_n = \ < \ , 
    \end{cases}
\end{equation}
and 
\begin{equation}
    \Delta_{\rm G}(a_n) =  
    \begin{cases} 
        \Delta_{\rm G}^> & \text{if } a_n = \ > \ , \\
        \Delta_{\rm G}^< & \text{if } a_n = \ < .
    \end{cases}\label{eq:101a}
\end{equation}
\end{subequations} 
It should be noted that the above equations could have been defined in terms of $|\zeta_n|$ rather than $\alpha_n$. However, because the difference between $\alpha^{\ar}$ and $\alpha^{\al}$ is more pronounced, it is more natural to use this as the variable on which $\Theta_{\rm G}$ and $\Delta_{\rm G}$ depend.

\begin{figure}[t!]
	\includegraphics[width=\columnwidth]{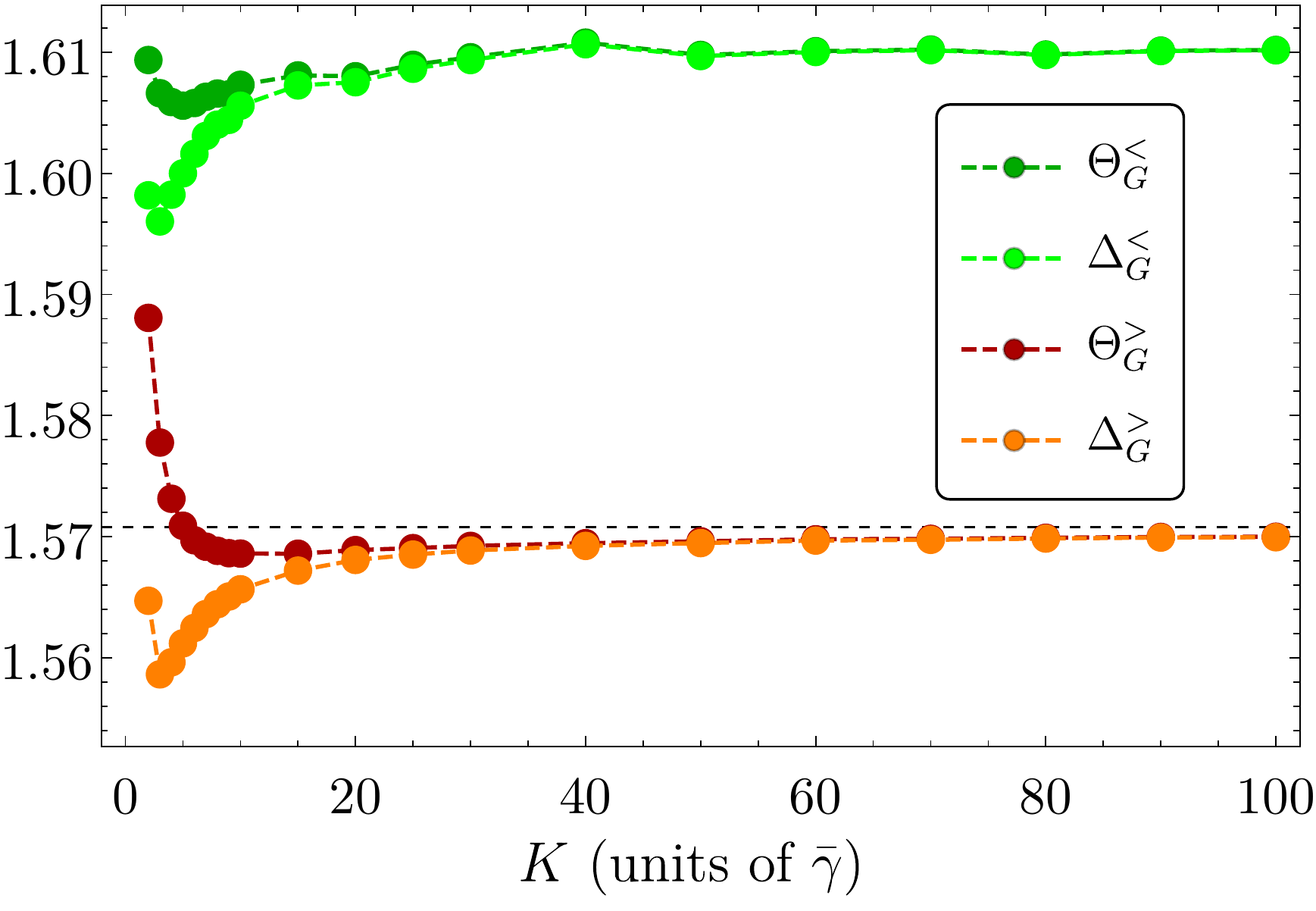}
\caption{Calculated values of the Greedy$_4$ parameters, from top to bottom, $\Theta_{\rm G}^{\al}, \Delta_{\rm G}^{\al}, \Theta_{\rm G}^{\ar}, \text{ and } \Delta_{\rm G}^{\ar}$, for a range of $\ks$. See text for details of calculation.   The horizontal dashed line shows the value of $\pi/2$; see Eq.~\eqref{ThetaG0asym}.}
\label{Fig:PairsvsK}
\end{figure}

We calculate numerical values of  $\{\Delta_{\rm G}^{\ar},\Delta_{\rm G}^{\al},\Theta_{\rm G}^{\ar},\Theta_{\rm G}^{\al}\}$ as follows. Each Greedy trajectory provides us with a sequence of Greedy chosen measurement angles $\{\theta_n\}$, a sequence of waiting times $\{\tau_n\}$, and an associated sequence of $\{\alpha_n\}$. First, we discard the transient data (the first few time steps) and then separate the chosen angles into two groups by checking if the associated $\alpha_n$ is larger or smaller than the threshold $\bar{\alpha}$. That is, we have two sets of angles defined as
\begin{subequations}
\begin{align}
\vartheta^{\ar} &= \{|\theta_n| : a_n = \ >  \}, \\
\vartheta^{\al} &= \{|\theta_n| : a_n = \ <  \}. 
\end{align}
\end{subequations}
Similarly for the waiting times, scaled by $K$, we define two sets as
\begin{subequations}
\begin{align}
\Upsilon^{\ar} &= \{\tau_n/K : a_n = \ > \}, \\
\Upsilon^{\al} &= \{\tau_n/K : a_n = \ < \},   
\end{align}
\end{subequations}
The above procedure is repeated for all 100 trajectories. Then, we take all members of $\vartheta^{\ar}$s and average them to be the numerical value of $\Theta_{\rm G}^{\ar}$. The numerical value of $\Theta_{\rm G}^{\al}$, $\Delta_{\rm G}^{\ar}$ and  $\Delta_{\rm G}^{\al}$ are calculated similarly, following the conditions of $a_n$ in Eqs.~\eqref{eq:GreedyBinary}. Fig.~\ref{Fig:PairsvsK} shows the values of $\Delta_{\rm G}^{\ar},\Delta_{\rm G}^{\al},\Theta_{\rm G}^{\ar},\Theta_{\rm G}^{\al}$ for a range of the sensitivity $K$ from $2$ to $100$. All four of these parameters are within 2.5\% of $\pi/2$ across the whole range. 

Now that we know the binary values of $\Theta_{\rm G}$ and $\Delta_{\rm G}$, we no longer need the full optimization part of the Greedy algorithm. As a result, we obtain a new adaptive algorithm in which an experimenter can simply use \eqref{eq:GreedyBinary} with the knowledge of $\alpha_n$ to decide measurement settings for the next step. We call this new adaptive algorithm ``Greedy$_4$" (Greedy with sub-index $4$), since it is defined by four parameters, namely $\Delta_{\rm G}^{\ar},\Delta_{\rm G}^{\al},\Theta_{\rm G}^{\ar}$ and $\Theta_{\rm G}^{\al}$. We emphasize again that in order to determine these four parameters, the full Greedy algorithm, including the optimization part, must be run. Once the four parameters are found, the experimenter could use them for future experiments.

To calculate the expected reward $\cohc(t_n)$ as in Eq.~\eqref{eq:GreedyCohAve} for Greedy$_4$, we  numerically generate all possible $Y_N$ and average the associated reward functions with the probability weight $\wp(Y_N)$ from Eq.~\eqref{eq:probPvec}. Each realisation of $Y_N$, like full-Greedy, is associated with a different sequence of waiting times $\tau_n$, which take two values. However, as shown in Fig.~\ref{Fig:PairsvsK}, the difference between $\Delta_{\rm G}^>$ and $\Delta_{\rm G}^<$ is not great. Moreover, as we have already seen in Fig.~\ref{fig:greedychosen}, most of the time the value $\Delta_{\rm G}^>$ is chosen. The other value is chosen only when the system detects that a jump has occurred, which happens at approximately the same rate as the jumps, $\bg$. Thus it is not a bad approximation to take the average as if every  $\tau_n$ were the same, an effective waiting time is given by 
\begin{equation}
\tau_{\rm eff} = \frac{\Delta^{\ar}}{\ks}(1-\bg\, \tau_{\rm eff}) + 
\frac{\Delta^{\al}}{\ks}\bg\, \tau_{\rm eff}. 
\end{equation}
Here $\tau_{\rm eff}$ appears on the right hand side because $\bg\tau_{\rm eff}$ is the probability, in one waiting time, that the RTP jumps. Solving for $\tau_{\rm eff}$ we obtain 
\begin{equation}
\tau_{\rm eff} = \frac{\Delta^{\ar}}{\ks -\bg(\Delta^{\al} - \Delta^{\ar})}.
\end{equation}
Using this $\tau_{\rm eff}$, we find the expected reward function for Greedy$_4$, which is shown with a black dashed line in Fig.~\ref{fig:GreedyDecoherence}. This line is an \emph{exact} average in the sense that it is not stochastic. Moreover, this line agrees with the stochastic average using the full Greedy, within the stochastic error, and confirms that the expected reward function calculated from the full Greedy is in agreement with Greedy$_4$.

\begin{figure*}[th!]
	\includegraphics[width=\textwidth]{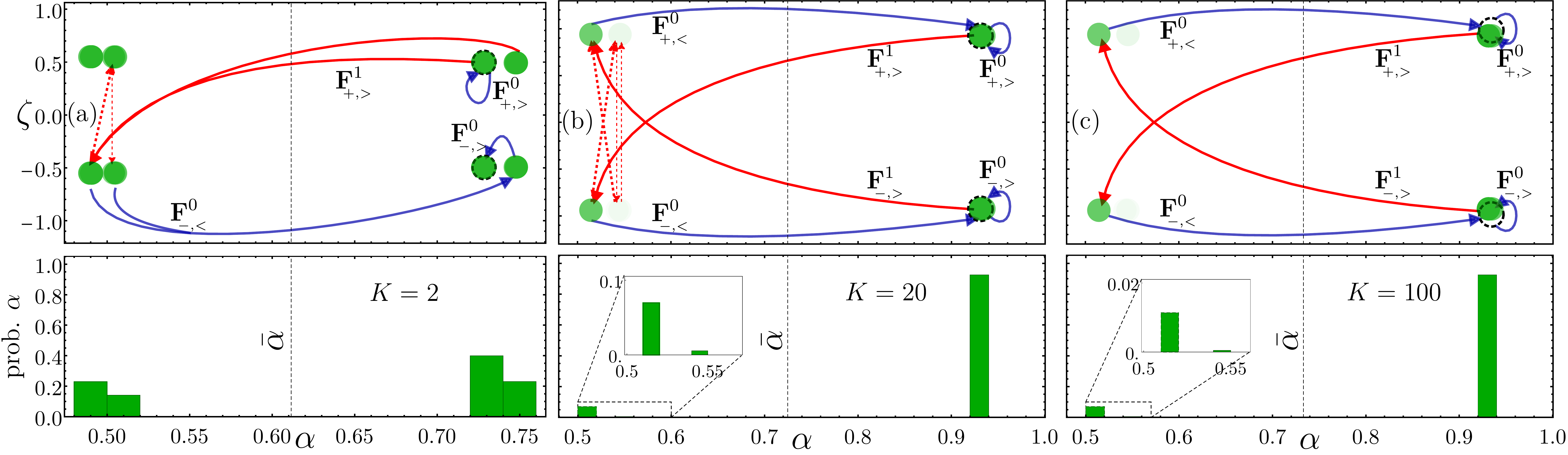}
	\caption{
	Dynamics of the Greedy$_4$ algorithm in phase space (top row) and probability of $\alpha$ (bottom row). We used $\kd=0.2$ in all plots, while $\ks=2, 20,\text{and}~ 100$, from left to right. The value of $\bar\alpha$ is indicated by a dashed line in each plot. The blue arrows represent maps due to null results ($y=0$), while the red arrows represent non-null outcomes ($y=1$). Less frequent transitions are shown by red dashed arrows. We have only shown half of the arrows in the instance of $\ks=2$; the reader can visualize the other half by mirroring the image around the $\zeta=0$ axis.
	}
	\label{fig:phase}
\end{figure*}

\subsection{Dynamics of Greedy\texorpdfstring{$_4$}{} in \texorpdfstring{$(\alpha, \zeta)$}{} phase space}
\label{subsec:Greedy_phase}
In this subsection, we use the sufficient statistics $(\alpha, \zeta)$ as a way to visualize the dynamics of the Greedy$_4$ algorithm in ``phase space''. The plots in Fig.~\ref{fig:phase} show phase space dynamics of 100 trajectories for $\ks=2,20\text{ and }100$. We use light green dots to represent all $(\alpha, \zeta)$ states and thus the green color saturation indicates the relative probability of being in that state. Blue arrows in the figure show the state transition due to the null result, \ie, $y=0$, while the red arrows show that of the non-null result, \ie, $y=1$. The maps used for the transitions are shown as labels. Since Greedy$_4$ chooses between the binary values of the measurement angle and the waiting time, we can simplify the notation for the $\mathbf{F}$ map introduced in Sec.~\ref{subsec:Fmap} as 
\begin{equation}\label{eq:compact_notation}
\mathbf{F}_{s,a}^y := \mathbf{F} \left( s \Theta_{\rm G}({a}), \Delta_{\rm G}({a})/ K, y \right), 
\end{equation}
where $\Theta_{\rm G}({a})$ and $\Delta_{\rm G}({a})$ are the functions defined in Eqs.~\eqref{eq:GreedyBinary}. Note that, as a consequence of taking $\gu = \gd$, the phase diagrams for $(\alpha,\zeta)$ are symmetric in reflection around the $\zeta=0$ line. Furthermore, the bottom panel of each sub-figure depicts the normalized histograms for different values of $\alpha$, which can be interpreted as the probability of the system being on that specific $\alpha$.

In these phase diagrams, we see that the statistical states $(\alpha, \zeta)$ have a tendency to be in only a few states, instead of spreading over the entire phase space. This was already noted in the discussion of Fig.~\ref{fig:greedychosen} for $K=20$, but here we visualize the correlated changes in $\alpha, \zeta$ more easily, as well as the transition from $K$ of order $1$ to $K\gg1$. In all cases, the two most occupied states turn out to correspond to the eigenstates with the largest absolute eigenvalues of $\mathbf{F}_{+,>}^0$ and $\mathbf{F}_{-,>}^0$ (Recall that these are $2\times2$ complex linear maps acting on $\ul{A}$, and it turns out that they each have two eigenstates). We denote these stable eigenstates (the ones with largest absolute eigenvalues) by $\ul{E}_{\,+,>}^0$ and $\ul{E}_{\, -,>}^0$, respectively. The $(\alpha,\zeta)$-values for these eigenstates are depicted as the dashed circles in the phase space plot. 

To understand the state dynamics, let us first consider a state being in  $\ul{E}_{\,+,>}^0$, the stable eigenstate of $\mathbf{F}_{\,+,>}^0$. In this case, any null-result $(y=0)$ measurement will map the state to itself and no effect on the system's phase space configuration. On the other hand, a non-null measurement ($y=1$) will map the state to a point on the left side, with $\alpha < \bar{\alpha}$ while switching the sign of $\zeta$ from positive to negative. Any subsequent non-null result, which is less common, will cause the sign of $\zeta$ to switch again, while $\alpha$ remains in the same region $\alpha < \bar{\alpha}$. However, since the probability of two consecutive non-null results is low, the subsequent measurement most likely yields a null result, mapping the system state to a point on the right side with $\alpha > \bar{\alpha}$, while keeping the sign of $\zeta$ unchanged. The following null results, as they are more likely, will move the state towards the other stable  eigenstate, $\ul{E}_{\,-,>}^0$.

The above description holds for any value of $\ks > \mg$. But there are differences as $\ks$ changes. First consider the case for a relatively low sensitivity, $\ks=2$, \ie, of the same order as $\mg=1$. In this case we observe in Fig.~\ref{fig:phase}(a) that the dynamics in the phase space can be grouped into eight locations. We understand the dynamics as follows. For small $\ks$, we find that the likelihood of consecutive non-null measurement results is higher compared to that of the larger $\ks$. So, if the system configuration has moved to the left side of the phase space, as a consequence of a  measurement result, any further non-null measurements yield maps that changes the sign of $\zeta$ but keeps the $\alpha$ on the left side. These maps as we showed them with red dashed arrows, reveals four main locations on the left side of the phase space. Furthermore, we observe that a map due to a null result, $(y=0)$, brings the system from the left to the right side of the phase space, but the resulting configuration is distanced from the eigenstates. Then, further null results bring the system to the eigenstate. As a result, there are four main locations on the right side. Note that we have not shown all the transitions in the $\ks=2$ case to avoid messiness (see caption).

As $\ks$ increases, the first thing that happens is that the right-most points on the phase diagram all converge to the eigenstates $\ul{E}^0_{\pm,>}$. That is, for $\alpha > \bar{\alpha}$ we have only these two eigenstates to consider. This is seen already in Fig.~\ref{fig:phase}(b). As $\ks$ increases even further, reaching the asymptotic regime $\ks \gg \mg $, the chance of two consecutive non-null results is negligible. This leads to there also being only two points on the left side, as 
seen in Fig.~\ref{fig:phase}(c). Thus, the dynamics are simplified and can be grasped by concentrating on just four points: $\ul{E}_{\,\pm,>}^0$, and $\mathbf{F}_{\pm,>}^1 \ul{E}_{\,\pm,>}^0$. Moreover, the system spends the majority of its time in one of the eigenstates and occasionally, with a non-null ($y=1$) measurement result, goes to the $a=\,<$ state while altering $s=\sign(\zeta)$. The next measurement almost always yields $y=0$, causing a jump back to an eigenstate while keeping $s$ the same.

\subsection{``Greedy\texorpdfstring{$_2$}{}" in the asymptotic regime}
\label{subsec:Greedy_2}

In the asymptotic regime $\ks\gg\mg$, we show that the Greedy$_4$ algorithm in Eqs.\eqref{eq:GreedyBinary} can be simplified even further. By inspecting Fig.~\ref{Fig:PairsvsK} carefully, we observe that in this regime the four parameters converge to only two, because we have the identities $\Theta_{\rm G}^{\ar} = \Delta_{\rm G}^{\ar}$ and $\Theta_{\rm G}^{\al} = \Delta_{\rm G}^{\al}$. From the original definitions in Eq.~\eqref{eq:Gthetatau}, it means that the Greedy-chosen measurement angles and waiting times are related via a simple relation: 
\begin{equation}\label{eq:theta_tau}
\theta_{n+1} = s_n K \tau_{n+1}.
\end{equation} 
As a result, we have a simpler algorithm with just two parameters: $\Theta_{\rm G}^{\ar}$ and $\Theta_{\rm G}^{\al}$. We refer to this algorithm as ``Greedy$_2$", which is valid only for large $K$ or the asymptotic regime. The algorithm can be summarized as
\begin{subequations}\label{eq:greedy2}
\begin{align}
\theta_{n+1} = &\,\,s_n \Theta_{\rm G}(a_n), \label{thetansn} \\
\tau_{n+1} = & \,\, \Theta_{\rm G}(a_n)/K,
\end{align}
\end{subequations}
where $\Theta_{\rm G}(a_n)$ is given by Eq.~\eqref{eq:GreedyBinary1}.

This choice has an intuitive explanation, as follows. The term $K \tau_{n+1}$ is simply a SQ's phase evolved (after the SQ's phase reset to zero) during the time interval $(t_n,t_{n+1})$, where $\tau_{n+1} = t_{n+1}-t_n$, given that the RTP stays unchanged  with a value of $s_n$ during that time. That is, under the approximation that $z(t)$ stays equal to our best estimate of its value, $s_n$ at the start of the  interval, the evolved SQ's phase is given by
\begin{equation}\label{eq:asymAngle}
    \phis(t_{n+1}) = \ks \int_{t_{n}}^{t_{n+1}}\!\! \rtp(t){\rm d}t \, \approx  \, { s}_{n} \ks \ddt_{n+1}.
\end{equation}
This approximation can be justified in the asymptotic regime, as our confidence in the value of $z_n$ is high ($|\zeta_n|\approx 1$) and the probability of a jump in the interval is low, $O(\bg/\ks)$. Therefore,  Eq.~\eqref{eq:theta_tau} simply says that the measurement angle should be chosen to be exactly the most likely SQ phase at the time of measurement, \ie, $\theta_{n+1} = \Phi(t_{n+1})$. In other words, the measurement probes whether the SQ is in the most likely state. A null result ($y_{n+1}=0$) is the answer `yes' while the other result ($y_{n+1}=1$) is the answer `no'. That is, in this aspect, Greedy$_2$ performs an optimal choice for minimum-error binary state discrimination~\cite{Helstrom,WisMil10}, where one state (which is pure) arises if and only if $z(t)=s_n$ for the entire interval, and the other state (which is mixed) arises if and only if $z(t)=-s_n$ for some of the interval. 

Note that the above simple intuition does not explain the value chosen by Greedy$_2$ for its two parameters, $\Theta_{\rm G}^{\ar}$ and $\Theta_{\rm G}^{\al}$. However, inspecting Fig.~\ref{Fig:PairsvsK} again, we see that one of these values does have a simple interpretation in the asymptotic limit:
\begin{align} \label{ThetaG0asym}
\lim_{\ks/\mg \to \infty}\Theta^{\ar}_{\rm G} = &\,\, \pi/2.
\end{align}
Note that this is the more important of the two parameters, as in the asymptotic  regime, the measurement result is almost always a null result, the system state is almost always in the region $\alpha_n > \bar{\alpha}$, and the measurement choice is almost always $\Theta_{\rm G}(a_n) = \Theta_{\rm G}^{\ar}$. The choice of $\pi/2$ for $s_n \theta_{n+1}$ and $K\tau_{n+1}$ implies that the most likely pure state (from $z(t)=s_n$ for the entire interval) and the second-most likely pure state (from $z(t)=-s_n$ for the entire interval) are orthogonal at the measurement time, and can be perfectly distinguished by the measurement. 

\section{Map-based Optimized Adaptive Algorithm for Asymptotic Regime (MOAAAR)}
\label{sec:MOAAAR}
While using the Greedy algorithm for the SQ offers a great reduction in the decoherence of the data qubit, it is only locally optimal. There is no reason to expect it to be globally optimal. In this section we introduce and analytically study a new algorithm that is plausibly optimal in the asymptotic regime, $\ks \gg \gud \gg \kd$. However, to design the algorithm we use the knowledge and intuitions gained from the numerical investigation of Greedy in this regime. 
 
First, recall from Sec.~\ref{subsec:Greedy_2} that in asymptotic regime, Greedy chooses measurement settings that are perfectly matched according to Eqs.~\eqref{eq:greedy2}.  Second, recall from Sec.~\ref{subsec:Greedy_phase} that  out of the two values of $\Theta_{\rm G}(a_n)$, the choice is almost always $\Theta_{\rm G}(>)$, which we called $\Theta_{\rm G}^{\ar}$.  Considering just the single parameter, $\Theta_{\rm G}^{\ar}$, one could use it to define what one might call a Greedylike$_1$ algorithm, in which $\Theta_{\rm G}(a)$ in Greedy$_2$ is replaced by the constant $\Theta_{\rm G}^{\ar}$. This would reproduce the performance of the Greedy algorithm in the asymptotic regime, although it would not actually approximate the Greedy algorithm itself, which always uses two distinctly different angles, even in the asymptotic regime, as per Greedy$_2$.

Like the Greedy algorithm, this Greedylike$_1$ algorithm would still be adaptive, in general, because of the $s_n$ in \erf{thetansn}.  However, we already know the value of $\Theta_{\rm G}^{\ar}$ in the asymptotic regime, \erf{ThetaG0asym}.  While this choice is intuitive, it in fact corresponds to a {\em non-adaptive} measurement, as $\theta = s \pi/2$ is measuring in the same basis for $s=\pm1$. It would be surprising if the  globally optimal performance in the asymptotic regime could be attained by a non-adaptive algorithm, and indeed we find that this is not the case. We do this by cleaving to the well-motivated relations between the measurement angles and the waiting times in Eqs. \eqref{eq:greedy2}, and to the understanding that the asymptotic performance can be expected to be reproduced with $\Theta$ fixed, but without cleaving to the particular choice of $\Theta$ made by the Greedy algorithm. That is, as expected, we can do better by choosing $\Theta$ in a  globally optimal manner rather than a locally optimal (greedy) manner. 

To be precise, we here propose an adaptive (in general) algorithm for measurement angles and waiting times, 
\begin{subequations}\label{eq:adtmoa}
\begin{align}
    \theta_{n+1} =& \, s_n \Theta,\\
    \tau_{n+1}=&\,  \Theta/\ks,
\end{align}
\end{subequations}
as in Eqs.~\eqref{eq:greedy2}, but with the function $\Theta_{\rm G}$  replaced by a fixed parameter $\Theta$. This 
 simplifies the calculation of the expected reward function in the asymptotic regime, as we need consider only the following set of four  single-parameter maps: 
\beq \label{single-paramap}
  \mathbf{F}_{s}^{y}(\Theta):=\mathbf{F}(s \Theta, \Theta/K, y),
\eeq
where $s \in \{ +1, -1\}$ and $y\in \{0,1\}$. Similar to the preceding section, we denote the stable eigenstates of these maps by $\ul{E}_{\,s}^y$.

In this section, we first show how the problem of maximizing the expected reward function is, in the regime of interest (\ref{eq:regime}), equivalent to minimizing an expected decoherence rate. We then apply the adaptive strategy, Eqs.~\eqref{eq:adtmoa}, and derive an analytical expression for the decoherence rate as a function of $\Theta$ in the asymptotic regime. This can be used to search for a {\em globally} optimal parameter value, $\Theta\opt$, which defines our Map-based Optimized Adaptive Algorithm for Asymptotic Regime (MOAAAR). We call it Map-based because of the central role of the Bayesian maps, as mentioned in the preceding paragraph. We emphasize that it is adaptive because we will find $\Theta\opt\neq \pi/2$, which means that, via $s_n$, previous measurement results do alter the basis in which the SQ is measured. (Recall that this is in contrast with the $\Theta_{\rm G}^{\ar} = \pi/2$ choice of the Greedylike$_1$ algorithm in this regime.) 

\subsection{Expected reward and decoherence rate}

Recall the original expected reward function Eq.~\eqref{eq:exprew} defined for the final time $T$, when the phase-correction control is to be applied. This is a sum over exponentially many terms, which makes it infeasible to do an analytic maximization. However, from the numerical results in the previous section (Greedy), we know that the system's state $(\alpha, \zeta)$ effectively jumps between only a finite number of states, and explores all of them over time. As we will see, this is also true for the globally optimized MOAAAR strategy.  This means that the stochastic evolution of our state of knowledge is ergodic~\cite{Gar85}, so the average decay of the coherence over a  typical trajectory over many steps ($T\gg \ks^{-1}$) is the same as an ensemble average over one step. There will be deviations from this due to initial [$t = O(1/\bg)$] and final [$T-t = O(1/\bg)$] transients, but these can be ignored for long times as we are interested in. Thus we can then safely use an expected reward function at an intermediate time $t_n$, $\cohc(t_n)$ as a proxy of the expected reward at the actual final time, $\cohc(T)$.

We also saw in the preceding section that in the regime $\ks \gg \mg$, the Greedy algorithm makes the system spend almost all of the time close to the two stable eigenstates $\ul{E}_{\,\pm}^{y=0}$. We find this to be even more true with MOAAAR; we will quantify deviations from this assumption in Sec.~\ref{subsec:apperappra}. Thus we can approximate the coherence (the reward function) at time $t_n$ as 
\begin{align}
\cohc(t_n) =&\,  \sum_{Y_n} \wp(Y_n) | \ccc_{|Y_n}| =  \sum_{Y_n}\, \big | \ul{I}^\top\,  \ul{A}_n \big| \nonumber \\
=& \sum_{\substack{Y_n| \\ s_n = +1}} \! |\beta_+(Y_n)\, \ul{I}^\top\, \ul{E}_+^0| + \!\! \sum_{\substack{Y_n| \\ s_n = -1}} \! |\beta_-(Y_n)\, \ul{I}^\top\, \ul{E}_-^0|.
\label{eq:cohdecoh}
\end{align} %
The first line follows from the definitions of the reward function in  \erf{eq:GreedyCohAve} or Eq.~\eqref{eq:exprew}. In the second line, we have separated the summation, $\sum_{Y_n}\big | \ul{I}^\top\,  \ul{A}_n \big|$, into two contributions associated with $s_n = +1$ and $s_n = -1$, and used the fact that the state $\ul{A}_n$  is almost always proportional to one of the two eigenstates ($\ul{E}_{\,+}^{0}$ and $\ul{E}_{\,-}^{0}$). Which of the two states pertains at time $t_n$ is determined solely by $s_n$ (which is a function of $Y_n$), while $\beta_+(Y_n)$ and $\beta_-(Y_n)$ are the appropriate proportionality scalars. 

Let us now consider the expected reward at the next SQ's measurement at $t_{n+1}= t_n + \tau$, which can be computed by applying one additional map, ${\mathbf F}_{\pm}^y(\Theta)$, to Eq.~\eqref{eq:cohdecoh} and summing over its measurement results, $y_{n+1}  \in \{ 0,1\}$, for which we use the dummy variable $y$. This gives 
\begin{align}
    \cohc(t_{n+1}) =&\sum_{y}  \sum_{\substack{Y_n| \\ s_n = +1}} \! |\beta_+(Y_n)\, \ul{I}^\top\, {\mathbf F}_+^y(\Theta)\,  \ul{E}_+^0| \nonumber \\ 
    &+ \sum_{y}\sum_{\substack{Y_n| \\ s_n = -1}} \! |\beta_-(Y_n)\, \ul{I}^\top\, {\mathbf F}_-^y(\Theta)\, \ul{E}_-^0|.\label{eq:cohcnext}
\end{align}
Note that the signs $s_n$ determine which $\mathbf{F}$-map is applied, and matches the subscript of the eigenstates.

Next, we  rearrange the formula Eq.~\eqref{eq:cohcnext} such that it is written in terms of the coherence at time $t_n$ in Eq.~\eqref{eq:cohdecoh}, with one (justified) approximation. We can interpret the two terms in Eq.~\eqref{eq:cohdecoh} as two weighted conditional coherence contributions,  namely $P_{\rm ss}(s_n=+1)\,  \cohc_{|s_n = +1}(t_n)$ and $P_{\rm ss}(s_n=-1)\,  \cohc_{|s_n = -1}(t_n)$, respectively. [Recall that we defined a general conditioned coherence in \erf{eq:cond_coh1}.] Since $s_n$ is our best guess for the RTP $z_n$, and since we are not in a transient regime, we can take these probabilities to be the same as the steady-state probabilities for the RTP. Also, in this long time  steady-state regime, we can safely say that the value that $s_n$ happens to have at this particular time has almost no bearing on the coherence at this time, which is determined by a long history of stochastic events, only the most recent of which is relevant to $s_n$. That is, we can make the approximation  $\cohc_{|s_n=+1}(t_n) \approx \cohc_{|s_n=-1}(t_n) \approx \cohc(t_n)$. Rewriting the terms involving $\beta_{\pm}(Y_n)$ in Eq.~\eqref{eq:cohdecoh} in terms of conditional coherences, we can simplify the expected reward at time $t_{n+1}$ to
\begin{align}
\label{eq:avgdec}
    \cohc(t_{n+1}) =&\, \sum_y \Bigg\{   \frac{P_{\rm ss}(s_n=+1) | \ul{I}^\top\, {\mathbf F}_+^y(\Theta) \ul{E}_+^0|}{|\ul{I}^\top\, \ul{E}_+^0|} \cohc_{|s_n=+1}(t_n) \nonumber \\
    &\quad + \frac{P_{\rm ss}(s_n=-1) | \ul{I}^\top\, {\mathbf F}_-^y(\Theta) \ul{E}_-^0|}{|\ul{I}^\top\, \ul{E}_-^0|}\cohc_{|s_n=-1}(t_n) \Bigg\} \nonumber \\
    \approx &\,  [1- {\bar \Gamma(\Theta)}\tau]\,  \cohc(t_n).
\end{align}
This describes an exponential decay with an expected decoherence rate defined as
\begin{equation} \label{eq:asymav}
\bar\Gamma(\Theta) := \sum_{s=\pm} {P}_{\rm ss}(s)
\frac{ | \underline{I}^\top \underline{E}_s^0 | - \sum_{y} | \underline{I} ^\top\, {\bf F}_s^y(\Theta) \, \underline{E}_s^0 |  }{\ddt | \underline{I} ^\top \underline{E}_s^0 |}.
\end{equation}
Here we have dropped the $n$ subscript entirely as it is no longer necessary.

The $\bar\Gamma(\Theta)\tau$ in \erf{eq:avgdec} can also be thought of as a steady-state average of the relative change of the conditional coherence, as the system's state evolves from $\ul{A}_n$ (conditioned on $Y_n$) at time $t_n$ to the next time $t_{n+1}=t_n+\tau$ with an unknown $y_{n+1}$. This relative change was called $\delta(\ul{A}_n)$ in CL~\cite{PRL} (where the vector $\ul{A}_n$ is its argument, not the scalar quantity which is changing). In terms of this, we have a conditional rate,
\begin{align} \label{alternate-Gamma}
\Gamma(\Theta,\ul{A}_n)&= \frac{\delta(\ul{A}_n)}{\tau}, \notag \\
&= \frac{1}{\ddt}
\frac{ \left|\ccc_{|Y_n} \right| 
- \sum_{y_{n+1}} \wp({y_{n+1}}|Y_n)\left|\ccc_{|Y_{n+1}} \right|}
{\left|\ccc_{|Y_n} \right|},
\end{align}
and the average rate in  Eq.~\eqref{eq:asymav} is obtained by averaging over $Y_n$. 
This average exponential-decay rate well approximates the coherence decay rate for a typical trajectory, as argued above. Therefore we can use $\bar\Gamma(\Theta)$ as a proxy for the global reward function (\ref{eq:exprew}). 
That is, we reformulate the problem of maximizing the expected final coherence as the problem of minimizing the average decoherence rate. This can be done essentially analytically, as we see in the following subsection. 

\begin{figure*}[t!]
	\includegraphics[width=\textwidth]{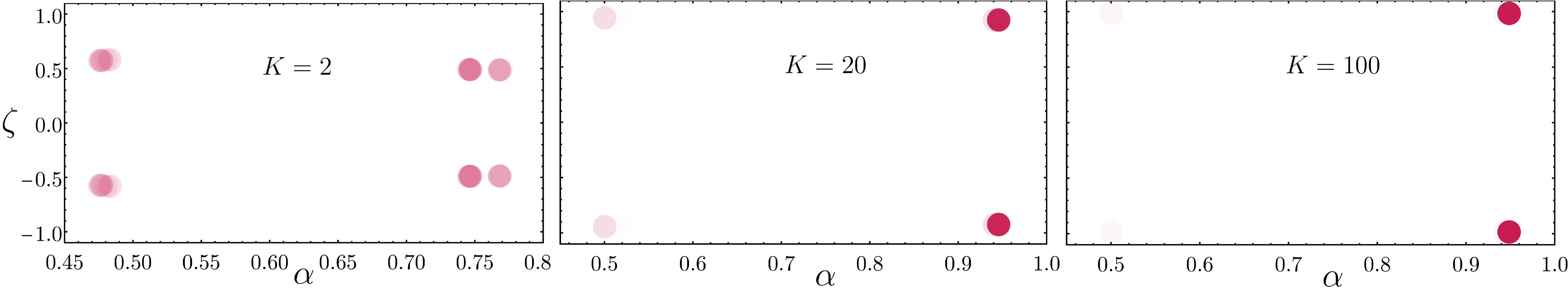}
	\caption{The $(\alpha, \zeta)$ phase-space plots for MOAAAR using optimal measurement angle $\Theta = \Theta\opt$ for different values of $K$. Similar to the Greedy case, we chose $\mg=1$ and $\kd=0.2$. It is clear that when $K$ is large ($K=100$), the system's states are almost always in only two stable states (corresponding to the  stable eigenstates of the two maps $\mathbf{F}_{\pm}^0(\Theta\opt)$.}
	\label{fig:phaseMOAAAR}
\end{figure*}

\subsection{Analytical expression for decoherence}

We wish to find the minimum of the function $\bar\Gamma(\Theta)$ in \erf{eq:asymav}. Recall that $\Theta$ is the single parameter that defines our adaptive algorithm as per Eqs.~\eqref{eq:adtmoa}. This in term defines the four maps in \erf{single-paramap} (from two choices of $s \in \{+1,-1\}$ and two choices of $y \in \{0, 1\}$), as a function of $\Theta$, via the original definition in Eq.~\eqref{eq:F_elements_def}. We are interested in the stable eigenstates $\ul{E}_\pm^0$ of the map ${\bf F}^0_\pm(\Theta)$. With the help of Mathematica, we analytically solve for these eigenvectors and their expansions in the asymptotic regime, taking $\gu, \gd \ll K$. Ignoring an irrelevant complex scalar multiple, these eigenvectors can be represented by the two sufficient statistics introduced in Sec.~\ref{subsec:suff_stat}. These are the parameters $\alpha_s^0$ and $\zeta_s^0$ found by substituting  $\ul{A}_n = \ul{E}_{s}^0$ into $\alpha_n(\ul{A}_n)$ \eqref{defalpha} and  $\zeta_n(\ul{A}_n)$ \eqref{defzeta} respectively. Keeping terms up to $O(1/K)$, we get 
\begin{subequations}\label{eq:alzeanalytic}
\begin{align}
    \alpha_{s}^0=&\,\, \frac{1}{12(2\Theta+\sin2\Theta)^2}(N_\Theta+s\frac{\gamma_{\downarrow}-\gamma_{\uparrow}}{ K}M_\Theta),\\ 
    \zeta_{s}^0=&\,\,s \times \bigg\{ 1-\frac{\gamma_{s}}{K}\big[\csc\Theta(\cos\Theta+\Theta\csc\Theta)\big]\bigg\}, 
\end{align}
\end{subequations}
where we have used $\gamma_s$ with the meaning  $\gamma_+ = \gamma_{\downarrow}$, $\gamma_- = \gamma_{\uparrow}$. The terms $N_\Theta$ and $M_\Theta$ are lengthy functions of $\Theta$ alone, and are shown in  Appendix.~\ref{App:NandM}.

By substituting $\alpha_{s}^0$ and $\zeta_{s}^0$ in Eqs.~\eqref{eq:alzeanalytic} for the eigenstates $\underline{E}_s^0$ in Eq.~\eqref{eq:asymav} and expanding terms to the lowest order of  $(\kd/\ks)$, we obtain, to lowest order, 
\be\label{eq:analymoa}
\bar\Gamma(\Theta) \approx H_{\Theta}  \bg \kd^2/(2\ks^2),
\ee
where
\begin{align}
H_{\Theta} = 
 3\Theta^2 \csc^4 \Theta - [2\Theta (\Theta - \cot\Theta) +1] {\rm csc}^2\Theta + \tfrac{1}{3}\Theta^2  - 1. \label{defH}
\end{align}
This is our most important result,  also shown in the  CL~\cite{PRL}, where we plot $H_{\Theta}$. We find that this function has an absolute minimum of $H\opt \approx 1.254$ at $\Theta = \Theta\opt \approx 1.50055$. This measurement angle $\Theta\opt$ defines our MOAAAR, and is plausibly the lowest possible decoherence rate using projective measurements on a SQ within the regime of Eq.~\eqref{eq:regime}.  
We show in Fig.~\ref{fig:phaseMOAAAR} the phase space plots for $K=2, 20, 100$ using the MOAAAR algorithm. 
This confirms that, even more so than for the Greedy algorithm in Fig.~\ref{fig:phase}, as $K/\mg$ grows, the system spends more and more time in the two stable eigenstates. 

\subsection{Approximation errors and approaching rates}
\label{subsec:apperappra}
With the analytical expressions for MOAAAR, we can analyze any errors that could occur from the approximations we have made. First, let us revisit the approximation, $z_n \approx s_n = \sign(\zeta_n)$, used prior to Eq.~\eqref{eq:avgdec} to justify using the steady state probability of $z_n$ in place of those for $s_n$. From the analytical expression in Eqs.~\eqref{eq:alzeanalytic}, we can see that  $\zeta_s^0 = s[1-O(\gamma_s/\ks)]$, for any value of $\Theta$. Using the result (correct to even higher order) for $\zeta_n$ in Eq.~\eqref{eq:zeta_z}, we can thus say that, 
\begin{align}
    \zeta_n \approx &\,\, \frac{\wp(Y_n, z_n = +1) - \wp(Y_n, z_n = -1)}{\wp(Y_n)} \nonumber \\
    = &\,\, \wp(z_n = +1 | Y_n) - \wp(z_n = -1 | Y_n).
\end{align}  
Combining the above, we have that, in the stable eigenstates, 
\beq \label{probdisz}
\wp(z_n \neq s_n) = O(\gamma_s/\ks).
\eeq 
Therefore, not only do the probabilities of $s_n$ and $z_n$ match on average, but the variables themselves are almost always identical when the system is in the stable eigenstates. However, this raises the question of the probability that the system is in one of these eigenstates, to which we now turn.

\begin{figure}[t!]
\centering
\includegraphics[width=\columnwidth]{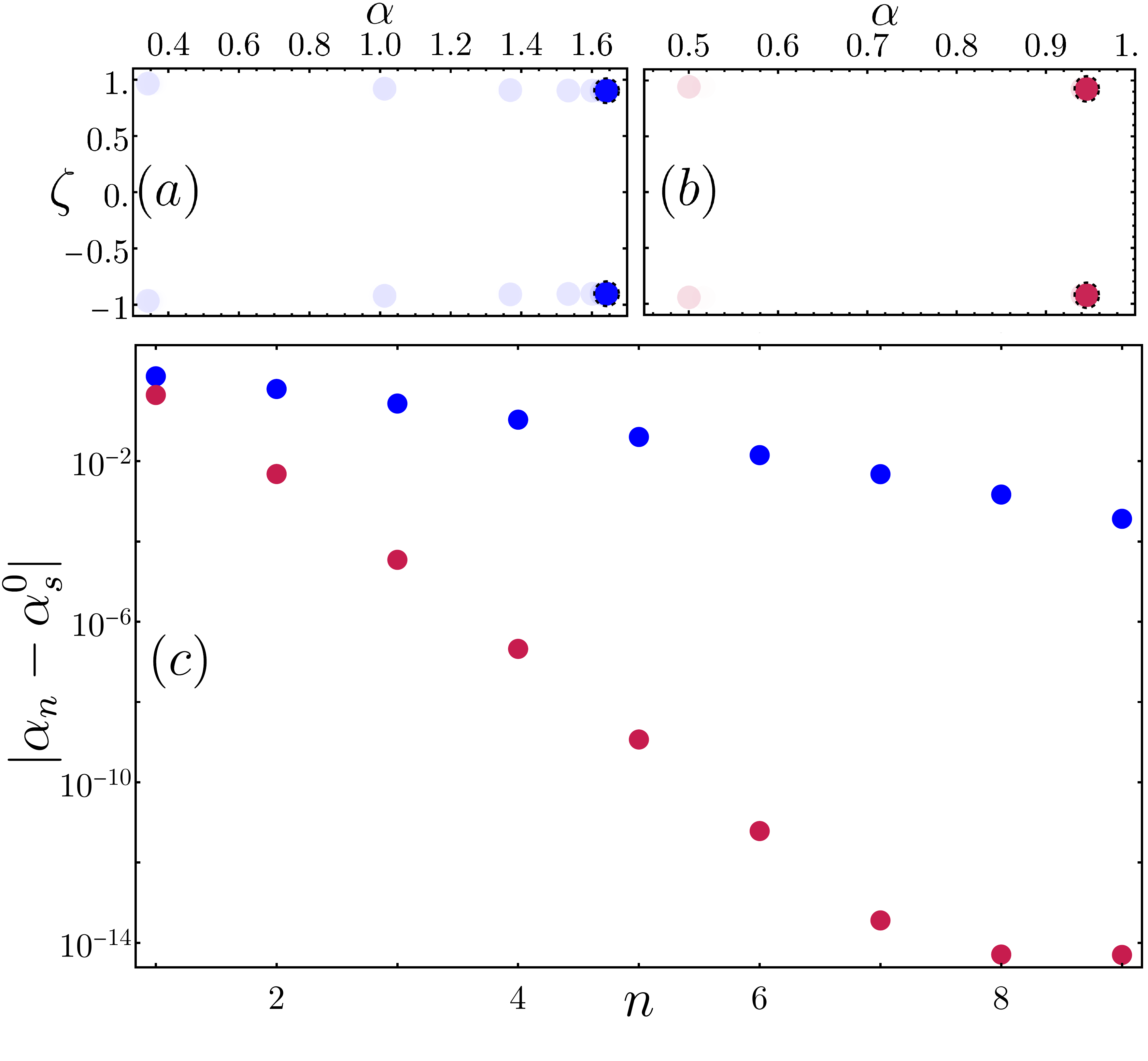}
\caption{Plots showing the exponential approach to the two stable eigenstates on the right side of the phase space. Subplots (a) and (b) are phase-space $(\alpha, \zeta)$ plots under the adaptive algorithm using $\Theta = 1.0$ (blue dots) and $\Theta = \Theta\opt$ (MOAAAR, maroon dots) respectively.  To generate these points, we use $\mathbf{F}_{s}^1 E^0_s$ for both $s \in \{ +1, -1\}$ as a starting point and let the state evolve for the number of time steps $n = 9$. In each case, the eigenstates are shown as dashed circles. (c) A Log-scale plot of distances from the eienstate's values, $|\alpha_n - \alpha_s^0| $, using the same data in (a) and (b). The last maroon data point (at $n = 9$) has reached the numerical machine precision in Mathematica. Here $K=20$, and $\kd=0.2$ as usual.}
\label{fig:exponential}
\end{figure}
 
Because the phase space variables,  $\alpha$ and $\zeta$, are continuous, we cannot strictly talk about the probability of the system being in a stable eigenstate. However, we can calculate the probability that it is arbitrarily close to such an eigenstate.  For a given $s$, if null results ($y=0$) keep occurring, the system's state moves towards the eigenstate $\ul{E}_{\,s}^0$ exponentially fast. This is shown  in Fig.~\ref{fig:exponential} (blue dots), which takes an arbitrary value $\Theta = 1.0$ as an example,  as well as the optimal value $\Theta\opt$. (We do this because we want the results in this subsection, used to justify the arguments of the preceding subsection, to hold for any $\Theta$, not just the optimal value  $\Theta\opt$ found via those arguments.)  The sign of $\zeta$ stays fixed, while the value of $\alpha$ moves towards $\alpha_{s}^0$ exponentially fast for both $s = \pm1$.  Moreover, we see that the transient values of $|\zeta|$ are even closer to $1$ than the values in the stable eigenstates. Thus, it is very safe to assume \erf{probdisz} for the whole evolution.
 
The approximation that the system state is almost always in a stable eigenstate $\ul{E}_\pm^0$ was also used in the first step, \erf{eq:cohdecoh}, towards finding the decoherence rate in Eq.~\eqref{eq:asymav}. Thinking about the alternate expression for the decoherence rate, \erf{alternate-Gamma}, we have to ask how much difference arises, when averaging over all records, as a result of the system state's deviation from the stable eigenstates. After a non-null result, $\alpha$ jumps a finite distance from $\alpha_s^0$. After $n$ time steps, the distance (in $\alpha$) between the system's state and the fixed point is an exponentially decreasing function, \ie, $|\alpha_n - \alpha_s^0| \sim e^{- k n}$, where $k$ is some dimensionless approaching constant, the slope of the curves in Fig.~\ref{fig:exponential}(c).  We can thus approximate the number of steps after a jump that the system is still more than  $\epsilon$-distanced away from the eigenstates as $n_\epsilon = O[\log(1/\epsilon)/k]$. Now, the proportion of its evolution that the system spends this far from the eigenstates is $n_\epsilon$ times the probability per time step of a jump away from the eigenstates. But the latter is just equal to the probability of a non-null result. This is the same order as \erf{probdisz}, the probability that the best estimate of $z_n$ is wrong, which is $O(\bg/K)$. Since a jump takes a system a finite distance away in the $\alpha$-direction, we make the pessimistic assumption that if the system is more than  $\epsilon$-distanced away from the eigenstates then the relative error in the quantity of interest (the decoherence rate) could be of order unity. If, on the other hand, the system is less than  $\epsilon$-distanced  away from the eigenstates (which is almost all the time), then if we are again pessimistic, we will say that the relative error in the quantity of interest is $O(\epsilon)$. Adding these two contributions together, we get the relative error to be $O[\log(1/\epsilon)\, \bg /\ks k] + O(\epsilon)$. The minimum value of this relative error is achieved when we choose $\epsilon$ to  scale as $\bg /\ks k \ll 1$, which gives a relative error that is small in the asymptotic regime. Thus the errors in the approximations leading to \erf{eq:analymoa} are insignificant in the regime (\ref{eq:regime}).

\subsection{Numerical comparison of performance of Greedy and MOAAAR and other algorithms}\label{sec:greedy-moaaar}

Having completed the analytical  analyses of MOAAAR, we turn to numerics to compare it with the Greedy algorithm. We  also compare it to some other, non-optimized, algorithms. 
In Fig.~\ref{fig:Cohvstime}, we plot the  data qubit's decoherence as a function of time. Although the time is not long compared to $\mg=1$, the decoherence for the optimized algorithms are very close to linear in time, which is a consequence of the very simple phase-space dynamics of those algorithms. All of the numerical results were calculated {\emph{exactly} using the coherence (or the expected reward) definition in Eq.~\eqref{eq:GreedyCohAve}, where all possible trajectories were generated using the ${\bf F}$ maps that correspond to the algorithm-chosen measurement angles and waiting times, for all possible realizations of $Y_n$. For the adaptive schemes we took the first measurement angle to be $\pi/2$, since in the absence of any initial information about $z_1$, there is no reason to choose one sign for $\theta$ rather than the other. This is the automatic choice of the full Greedy algorithm, but not of Greedy$_4$. We also do this for later plots. 

From worst-to-best, the curves in Fig.~\ref{fig:Cohvstime} are as follows: (i) the no-control case, Eq.~\eqref{uncond_exp}. (ii) a non-adaptive algorithm with $\theta_{n+1} = \pi/2$ and $\tau_{n+1} = 1/K$. Even this completely unoptimized algorithm already does much better. (iii) the adaptive algorithm of Eqs.~\eqref{eq:adtmoa} with the choice $\Theta=1$, giving $\theta_{n+1} = s_n$ and $\tau_{n+1} = 1/K$, this being the same waiting times as in (ii). Here we see that making the measurement angle adaptive significantly improves the performance. 
(iv) the Greedy$_4$ algorithm of  Eqs.~\eqref{eq:GreedyBinary}, which does much better again. (v) MOAAAR, which like (iii) uses Eqs.~\eqref{eq:adtmoa}, but this time  with the optimized $\Theta\opt = 1.50055$. This is better than Greedy, but only by a small amount.

\begin{figure}[t!]
\includegraphics[width=\columnwidth]{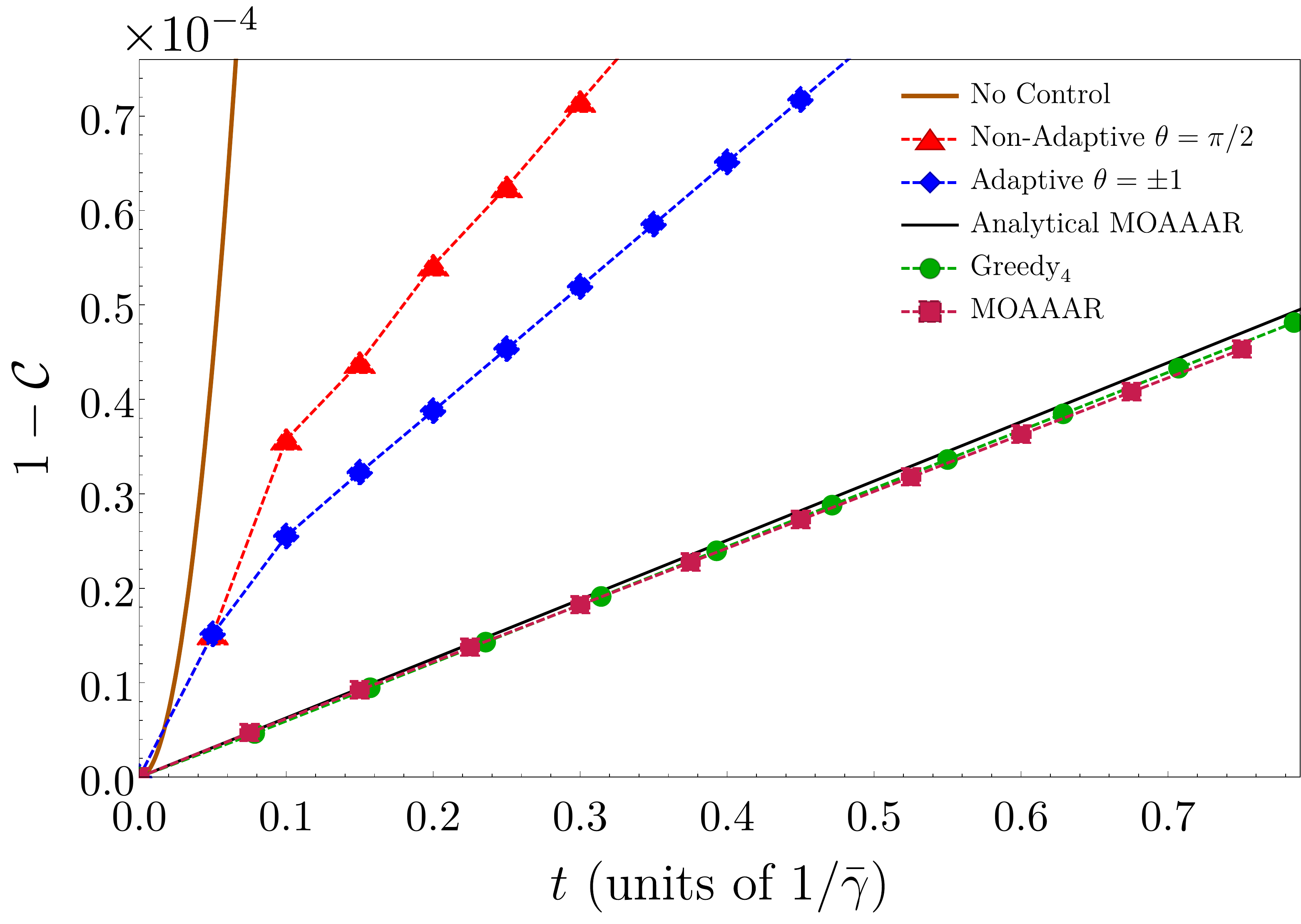}
\caption{Plots of data qubit's decoherence using different SQ's measurement and control strategies. Analytical results are shown with solid curves, while numerical results are shown with points connected via dashed lines. From top to bottom: the no-control case  (solid brown), a non-adaptive algorithm with $\theta_{n+1} = \pi/2$ and $\tau_{n+1} = 1/K$ (red squares), an adaptive algorithm  with $\theta_{n+1} = s_{n}$ and $\tau_{n+1} = 1/K$ (blue squares), the analytical MOAAAR from Eq.~\eqref{eq:analymoa} (solid black), the Greedy$_4$ algorithm (green dots), and the MOAAAR with $\Theta\opt = 1.50055$ (maroon square). In all calculation, we used $\gu=\gd=1$, $\kd=0.2$, and $\ks=20$. }
\label{fig:Cohvstime}
\end{figure}

Finally in Fig.~\ref{fig:Cohvstime}, we also plot the analytical result for MOAAAR from Eq.~\eqref{eq:analymoa}. This agrees quite well with the numerics, but not perfectly. Indeed, the predicted decoherence is worse than what is achieved from the exact simulations, and in fact is worse even than the decoherence achieved by Greedy. This may surprise the reader, since we claimed in the preceding section that all the approximations leading to Eq.~\eqref{eq:analymoa} were justified in the asymptotic regime, \erf{eq:regime}. The issue is that $\ks/\mg = 20$ is not far enough into the asymptotic regime to see the accuracy of the asymptotic analytical result. 

We address this in the next plot, Fig.~\ref{fig:slopes}, which compares the decay rates of the two algorithms for various values of $K$. These rates are obtained  as the slopes of straight lines fitted to data such as those  shown in Fig.~\ref{fig:Cohvstime}, throwing away the first two data points to avoid  transient effects. We see that, using the rate in Eq.~\eqref{eq:analymoa} and $H_{\Theta}$ in Eq.~\eqref{defH}, the MOAAAR and Greedy decoherence rates, scaled by $2\ks^2/(\bg\kd^2)$, do appear to approach the analytic asymptotic values of $H\opt \approx 1.254$ and $H_{\pi/2} \approx 1.290$, respectively, for large $K$. This plot better shows that the decoherence rates for MOAAAR are always smaller than those for Greedy, even for relatively small $K$, where both algorithms give almost the same decoherence rates, which can be seen in the inset of Fig.~\ref{fig:slopes}.

In Fig.~\ref{fig:slopes}, we also plot the decoherence rates calculated with an analytical closed-form expressions, which is derived in {Section~\ref{sec:CF}}. This closed-form expression agrees very well with the numerical results, especially for MOAAAR, over almost two orders of magnitude of variation in $K$. Moreover, the closed form expression can be shown to approach the asymptotic values as $K\to\infty$, with a relative deviation scaling as $\mg/K$. 

\begin{figure}[t!]
\includegraphics[width=\columnwidth]{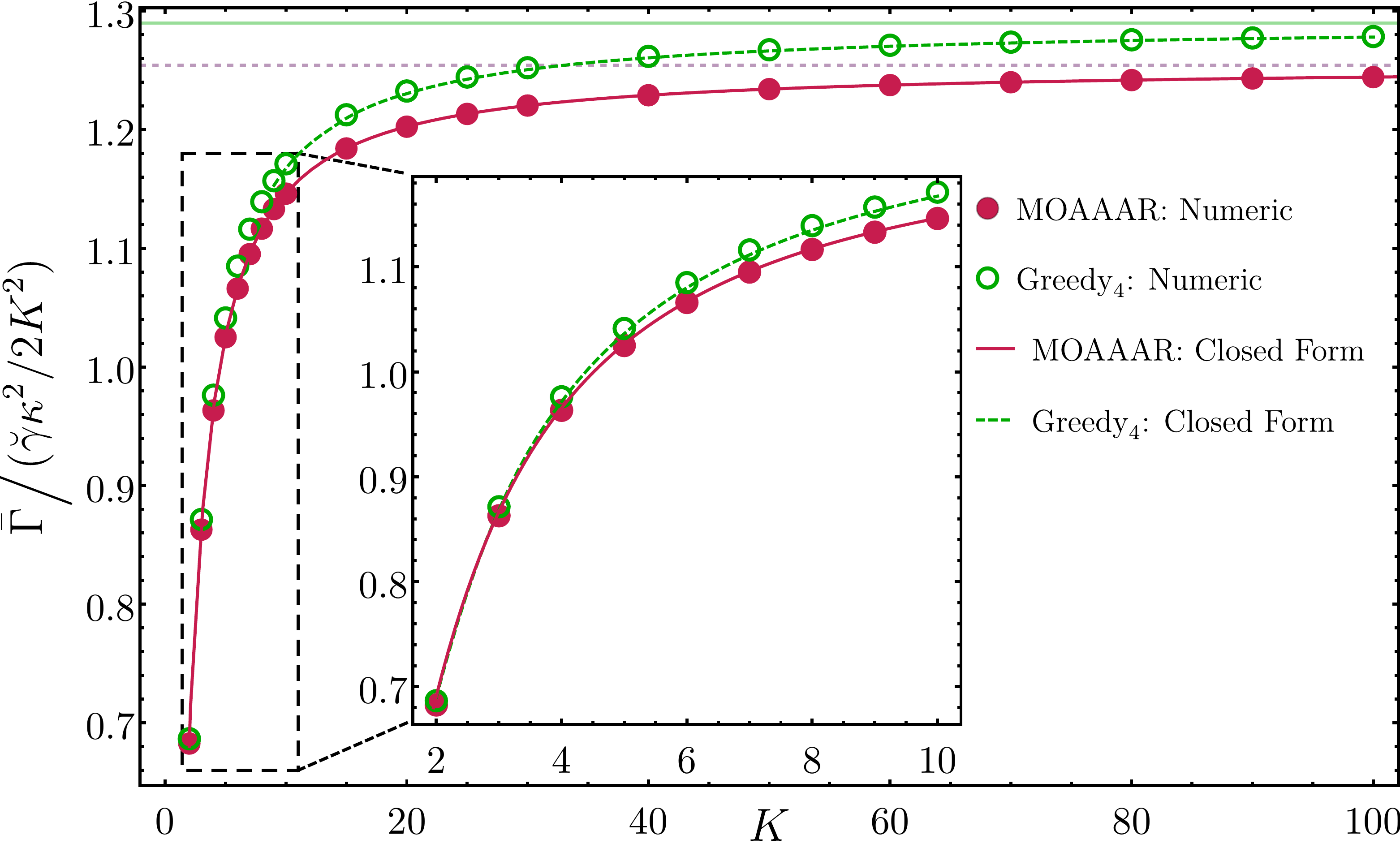}
\caption{The decoherence rates of the data qubit for the two competing strategies: the MOAAAR (maroon) and the Greedy$_4$ algorithm (green), for different values of $\ks$, setting $\gu=\gd=1.0$ and $\kd=0.2$. Numerical results for the rates, shown as maroon disks and the green circles, are calculated from the slopes of curves in Fig.~\ref{fig:Cohvstime} (throwing away the first two points with transient effects). Here, the solid maroon and dashed green curves are not simply lines connecting data points, but rather the closed-form approximation of the rates in  Eq.~\eqref{eq:closedformMOAAAR} for MOAAAR and Eq.~\eqref{eq:closedform} for Greedy$_4$, respectively. The fainter maroon horizontal dashed line near the top shows the analytical result of the MOAAAR decoherence rate in the asymptotic regime Eq.~\eqref{eq:analymoa} with $\Theta\opt$, while the fainter green solid line above it shows the same for Greedy, with $\Theta = \pi/2$ as in Eq.~\eqref{ThetaG0asym}.}
\label{fig:slopes} 
\end{figure}

\subsection{Further numerical evidence for optimality of MOAAAR} 

In the previous section, we showed that MOAAAR performs better than other discussed algorithms. Of course, there are infinitely many possible algorithms that one could compare to, even restricting to the form of strategy in \erf{defnumun}, derived from the separation principle (Sec.~\ref{sec:ASASP}) applied to the sufficient statistics of Sec.~\ref{subsec:suff_stat}. It is because we do not know any way to make such a comparison that we have claimed MOAAAR only to be {\em plausibly} optimal, in the asymptotic regime. The claim of  plausibility is, naturally, subjective, but is based on our detailed study of the problem. Nevertheless, the reader may ask whether more evidence for the optimality of MOAAAR can be given. To address this, in this subsection we consider a more general algorithm.

As the reader will recall, MOAAAR's best measurement angle $\Theta = \Theta\opt \approx 1.50055$ was found under the constraint in Eqs.~\eqref{eq:adtmoa}, which assumed the relationship $\tau=\Theta/K$ between the magnitude of the measurement angle and   the waiting time. This was chosen based on an analysis of Greedy$_2$ in the asymptotic regime, and has an intuitive interpretation in terms of optimal state discrimination; see Sec.~\ref{subsec:Greedy_2}. In this subsection, we relax this constraint and numerically calculate the decoherence rate over a wide range of parameters: $\Theta \in [0,\pi]$ and $\tau \in [1/K,6/K]$. We keep the natural adaptive strategy for the measurement angle, $\theta_{n+1} =s_n \Theta$, where $s_n = \sign(\zeta_n)$, as any other choice would break the symmetry between the RTP states.

\begin{figure}
    \includegraphics[width=\columnwidth]{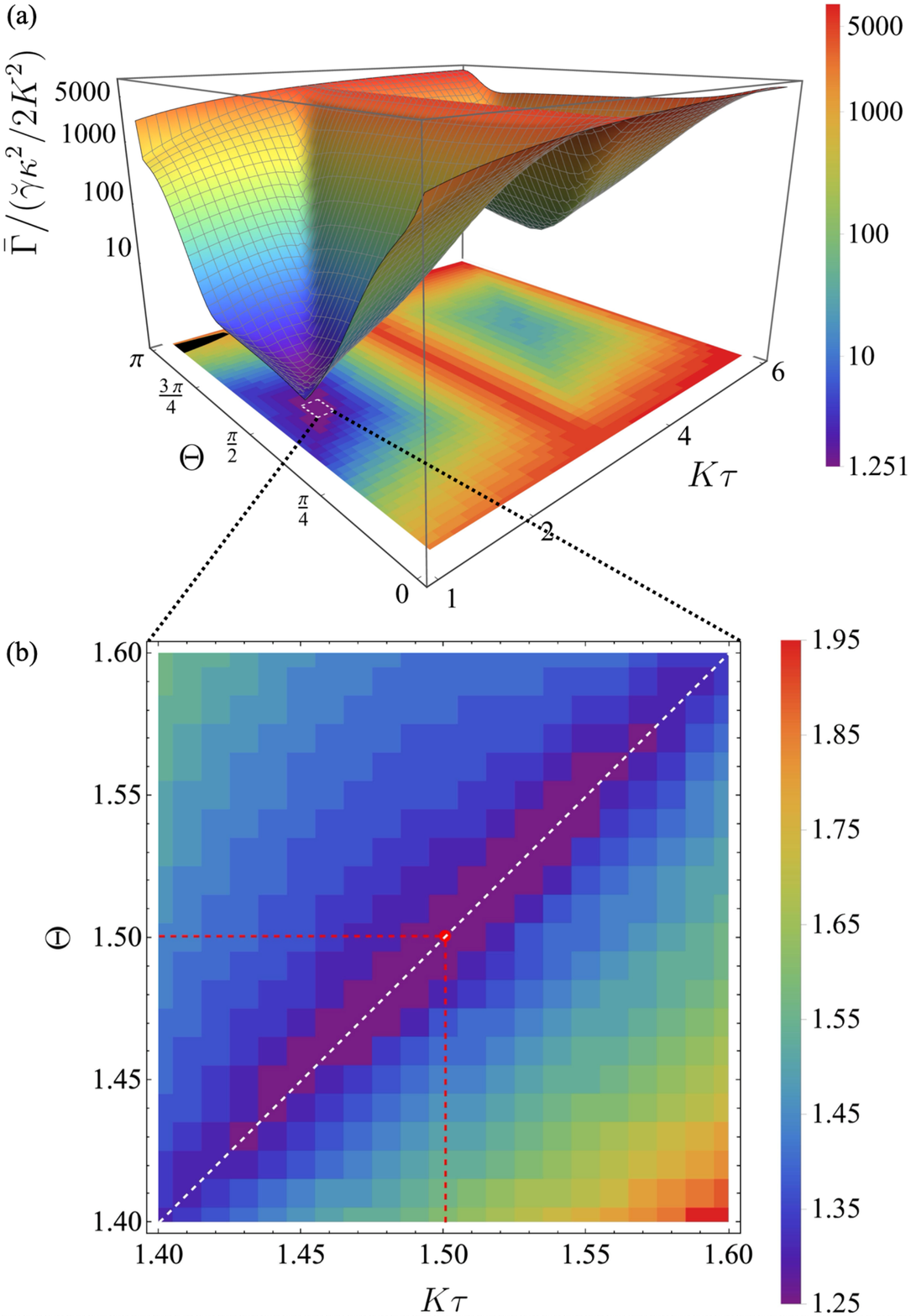}
    \caption{Numerical decoherence rates for different SQ's adaptive measurement strategies defined by measurement angles $\theta_n = s_n \Theta$ and waiting times $\tau_n = \tau$, with $\gu=\gd=1,\kappa=0.2$, and $K=100$.  Subplot (a) shows the decoherence rate as a 3D colored surface plot with a colored contour plot (projection of the surface plot on the $\Theta$--$K\tau$ plane) for $\Theta\in[0,\pi]$ and $\tau\in[1/K,6/K]$. Different colors (on the color bar) indicate different values of the decoherence rate divided by the factor $\bg\kd^2/(2\ks^2)$. Note the log scale for this axis and colour spread. The tiny black area of the plane at $\Theta \approx \pi$, $K\tau \approx 1$ shows where we do not have extreme confidence in the calculated rate. Subplot (b) shows a zoomed-in region,  $\Theta\in[1.4,1.6]$ and $\tau \in[1.4/K,1.6/K]$, which includes a segment of the line $\tau=\Theta/K$ (white dashed line), which was the ansatz adopted to find MOAAAR, including the point $\{\Theta\opt, \tau\opt\} =\{ 1.50055, 1.50055/K\}$ (red dot) defining MOAAAR itself.}
    \label{fig:optimalityandzoom}
\end{figure}

We first investigate the phase-space dynamics and find that, outside the regime where $\tau \approx  \Theta/K$, the very simple dynamics as shown in Fig.~\ref{fig:phaseMOAAAR} or Fig.~\ref{fig:exponential} are no longer evident. Instead, these strategies generally result in highly scattered state dynamics in the phase space. 
This means that we can no longer use eigenstates of the ${\bf F}$ maps to calculate the decoherence rate. We therefore need to use the brute-force calculation following Eq.~\eqref{eq:GreedyCohAve} to compute the exact coherence for a given strategy from averaging over all possible realizations of measurement results $Y_n$. From the coherence data over time, similar to the plots in Fig. \ref{fig:Cohvstime}, we then compute a decoherence rate by fitting the data with a linear function and extracting its slope. 

We also note that the further the measurement strategies are from the regime $\tau \approx \Theta/K$, the longer time it takes for the  coherence process to reach its asymptotic  behaviour of exponential decay. (The ``long" times we consider are still much shorter than the reciprocal of the decay rate, which is why using a linear fit for the long-time decoherence curve is permissible.) This is expected as, if more points in phase space are accessible, it may take longer to reach the ergodic regime. Thus, to avoid the transient effects, we perform the numerical calculations for more steps than previously (in this case, $T/\tau = 15$ steps) and only use the last $5$ steps for the linear-model regression. 
Also, in order to be confident about the decoherence rate extracted from the slope, we calculate the coefficient of determination, known as the $R^2$ value~\cite{DraSmi1998}, of the linear regression fit. Its value ranges from $0$ to $1$, with $R^2 = 1$ representing perfect fitting. We find that we can have extremely high confidence in our linear fits ($R^2>0.998$) except in a tiny region of the parameter space we explore, which is very far from the optimal regime; see Fig. \ref{fig:optimalityandzoom}(a). 

We present in Fig.~\ref{fig:optimalityandzoom} our numerical results for the decoherence rate divided by the factor  $\bg\kd^2/(2\ks^2)$, the same as in Fig.~\ref{fig:slopes}. In Fig.~\ref{fig:optimalityandzoom}(a), we show this 
for the full range of parameters. 
There is one global minimum and another local minimum, which correspond to $\{ \Theta, \tau\} \approx \{ 1.5, 1.5/K\} $ and $\{ \Theta, \tau\} \approx \{ 1.5, (1.5+\pi)/K \}$, respectively. (The pattern of local minima is presumably repeated for $\tau \approx (1.5+n\pi)/K $ for all integers $n$.) 
%
%
In order to see clearly the optimal point, we zoom into the region $\Theta\in[1.4,1.6]$ and $\tau \in[1.4/K,1.6/K]$ in Fig.~\ref{fig:optimalityandzoom}(b), where the decoherence rates are calculated with finer grid values of $\Theta$ and $\tau$. The results show that the smallest values of the decoherence rate all occur around the line $\Theta= K\tau$ (shown as the white dashed line). We also find that $\Theta=1.50055$ and $\tau=1.50055/K$ (shown as the red dot) indeed gives the smallest value of $\bar{\Gamma}/(\check{\gamma}\kappa^2/2K^2)=1.244$, which agrees perfectly with the optimal measurement strategy chosen by MOAAAR. These numerical results thus do offer more evidence that MOAAAR is a plausibly optimal adaptive algorithm for measuring the SQ.\\


\section{Closed-form expression for decoherence rate in non-asymptotic regime}
\label{sec:CF}


In Section~\ref{sec:greedy-moaaar},  we defined a simple adaptive strategy, MOAAAR, by finding the optimal value $\Theta\opt$ that minimizes the average decoherence rate, Eq.~\eqref{eq:asymav}, computed from the stable eigenstates of the two $\bf F$ maps in the asymptotic limit $\ks\gg\mg$. In this section, we show that we can generalize this to find a closed-form expression for the average decoherence  rate even in the non-asymptotic regime. Moreover, it will apply for Greedy$_4$ and Greedy$_2$ as well as MOAAAR. 

We do this by taking into account the small probabilities that the system's states can be outside of the eigenstates.  
In particular, guided by the numerical results in Fig.~\ref{fig:phase} and Fig.~\ref{fig:phaseMOAAAR}, we assume that as well as the two stable eigenstates (on the right side of the phase space) the system also spends time in another two less likely states on the left side of the phase space. Using all the four states and considering the probability transfers among them, we can then recalculate the average decoherence rate in closed form. For simplicity, we consider a symmetric-jump case, \ie, $\gu=\gd$ (though  all of this can be simply adapted to the asymmetric case), where there is a symmetry in swapping between $s=+1$ and $s= -1$. 

We first define the four states, using the maps defined in \erf{eq:compact_notation}, which we reproduce here for convenience: 
\begin{equation}\label{eq:compact_notation1}
\mathbf{F}_{s,a}^y := \mathbf{F} \left( s \Theta({a}), \Delta({a})/ K, y \right).
\end{equation}
Recall that $a =\ \al$ refers to the left side of phase space and $a=\ \ar$ to the right side. The two right states (with $s=\pm 1$) are the stable (largest absolute eigenvalue) eigenvectors, $\ul{E}_{s}^0$, of ${\bf F}_{s, \ar}^0$. The left states are the states that these jump to when a non-null result occurs, ${\bf F}_{s, \ar}^1 \ul{E}_{s}^0$. Because we are assuming the symmetry $\gu=\gd$, we need only one right and one left state, and we define two \emph{normalized complex vectors} for these  as
\begin{subequations}\label{eq:ab_vectors}
\begin{align}
    \ul{r}&=\ul{E}_+^0 \, /\, \ul{I}\tp \ul{E}_+^0\, ,  \label{defa}\\
    \ul{l}&={\bf F}_{\!\!+, \ar}^1 ~ \ul{r}\, / \,  \ul{I}\tp  {\bf F}_{\!\!+, \ar}^1~\ul{r}\, . \label{defb}
\end{align}
\end{subequations}
Note that $\ul{l}$ is a state with $s=-1$ and $a=\ \al$. 

In order to calculate the average decoherence rate, we also need to know the probability that the system be in each of the four states. In the symmetric case, we can reduce the problem to only finding probabilities of the system being either on the right or left sides of the phase space, of which we denote the probabilities by $p^{\ar}$ and $p^{\al} = 1- p^{\ar}$, respectively. To compute the probabilities, we need the real matrices defined in Eq.~\eqref{eq:Fcheck}. Using the same compact notation as in Eq.~\eqref{eq:compact_notation1}, we write these as  $\check{\mathbf{F}}_{s,a}^y$. These matrices satisfy 
\begin{equation} 
\forall s,a, \ \sum_y \ul{I}\tp \check{{\bf F}}_{s,a}^y = \ul{I}\tp.
\end{equation}
That is, for all four cases $s \in \{+1,-1\}$ and $a\in \{\ar,\al\}$,  $\sum_y \check{\bf F}_{s,a}^y$ is a transition matrix, known in stochastic process theory as a column-stochastic matrix, with $\check{\bf F}_{s,a}^1$ and $\check{\bf F}_{s,a}^0$ known as sub-stochastic matrices~\cite{Paz1971,shawan2001}. From these we define \emph{normalized probability vectors}, $\check{\ul{r}}$ and $\check{\ul{l}}$ similar to Eqs.~\eqref{eq:ab_vectors} as
\begin{subequations}\label{eq:check_ab_vectors}
\begin{align}
    \check{\ul{r}}&=\ul{\check E}_+^0 \, /\,  \ul{I}\tp \ul{\check E}_+^0\,, \label{def_checka}\\
    \check{\ul{l}}&={\check{\bf F}}_{\!\!+, \ar}^1\check{\ul{r}}\, / \,  \ul{I}\tp  {\check{\bf F}}_{\!\!+, \ar}^1~\check{\ul{r}}\, , \label{def_checkb}
\end{align}
\end{subequations} 
where $\check{\ul{E}}_+^0$ is the eigenvector of $\check{\bf F}^0_{+, \ar}$ with the largest absolute eigenvalue. Note that the divisors here ensure that $\check{\ul{r}}$ and $\check{\ul{l}}$ have positive elements, which are the conditioned Bayesian probabilities for $\rtp=\pm 1$ when the system has normalized coherence vector $\ul{r}$ and $\ul{l}$, respectively. This is because the sequence of results, $Y_n$, leading to $\ul{r}$ or $\ul{l}$ via the ${\bf F}$-maps will lead to $\check{\ul{r}}$ or $\check{\ul{l}}$ via the $\check {\bf F}$-maps. 

Since a sum of elements (\ie, a 1-norm) of a probability vector,  as defined in Eq.~\eqref{eq:Pn}, tells us a probability of any stochastic measurement records leading up to that state, a norm of the form  $\ul{I}\tp {\check{\bf F}}_{+,\ar}^y \, \check{\ul{r}}$ is the probability that a system in a right state $\ul{r}$ will yield the result $y$. If the result is $y=0$ then it will remain in the same state. If it is $y=1$ then it will jump to the left state $\ul{l}$. Similarly, the norm $\ul{I}\tp {\check{\bf F}}_{-,\al}^y \, \check{\ul{l}}$ is the probability that a system in a left state $\ul{l}$ will yield the result $y$. (Note the change in $s$ and $a$, reflecting the adaptive measurement.) Therefore, at any particular time in the steady-state regime, we can find a probability of the system being in the right states from the two contributions: (1) the right states in the past step (Eq.~\eqref{def_checka}, with the weight $p^{\ar}$) were mapped to themselves via the maps with null results and (2) the left states in the past step (Eq.~\eqref{def_checkb}, with the weight $p^{\al}= 1-p^{\ar}$) were mapped to the right states via the maps with null results. This can be translated into an equation for $p^{\ar}$ as, 
\begin{align}
    p^{\ar} &=    p^{\ar}\,  \left| \ul{I}\tp{\check{\bf F}}_{+,\ar}^0\,\, \check{\ul{r}}\, \right| + (1-p^{\ar})\,\left| \ul{I}\tp{\check{\bf F}}_{-,\al}^0 \,\, \check{\ul{l}}\, \right|,
\end{align}
which has a trivial exact solution. 

\begin{figure*}[th!]
\includegraphics[width=\columnwidth]{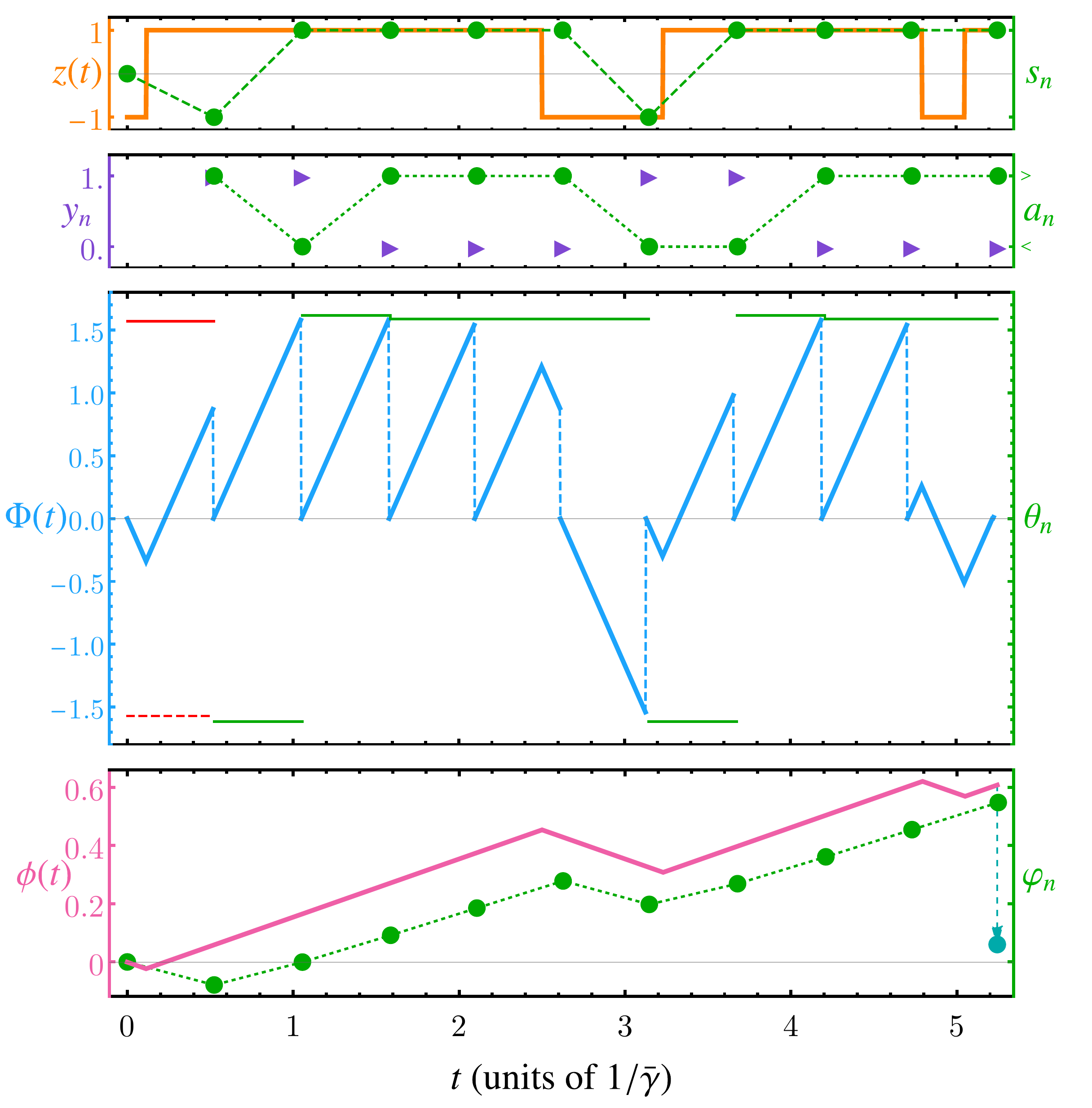}
\includegraphics[width=\columnwidth]{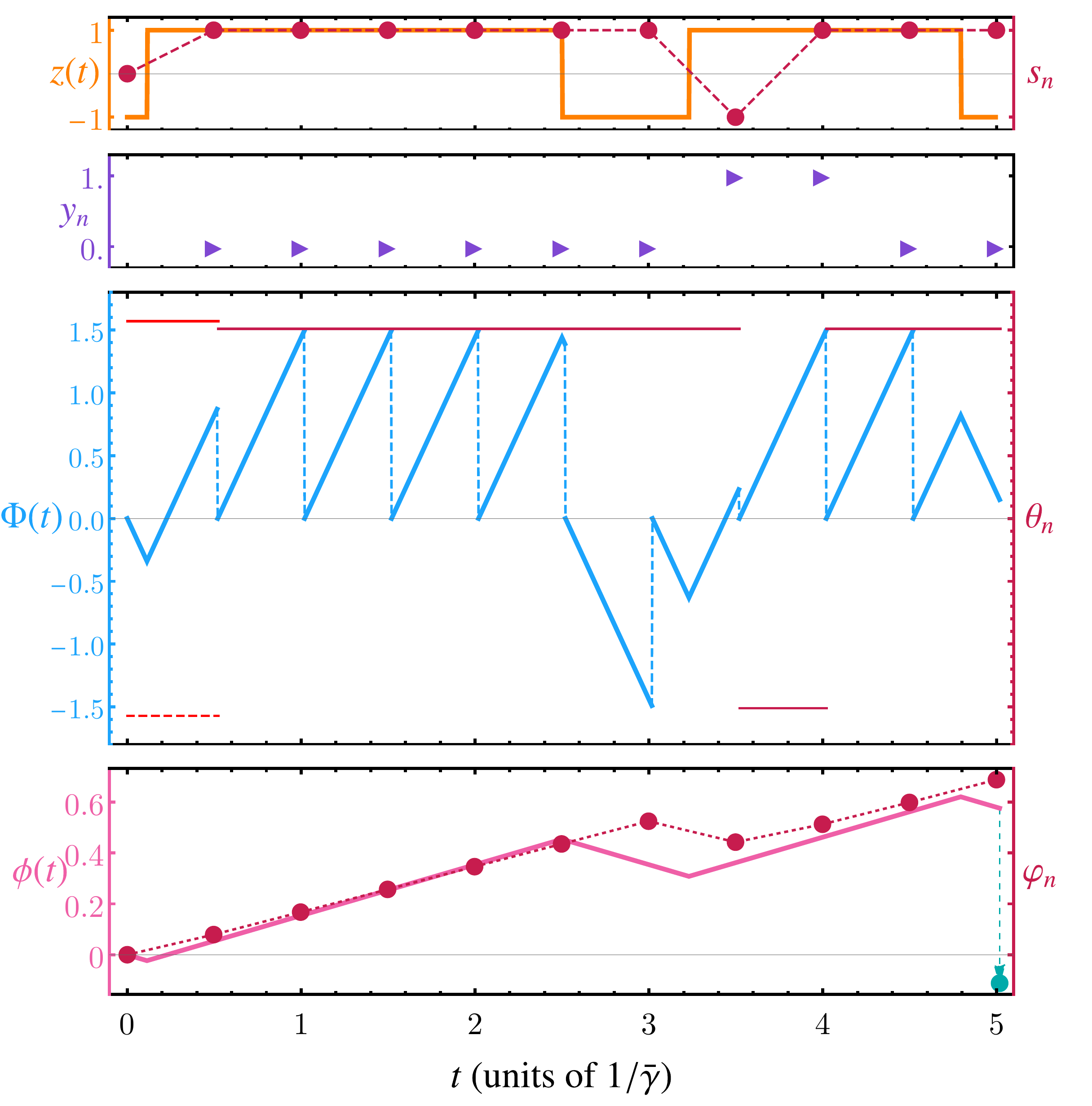}
\caption{Comparison of typical behaviour for Greedy$_4$ (left panels) and the MOAAAR (right panels) for $\ks=3$, $\gu = \gd =1$, and $\kd=0.2$. The top row shows a  typical  realization of the RTP, $z(t)$ (orange solid line), which is the same in both panels, apart from the fact that the total time on the left is greater because, although 10 measurements are shown in both cases, Greedy chooses longer measurement waiting times. Also in the top row, $s_n = \sign(\zeta)$ (circles at the measurement times) is the maximum likelihood estimate of $z(t)$ given $Y_n$ (all measurement results up to and including $y_n$). The second row shows the results $y_n$ (purple  triangles). The left panel also shows $a_n$ (circles), which is the second statistic derived from $Y_n$ that the Greedy algorithm uses (together with $s_n$) to determine the waiting time $\tau_{n+1}$, and angle $\theta_{n+1}$ for, the $(n+1)$th measurement. MOAAAR uses only $s_n$ for its adaptive choice of $\theta_{n+1}=s_n\Theta\opt$ while keeping $\tau_{n+1}=\Theta\opt/\ks$. The third row shows the phase of the SQ, $\Phi(t)$, which is reset to 0 after each measurement. The solid horizontal line segments from $t_n$ to $t_{n+1}$ shows the measurement angle $\theta_{n+1}$ chosen by the algorithms [except the first one (red) which is chosen to be $\pi/2$ (or, equivalently, $-\pi/2$, dashed)].  Having $\Phi(t_{n+1}) = \theta_{n+1}$ guarantees a null result, $y_{n+1}=0$. The last row shows the phase of the data qubit, $\phi(t)$ (solid pink line) and the best estimate of it, $\varphi_n$ (circles), conditioned on $Y_n$. At the final time the estimated phase is used to correct (teal arrow) the unwanted accumulated phase, taking $\phi(t_N)$ to $\phi(t_N)-\varphi_{N}$ (teal circle). 
\label{fig:last}}
\end{figure*}

Now that we have the probabilities $p^{\ar}$ and $p^{\al}$, we can modify the formula of the average decoherence rate in Eq.~\eqref{eq:asymav} for the left and right states with the probability weights. The sum over $s \in \{+1,-1\}$  in Eq.~\eqref{eq:asymav} (which would be equally weighted in the symmetric case) can be replaced with an unequally weighted sum over the right and left states, while the sum over $y$ stays unchanged. For the waiting time in Eq.~\eqref{eq:asymav}, we replace it with a weighted average time $p^{\ar}\tau^{\ar} + p^{\al}\tau^{\al}$.  (Recall that $\tau^a = \Delta^a/\ks$.) Thus, we have the 4-state approximation of the decoherence rate,
\begin{align}\label{eq:closedform}
\bar\Gamma_{4} = 
\frac{1 - 
\sum_{y} \Big[ \left| 
p^{\ar}\,\ul{I}\tp {\bf F}_{+,\ar}^y \,\, \ul{r} \,\right| 
+ \left| p^{\al}\, \ul{I}\tp {\bf F}_{-,\al}^y \,\, \ul{l} \, \right| \Big]}{{p^{\ar} \ddt^{\ar} + p^{\al}\ddt^{\al}}}.
\end{align}
Here the subscript 4 is used to indicate that this is a 4-state approximation, though only two appear because of the symmetry assumed for simplicity in this section.  

For Greedy$_4$, the measurement angles and the waiting times to be used in the ${\bf F}$ maps in \erf{eq:compact_notation1} are as defined in Eqs.~\eqref{eq:GreedyBinary}, namely $\Theta(a) = \Theta_{\rm G}^{a}$, and $\Delta(a)=\Delta_{\rm G}^a$. In the asymptotic regime, where the two parameters of Greedy$_2$ are sufficient, we have $\Theta(a) = \Delta(a)=\Theta_{\rm G}^{a}$. For MOAAAR, we have just the single parameter: $\Theta(a)=\Delta(a)=\Theta\opt$. In this last case we can drop the $a$-dependence on the ${\bf F}$ maps, and 
the average decoherence rate can be further reduced to
\begin{align}\label{eq:closedformMOAAAR}
\bar\Gamma\opt_{4} = \frac{\ks}{\Theta\opt} \bigg(
1 - \sum_{y} \Big[ \left| 
p^{\ar}\,\ul{I}\tp {\bf F}_+^y \, \ul{r} \right|  
+ \left| p^{\al}\, \ul{I}\tp {\bf F}_-^y \, \ul{l} \right| \Big]\bigg).
\end{align}

As already discussed, Fig.~\ref{fig:slopes} shows that \erf{eq:closedform} gives an excellent approximation, for both MOAAAR and Greedy, even far from the asymptotic regime. It is worth noting that we also used this closed-form approximation in Fig.~3 of CL~\cite{PRL}. In the regime of relatively small $K$, specifically for $K\lesssim 5\mg$, the decoherence due to Greedy and MOAAAR are practically identical according to the closed-form approximation \erf{eq:closedform}, and extremely close according to slope-fitting to the exact numerical coherence calculation. 

We close this section (and the results of this paper) by giving a comparison of the workings of Greedy and MOAAAR in the regime of relatively small $K$, in Fig.                                                                                                                                                                                                                                                                                                                                                                                                                        ~\ref{fig:last}. In particular, we choose $\ks=3\mg$, to accentuate the differences between the two algorithms. We plot a typical trajectory, both for the RTP, $\rtp(t)$, which is the same in both cases, and for the measurement results $Y_n$, which are different in the two cases, because obviously the measurement strategy $\mu_n=(\theta_n,\tau_n)$ differ. We also plot the Bayesian parameters $s_n$ and $a_n$ which determine $\mu_n$ in the Greedy case. We omit $a_n$ in the MOAAAR case because only $s_n$ is used in its adaptive algorithm. 

Note that in the left panel of the second row of Fig.~\ref{fig:last}, $a_n$ is perfectly correlated with $y_n$ [except for the very first measurement where $\theta_1=\pm \pi/2$ and initial coherence vector $\ul{A}_0 = \ul P_{\rm ss}$ as in Eq.~\eqref{eq:intstateA}]. Thus Greedy$_4$ does not actually need to calculate $a_n$ in order to implement its algorithm. That is, $a_n$ equals $\ar$ whenever there is a null result, $y_{n+1}=0$, and equals $\al$ whenever there is a non-null result $y_{n+1}=1$. A null result is guaranteed if $\Phi(t_{n+1}) = \theta_{n+1}$. This condition can (and, in fact, usually does) occur for MOAAAR, but it never occurs for Greedy because Greedy always chooses $\ks\tau_{n+1} < |\theta_{n+1}|$. This is as shown in Fig.~\ref{Fig:PairsvsK} and can be seen upon close inspection of Fig.~\ref{fig:last}: $\Phi(t)$ never quite reaches the green horizontal-line segments on the Greedy side, while it often reaches the maroon line segments on the MOAAAR side. 

It is also interesting to look at the data qubit phase $\phi(t) = \kd \int_0^t z(t)$, and its best estimate, $\varphi_n$ at time $t_n$. The absolute value of the slopes of the line segments for the former equal $\kd$, and this is greater than the absolute values of the slopes of the line segments of the latter. That is, the Bayesian algorithms do not estimate $\phi(t)$ by simply integrating $s_n$, their most likely estimates of $z(t)$ at time $t_n$; in equation form, $\varphi_n \neq \sum_{j=1}^n s_j \, \kappa \tau_{j} .$ The plot shows that the optimal Bayesian estimate and final control of the data qubit phase can be surprisingly effective, even when the algorithms are quite often wrong about the RTP, with $s_n\neq z(t_n)$. Though it is not precisely true that $\varphi_N = \kd \tau \sum_{n=1}^N s_n$, this approximation is good enough for a rough derivation of the power-law scaling, $\bg \kd^2 / \ks^2$, for the decoherence rate arising for both MOAAAR and Greedy. The simple argument is as follows. 

From Sec.~\ref{sec:wthctrl}, by  approximating circular statistics by linear statistics, one sees that when the decoherence, ${\cal D} := 1-{\cal C}^{\rm c}$,  is small, it equals the mean square error (MSE) in the final phase estimate. That is, 
\beq \label{Dpenult}
{\cal D} \approx \left\langle \left[\varphi_N-\phi(T)\right]^2 \right\rangle.
\eeq 
Now, from Fig.~\ref{fig:last}, there will be a substantial contribution to the error in the final estimate $\varphi_N$ from the cases where the algorithm's guess, $s_n$, for $z_n$ is wrong. Thus, since we are only interested here in the scaling of the MSE, we can approximate the MSE in \erf{Dpenult} as 
\begin{align} \label{Dult}
{\cal D} &\sim  (\kappa \tau)^2 \left\langle \left[ 
\smallsum{n=1}{N} (s_n - z_n) \right]^2 \right\rangle ,
\end{align}
where the contribution in the sum is non-zero only when the guess is wrong. The proportion of intervals where $s_n$ and $z_n$ differ is just the error probability of \erf{probdisz}, which is of order $\gamma/K$. This scaling of the error probability, $O(\gamma/K)$, is also intuitive to understand, since the probability for a transition in the RTP between SQ measurements scales like this when the waiting times scale as $1/K$. Note that, for simplicity, we are not worrying about the distinction between $\gu$ and $\gd$, so $\bg = \gd = \gu = \mg = \gamma$.

Thus the sum in \erf{Dult} is just the sum of a large number ($\tilde N =O( N \gamma/K)$) of  independent variables $e_{\tilde{n}}$ of random signs and magnitude 2:  
\beq
{\cal D} \sim (\kappa \tau)^2 \left\langle \left[ 
\smallsum{\tilde{n}=1}{\tilde{N}} e_{\tilde{n}}  \right]^2 \right\rangle.
\eeq
Since $\langle e_{\tilde{n}} \rangle=0$ and the variance of a sum of independent random numbers is the sum of the variances, the mean square of that sum is $4\tilde{N}$. Now $N \sim T/\tau$, and $\tau \sim 1/K$ so we obtain finally 
\beq
{\cal D} \sim T \gamma (\kappa / K)^2, 
\eeq
as expected for a decoherence rate $\Gamma \sim \gamma \kappa^2 / K^2$.

\section{Conclusion}
\label{sec:conclusion}
In this paper, we looked into noise mitigation of a data qubit using a spectator qubit (SQ). We considered dephasing in a data qubit caused by a random telegraph process (RTP) with  transition rates $\gu$ and $\gd$. We assumed that the SQ  is affected by the RTP in the same way as the data qubit, but with much greater sensitivity, $\ks \gg \kd$. We then investigated various adaptive measurement and control strategies to reduce the data qubit dephasing. A locally optimal (Greedy) algorithm greatly reduced the no-control decoherence by a factor of order $(\mg/\ks)^2$ in the asymptotic regime $\ks\gg \mg \equiv (\gu + \gd)/2$. This enables us to develop a globally optimized algorithm that could reduce the decoherence rate even more, a Bayesian Map-based Adaptive Algorithm in the Asymptotic Regime (MOAAAR), with a reduction factor of $1.254 (\mg/\ks)^2$.  

Our results show that, in the right regime, the SQ, like techniques such as Dynamical Decoupling and Quantum Error Correction, can mitigate noise arbitrarily well.  The fact that we have a data qubit decoherence rate that is, plausibly, the minimum possible achievable within our model, makes it a useful benchmark for comparison with other techniques for optimizing control, such as machine learning. Numerical and heuristic techniques may be necessary for more complex scenarios, as discussed below, so knowing how quickly, or closely, they can approach optimal performance is useful. 

In our analysis, we assumed that the controller has full knowledge of the dynamical parameters $\kappa$, $\gu$, $\gd$, and $K$. In practice, one would need to characterize these parameters prior to applying noise mitigation via the SQ. This characterisation could be done via a suitable multiqubit  \textit{quantum noise spectroscopy} protocol such as those described in~\cite{paz2017multiqubit, von2020two, chalermpusitarak2021frame}.  Interestingly, this can be done using SQ itself~\cite{youssry2021noise, youssry2020characterization}. It would also be useful to study  how the uncertainty in characterisation would affect our algorithm's performance, and evaluate what the error thresholds in characterisation are for our method to work well, and which parameters are most important. We expect that the least important parameters are $\gu$ and $\gd$, followed by $\ks$, and then $\kd$. For practical applications we would also need to model imperfections including: finite SQ's measurement readout times; imperfect projective measurement of the SQ; and additional decoherence of the SQ.

Beyond this first tranche of future work, several  generalizations of our work are possible and may be useful. Typically, qubits are not affected by a single 2-level RTP, but by several. This could be modelled by a multi-level RTP, and the generalisation should be relatively straightforward. Beyond that, there is the question of general types of classical noise.  Moreover, we assumed that SQ and data qubit feels the same noise, but this assumption could be relaxed;  the SQ is still potentially useful as long as these noises are correlated. With regard to this topic,  Youssry \ea~\cite{youssry2021noise} showed that machine learning can effectively find the correlation between the noises on the SQ and on the data qubit. Finally, we could consider continuous weak measurement of the SQ~\cite{KorPRB99,GoaMilWisSunPRB01} rather than discrete-time projective measurements.  

\section*{Acknowledgments} This work was supported by the Australian Government via the Australia-US-MURI grant
AUSMURI000002, by the Australian Research Council via the Centre of Excellence grant CE170100012, and by National Research Council of Thailand (NRCT) grant N41A640120. A.C.~also acknowledges the support from the NSRF via the Program Management Unit for Human Resources and Institutional Development, Research and Innovation (Thailand) grant B05F640051. We thank members of the AUSMURI collaboration, especially Gerardo Paz Silva, Andrea Morello, and Ken Brown, for feedback on this work. We acknowledge the Yuggera people, the traditional owners of the land at Griffith University on which this work was undertaken. \\ 

\appendix

\section{Derivation of \texorpdfstring{$\mathbf{H}$}{} matrix} \label{app:H}
Here, we derive the matrix elements Eq.~\eqref{eq:Hmatrix} which are described by Eq.~\eqref{eq:H_element_def}, which we rewrite it here:
\begin{equation}\label{eq:H_element_def_APP}
H_{z_0}^{z_t}(t,\kappa) := \int e^{i\kappa X} \wp(X,z_t|z_0) \dd X.
\end{equation}
The above equation is just a Fourier transform of $\wp(X,z_t|z_0)$ evaluated at a conjugate variable $\kappa$. We note that the final result in \eqref{eq:FR-pofX} can be derived via many different techniques. In what follows, we derive the elements of matrix $\mathbf{H}$ using matrix diagonalization.

It is more convenient to work with discrete times and we can take the time-continuum limit at the end. We first divide the time duration of interest, $t\in [t_0, t]$, into $M$ steps: $\{t_0,t_1,\dots, t_m,\dots,t\}$, where $t_m = t_{m-1} + \Delta t$ and $\Delta t$ is an infinitesimal time. We note that $t_m$ is not the same as the measurement time $t_j$ in Eq.~\eqref{eq:measurementtime}. We also assume that the RTP value does not change during the infinitesimal time, \ie, $z(t \in (t_{m-1}, t_m]) = z(t_m)$. Later, we will take the limit of $M \rightarrow\infty$ to recover the time-continuous limit.

The conditional probability in Eq.~\eqref{eq:H_element_def_APP} depends on $z_0$ and $z_t$, as well as the accumulated noise, $X$. Since we did break the total duration time into many infinitesimal steps, we can write the conditional probability, $\wp(X,z_t = z_M |z_0)$, in terms of the ones for the infinitesimal step, which we can solve analytically. We write the conditional probability as
\begin{align}
    \wp(X,z_t &= z_M |z_0) = \!\sum_{z_1, \dots, z_{M-1}}\! \delta\big( X\!- \sum_{m=1}^M z_m \Delta t \big) \wp(z_1,\dots, z_M | z_0)\nonumber \\
    &= \! \sum_{z_1, \dots, z_{M-1}} \! \delta\big( X\!- \sum_{m=1}^M z_m \Delta t \big) \wp(z_M|z_{M-1}) \cdots \wp(z_1|z_0) ,
\end{align}
where the first summation is over RTPs at multiple times and we have used $\delta(\cdots)$ as the Dirac $\delta$-function. We then substitute the above equation into Eq.~\eqref{eq:H_element_def_APP} and find 
\begin{widetext}
\begin{align}
    H_{z_0}^{z_t}
    &= \! \int \! \dd X e^{ i \kd X} \sum_{z_1, \dots, z_{M-1}}\! \delta\big( X\! - \sum_{m=1}^M z_m \Delta t \big) \wp(z_M|z_{M-1}) \cdots  \wp(z_1|z_0) \notag \\
    &= \sum_{z_1, \dots, z_{M-1}} \exp\left[i \kd \sum_{m=1}^M z_m \Delta t \right] ~ \wp(z_M|z_{M-1}) \cdots \wp(z_1|z_0)
\end{align}
\end{widetext}
where in the second line we have used the integration property of Dirac $\delta$-function. We then distribute the exponential terms to accompany all the conditional probabilities and obtain
\begin{widetext}
\begin{align}\label{eq:Hzz_APP}
    H_{z_0}^{z_t}
    &=\sum_{z_1, \dots, z_{M-1}} e^{ i \kd z_M \Delta t/2}
   \Big( e^{ i \kd z_M \Delta t/2} 
    \wp(z_M|z_{M-1})
    e^{ i \kd z_{M-1} \Delta t/2} \Big)
    ~ \cdots ~
    \Big( e^{ i \kd z_1 \Delta t/2} 
    \wp(z_1|z_0)
    e^{ i \kd z_0 \Delta t/2} \Big)
    e^{ - i \kd z_0 \Delta t/2} \notag \\
    &= e^{ i \kd z_M \Delta t/2} \left(\sum_{z_1, \dots, z_{M-1}} M_{z_M,z_{M-1}} \cdots M_{z_1,z_0} \right) e^{ -i \kd z_0 \Delta t/2}
\end{align}
\end{widetext}
where in the second line we have defined
\begin{align}
    M_{z,z'} := e^{ i \kd z \Delta t/2}
    \wp(z|z')
    e^{ i \kd z' \Delta t/2}.
\end{align}
This $M_{z,z'}$ can be thought of as an element of a $2\times2$ matrix, where $z, z' \in \{ -1, +1 \}$. We can then use the results of Eqs.~\eqref{eq:J_matrix} and \eqref{eq:Pt_RTP} and, and write 
\begin{equation}
    \wp(z|z') = \left( 1+ \frac{1-e^{-2\mg \Delta t}}{2\mg}\mathbf{J} \right)_{z,z'}
\end{equation}
which is simply the matrix term in Eq.~\eqref{eq:Pt_RTP}. Therefore, the matrix $\mathbf{M}$ becomes
\begin{align}
\mathbf{M} =\dfrac{1}{2\mg}
\left(
\begin{matrix}
\left(\gu + \gd e^{-2\mg \Delta t} \right) e^{i \kd \Delta t} &
\gu - \gu e^{-2\mg \Delta t} \\
\\
\gd - \gd e^{-2\mg \Delta t} & 
\left(\gd + \gu e^{-2\mg \Delta t} \right) e^{-i \kd \Delta t}
\end{matrix}
\right).
\end{align}
Using the above matrix, we write the elements $H_{z_0}^{z_t}$ as
\begin{align}
    H_{z_0}^{z_t} = e^{ i \kd z_M \Delta t/2} \left(\mathbf{M}^M\right)_{z_M,z_0} e^{ -i \kd z_0 \Delta t/2}.
\end{align}
Thus, the matrix $\mathbf{H}$ becomes 
\begin{equation}\label{eq:H_Mton_APP}
    \mathbf{H} = \lim_{M \rightarrow \infty} \left(\begin{matrix}
    e^{i\kd \Delta t} & 0 \\ 0 & 1
    \end{matrix}\right)
    \mathbf{M}^M 
    \left(\begin{matrix}
    e^{-i\kd \Delta t} & 0 \\ 0 & 1
    \end{matrix}\right),
\end{equation}
where we have included the limit of $M \rightarrow \infty$ to reach the continuous time limit (since $\Delta t = (t-t_0)/M$,  in the limit of $M \rightarrow \infty$, we have $\Delta t \rightarrow 0$). We can calculate ${\bf M}^M$ by decomposing $\mathbf{M}$ into $\mathbf{M} = \mathbf{VDV}^{-1}$, where
\begin{align}
    \mathbf{D} = 
    \left(
    \begin{matrix}
    L^{+}-S & 0 \\
    0 & L^{+}+S
    \end{matrix}
    \right)
\end{align}
and
\begin{align}
    \mathbf{V} = 
    \left(
    \begin{matrix}
    \dfrac{2\mg}{\gd}\dfrac{L^{-}-S}{1-e^{-2\mg \Delta t}} & \dfrac{2\mg}{\gd}\dfrac{L^{-}+S}{1-e^{-2\mg \Delta t}} \\
    \\
    1 & 1
    \end{matrix}
    \right),
\end{align}
with
\begin{align}
    L^\pm &:= \dfrac{1}{4\mg} \Big[
    \left(\gu + \gd e^{-2\mg \Delta t} \right)e^{i\kd \Delta t} \pm \left(\gd + \gu e^{-2\mg \Delta t} \right) e^{-i\kd \Delta t}
    \Big], \\
    S &:= \sqrt{(L^{-})^2+ \frac{\bg}{4\mg}\left(1-e^{-2\mg \Delta t}\right)^2}.
\end{align}
Since $\mathbf{D}$ is a diagonal matrix, we have 
${\bf M}^M ={\bf V D}^M {\bf V}^{-1}$, with 
\begin{align}
\mathbf{D}^M = \left(\begin{matrix}
(L^+-S)^M & 0 \\ 0 & (L^++S)^M
\end{matrix}\right).
\end{align}
We then substitute these results in Eq.~\eqref{eq:H_Mton_APP} and after taking the limit, we find the final result as
\begin{widetext}
\begin{align}
    \mathbf{H}(t,\kd) = \exp[-\mg t] 
    \left(\begin{matrix}
    \cosh\left(\dfrac{\mc(\kd)}{2} t\right)- \dfrac{\nc(\kd)}{\mc(\kd)}\sinh\left(\dfrac{\mc(\kd) }{2}t\right) & \dfrac{2 \gu}{\mc(\kd)} \sinh\left(\dfrac{\mc(\kd)}{2}t\right) \\
    \\
    \dfrac{2 \gd}{\mc(\kd)} \sinh\left(\dfrac{\mc(\kd)}{2}t\right) & \cosh\left(\dfrac{\mc(\kd)}{2}t\right)+\dfrac{\nc(\kd)}{\mc(\kd)}\sinh\left(\dfrac{\mc(\kd) }{2}t\right)
    \end{matrix}\right)
\end{align}
\end{widetext}
with
\begin{subequations}
\begin{align}
\mc(\kd) &= \sqrt{(\gd+\gu)^2-4i\kd(\gd-\gu) -4\kd^2},\\
\nc(\kd) &= (\gd-\gu) -2i\kd.
\end{align}
\end{subequations}
This result is in agreement with the Fourier transform introduced in Eq.~\eqref{eq:FR-pofX}.

\section{Derivation of \texorpdfstring{$\mathbf{F}$}{}}\label{app:F_Derivation}
In this section, we derive the elements of the matrix $\mathbf{F}$. We start with the definition in Eq.~\eqref{eq:F_elements_def} which we rewrite here,
\begin{align}
    F_{z_{n-1}}^{z_n}\!\left(\mu_n, y_n\right) :=\!\!
    \int\!\! dx_n \,\wp_{\meas_n}(y_n|x_n)\,\wp(x_n,z_n|z_{n-1})e^{i \kd x_n}.
\end{align}
We recognize that the above equation is a Fourier transform of $\wp_{\meas_n}(y_n | x_n) \wp(x_n, z_n | z_{n-1})$, where formally we write
\begin{align}\label{eq:F_elements_Fourier}
F_{z_{n-1}}^{z_n}  = & \mathcal{F}_{x_n \rightarrow \kd} [ \wp_{\meas_n}(y_n | x_n) \wp(x_n, z_n | z_{n-1})],
\end{align}
where $\mathcal{F}_{x\rightarrow k}[f]$ is defined as
\begin{equation}
\mathcal{F}_{x\rightarrow k}[f]:= \int \dd x\, e^{i k x}\, f.
\end{equation}
We then use the convolution theorem, where a Fourier transform of a product of two functions is equal to a product of their individual Fourier transforms. That is, 
\begin{equation}
    \mathcal{F}_{x\rightarrow \kappa}[f\cdot g] = \int \dd k \,  \mathcal{F}_{x\rightarrow k}[f] \cdot \mathcal{F}_{x\rightarrow(\kappa - k)}[g].
\end{equation}
Applying the convolution theorem to Eq.~\eqref{eq:F_elements_Fourier}, we obtain
\begin{align}
&F_{z_{n-1}}^{z_n} =  \\
&\int \!\! \dd k \, \mathcal{F}_{\! x_n \rightarrow k} [ \wp_{\meas_n}(y_n | x_n)] {\cal F}_{x_n \rightarrow (\kd-k)} [ \wp(x_n, z_n | z_{n-1})], \notag
\end{align}
where each term in the integrand can be computed as the following. For the first one, we use Eq.~\eqref{eq:forwardP}, which we rewrite it here
\begin{align}
\wp_{\meas_n}(y_n|x_n) &= y_n + (-1)^{y_n}\cos^2\sq{\half (\theta_n - \ks x_n ) },
\end{align}
where its Fourier transform is 
\begin{align}
\mathcal{F}_{x_n \rightarrow k} [ \wp_{\meas_n}(y_n | x_n)] = \frac{1}{4} &\big[2 \delta_{k,0} +(-1)^{y_n} e^{-i \theta_n} \delta_{k, -\ks} \notag \\
&+(-1)^{y_n} e^{+i \theta_n} \delta_{k, +\ks} \big],
\end{align}
where $\delta$ here is the Kronecker delta function. For the second Fourier transform, we have
\begin{align}
    \mathcal{F}_{x_n \rightarrow(\kd-k)} &[\wp(x_n,z_n|z_{n-1})] \notag\\
    &=\!\! \int \!\! \dd x_n\,  \wp(x_n,z_n|z_{n-1}) e^{i(\kd - k)x_n} \notag \\ 
    &=H_{z_{n-1}}^{z_n}(\tau_n,\kd -k),
\end{align}
which we have calculated in the previous section. Combining these two terms and performing the integration over $k$, we find
\begin{align}
    F_{z_{n-1}}^{z_n} =  \frac{1}{4} &\big[2 H_{z_{n-1}}^{z_n}(\tau_n,\kd) \notag \\
    &+(-1)^{y_n} e^{-i \theta_n} H_{z_{n-1}}^{z_n}(\tau_n, \kd+K) \notag \\
    &+(-1)^{y_n} e^{+i \theta_n} H_{z_{n-1}}^{z_n}(\tau_n,\kd -K) \big].
\end{align}
Since, in the above equation, all the indices are the same, we can write a matrix form of the above equation as
\begin{align}\label{eq:FH_APP}
    \mathbf{F} = \frac{1}{4} &\big\{ 2 \mathbf{H}(\tau_n,\kd) +(-1)^{y_n} \times \notag \\
    & \hspace{3ex} \big[ e^{-i \theta_n} \mathbf{H}(\tau_n,\kd+K) +  e^{+i \theta_n} \mathbf{H}(\tau_n, \kd -K) \big] \big\}.
\end{align}

\section{Greedy optimal angles}\label{app:BerryWiseman}
Here we explain  details of the optimization process of the Greedy algorithm. The task is to find an optimum value for the measurement angles $\Greedythi, \Greedythii$ and  $\Greedythii'$, which maximize the coherence $\cohc_{\rm (i)}$ and $\cohc_{\rm (ii)}$ as discussed around Eqs.~\eqref{eq:GreedyCoh} and \eqref{eq:GreedyOpt}. In general, this optimization task can be computationally heavy since it needs to search over all possible values of the three angles, each in the range of $[-\pi/2, \pi/2]$. However, we can significantly simplify the task by using analytical results shown in Berry and Wiseman, Ref.~\cite{BerWisBre01}, where our coherence can be rewritten in their particular form of a cost function which has only three possible optimal values. Following the method presented in Ref.~\cite{BerWisBre01}, we find that the optimum of $\theta_n$ should be one of these three values:
\begin{subequations}
\begin{align}
\theta^{0}(\tau) &=\arg(ba^*-c^*a),\\
\theta^{\pm}(\tau) &=\arg\sqrt{\frac{c_2\pm \sqrt{c_2^2+|c_1|^2}}{c_1}},
\end{align}
\end{subequations}
where
\begin{subequations}
\begin{align}
a &= 2 ~ \underline{I}\tp \mathbf{H}(\tau,\kappa)\underline{A}_n,\\
b &= \underline{I}\tp \mathbf{H}(\tau,\kappa+K)\underline{A}_n, \\
c &= \underline{I}\tp \mathbf{H}(\tau,\kappa-K)\underline{A}_n,
\end{align}
\end{subequations}
and
\begin{subequations}
\begin{align}
c_1 &= (a^* c)^2-(a b^*)^2 + 4(|b|^2 - |c|^2) b^* c,\\
c_2 &= -2i~\text{Im}(a^2 b^* c^*).
\end{align}
\end{subequations}
We note that these angles are functions of the prospect waiting (to measure) time, $\tau$. To implement the Greedy algorithm, we only need to check these three values at each time step, which is a much simpler task  computationally. As an example, we show in Fig.~\ref{fig:C_three_theta} the coherence $\cohc_{\rm (i)}$ for the three different Berry-Wiseman angles as functions of $\tau$. The starting point is a value of $\cohc(t_n)$, right after the n$th$ measurement, choosing $n=2$.  We show the optimal value, $\coh^{\rm op}_{\rm (i)} = \max_{\Greedythi} ~\cohc_{\rm (i)} (\Greedythi)$, in the dashed curve as a function of $\tau$.

\begin{figure}[t]
\includegraphics[width=0.95\linewidth]{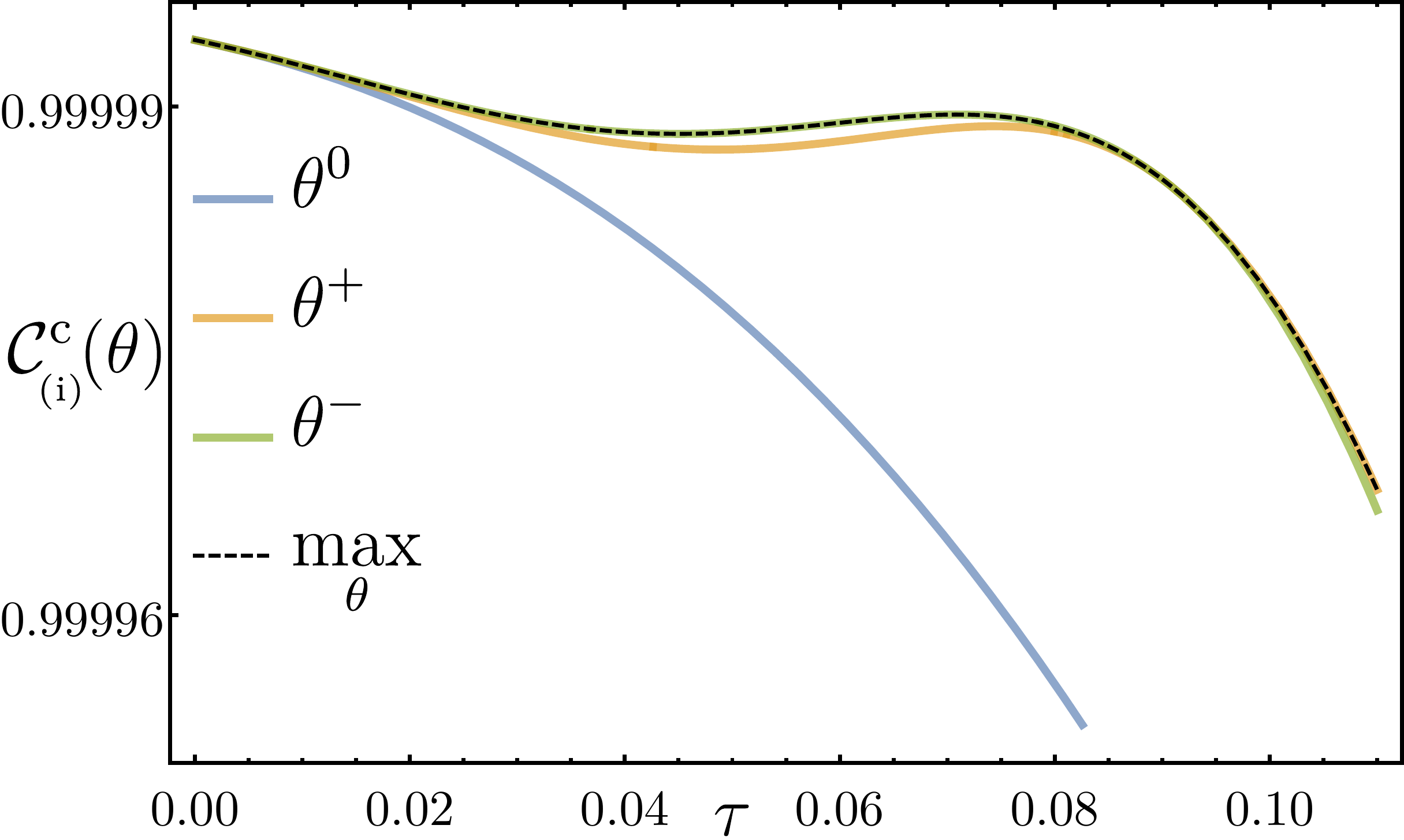}
\caption{\label{fig:C_three_theta} An example of maximizing the coherence (reward function), $\cohc_{\rm (i)}(\theta)$, over the three Berry-Wiseman angles in the Greedy algorithm. We can see that $\cohc_{\rm (i)}(\theta^0)$ is always the smallest, while the other two are competing. The dashed line is the maximum of these three curves. The parameters for this plot are, $\ks=20, \kd=0.2$ and $\gu=\gd=1$.}
\end{figure}

\begin{figure}[t]
\includegraphics[width=0.95\linewidth]{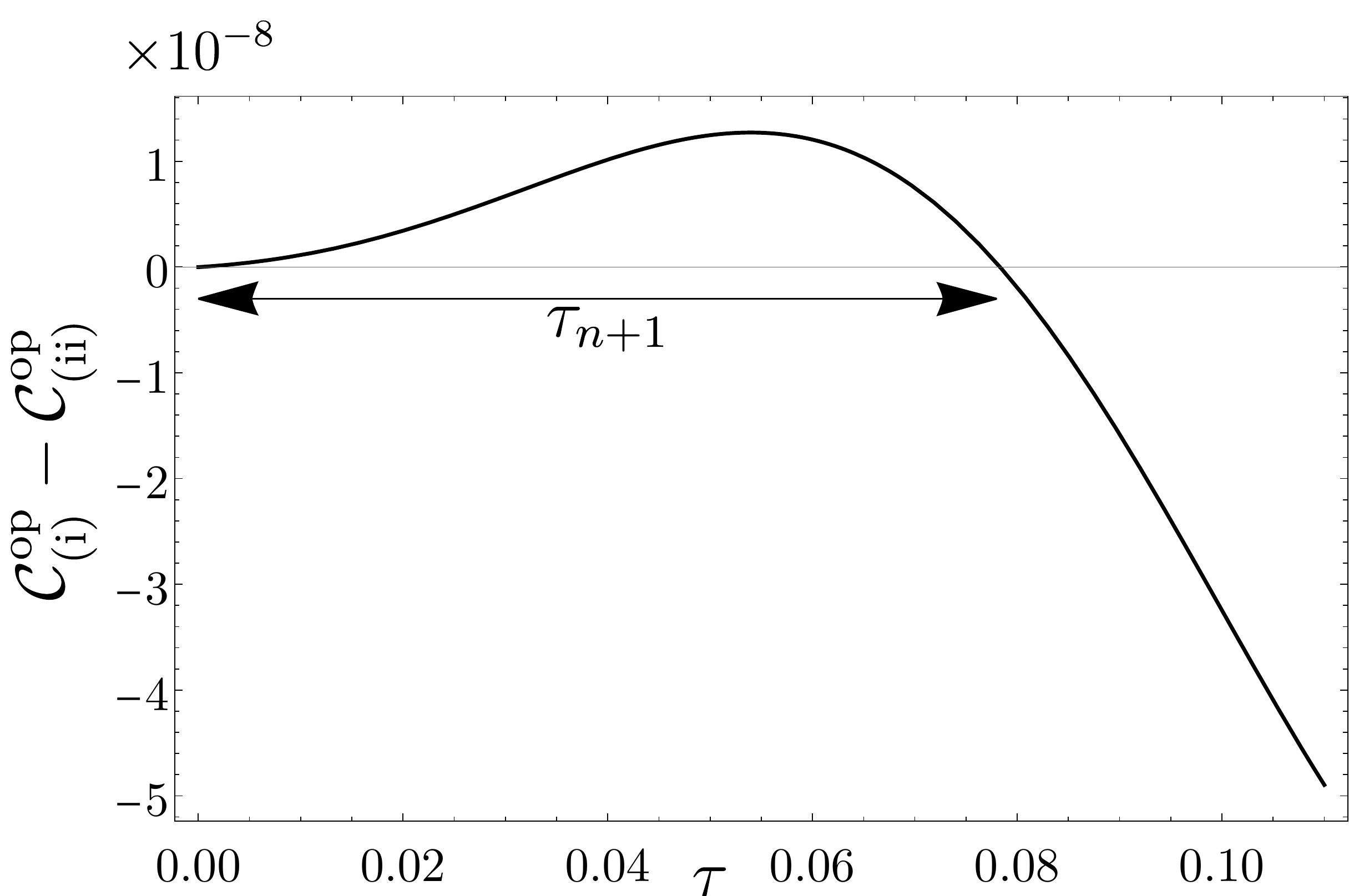}
\caption{\label{fig:Greedy_tau} An example of the process of choosing the next measurement (waiting) time $\tau_{n+1}$ in the Greedy algorithm. The plot shows the difference between $\coh^{\rm op}_{\rm (i)}$ and  $\coh^{\rm op}_{\rm (ii)}$ as a function of $\tau$. The waiting time is chosen when this function crossed the zero line. We used the same parameters as in Fig.~\ref{fig:C_three_theta} }
\end{figure}

In the Greedy algorithm in Eq.~\eqref{eq:GreedyCoh}, we also have to optimize the coherence with two measurement angles, \ie, optimizing the function $\cohc_{\rm (ii)}(\theta_{\rm(ii)}, \theta'_{\rm(ii)})$ over the total of 9 pairs of $\theta_{\rm(ii)}, \theta'_{\rm(ii)}$(3 options for each). We can simplify this task even further by realizing that the second measurement has a waiting time equal to $\dt$ after the first measurement. Thus we expect that during that infinitesimal time the absolute value of the accumulated phase by the spectator qubit is simply $K \dt$.  Then, we know the best angle should be $\Greedythii'= \sign (\zeta_n) K \dt$, which we also numerically checked that this angle does maximize the reward for the second measurement. So the optimization task only needs to search over the 3 possible angles of $\Greedythii$. We note here that  we do not show plots for the two measurement case here, because it is very similar to the one measurement case, with the difference of the order $10^{-8}$.

The maximization of the coherence (reward function) at time $t_n$ can then be used in choosing the next waiting time $\tau_{n+1}$ and angle $\theta_{n+1}$. From the results above, we find the optimum angles for the one and two measurement scenarios as
\begin{align}\label{eq:chooseangleAPP}
    \theta_{\rm(i)}^{\rm op}(\tau) := \argmax_{\Greedythi} \, \cohc_{\rm (i)}(\Greedythi), \\
    \theta_{\rm(ii)}^{\rm op}(\tau) := \argmax_{\Greedythii} \, \cohc_{\rm (ii)}(\Greedythii),
\end{align}
which are still functions of arbitrary $\tau$. To choose the next measurement setting, we first need to calculate to compare the reward between the two scenarios. That is, we compute the difference, $ D(\tau) := \coh^{\rm op}_{\rm (i)} - \coh^{\rm op}_{\rm (ii)}$, which is a function of $\tau$, shown in Fig.~\ref{fig:Greedy_tau}. So long as $D(\tau)$ is positive (\ie, $\coh^{\rm op}_{\rm (i)} > \coh^{\rm op}_{\rm (ii)}$) no measurement is required. On the other hand if $D(\tau) < 0 $, we conclude that a measurement should have been done to increase the coherence. Thus, the next waiting time, $\tau_{n+1}$ is chosen when $D(\tau)$ crosses the zero line, \ie,
\begin{align}
    D(\tau_{n+1})=0.
\end{align}
Now that we found $\tau_{n+1}$, we can use it to identify the choice of the next measurement angle,
\begin{equation}
    \theta_{n+1} = \theta_{\rm(ii)}^{\rm op}(\tau_{n+1}),
\end{equation}
using the maximized angles in Eq.~\eqref{eq:chooseangleAPP}.\\

\section{Expressions for \texorpdfstring{$N_\Theta$}{} and \texorpdfstring{$M_\Theta$}{}}\label{App:NandM}
Here, we present the lengthy expressions of $N_\Theta$ and $M_\Theta$ that we used in Eq.~\eqref{eq:alzeanalytic},
\begin{align}
N_\Theta &= 24 \bigg[\Theta - 2\Theta^3 + 3\Theta\cos 2\Theta  + \frac{1}{4} \sin 4 \Theta -\frac{1}{2} \sin 2\Theta \hspace{20pt} \notag \\
         & \hspace{5ex}+ \Theta^2 \big(8 \cot \Theta +4 \Theta \csc^2 \Theta -\sin 2\Theta\big)  \bigg], \\
         \notag \\
M_\Theta &=-15 + 8\Theta^2 (\Theta^2-15) \notag \\
         & \hspace{2ex}-12(2\Theta^2-1) \cos2\Theta +3\cos4\Theta \notag \\ 
         & \hspace{2ex}+192\Theta^3\cot^3\Theta -96\Theta^2(\Theta^2-1)\csc^2 \Theta \notag \\
         & \hspace{2ex}+96\Theta^4\csc^4\Theta+16\Theta^3\sin2\Theta.
\end{align}

\bibliographystyle{apsrev4-2}

\begin{thebibliography}{63}%
\makeatletter
\providecommand \@ifxundefined [1]{%
 \@ifx{#1\undefined}
}%
\providecommand \@ifnum [1]{%
 \ifnum #1\expandafter \@firstoftwo
 \else \expandafter \@secondoftwo
 \fi
}%
\providecommand \@ifx [1]{%
 \ifx #1\expandafter \@firstoftwo
 \else \expandafter \@secondoftwo
 \fi
}%
\providecommand \natexlab [1]{#1}%
\providecommand \enquote  [1]{``#1''}%
\providecommand \bibnamefont  [1]{#1}%
\providecommand \bibfnamefont [1]{#1}%
\providecommand \citenamefont [1]{#1}%
\providecommand \href@noop [0]{\@secondoftwo}%
\providecommand \href [0]{\begingroup \@sanitize@url \@href}%
\providecommand \@href[1]{\@@startlink{#1}\@@href}%
\providecommand \@@href[1]{\endgroup#1\@@endlink}%
\providecommand \@sanitize@url [0]{\catcode `\\12\catcode `\$12\catcode
  `\&12\catcode `\#12\catcode `\^12\catcode `\_12\catcode `\%12\relax}%
\providecommand \@@startlink[1]{}%
\providecommand \@@endlink[0]{}%
\providecommand \url  [0]{\begingroup\@sanitize@url \@url }%
\providecommand \@url [1]{\endgroup\@href {#1}{\urlprefix }}%
\providecommand \urlprefix  [0]{URL }%
\providecommand \Eprint [0]{\href }%
\providecommand \doibase [0]{https://doi.org/}%
\providecommand \selectlanguage [0]{\@gobble}%
\providecommand \bibinfo  [0]{\@secondoftwo}%
\providecommand \bibfield  [0]{\@secondoftwo}%
\providecommand \translation [1]{[#1]}%
\providecommand \BibitemOpen [0]{}%
\providecommand \bibitemStop [0]{}%
\providecommand \bibitemNoStop [0]{.\EOS\space}%
\providecommand \EOS [0]{\spacefactor3000\relax}%
\providecommand \BibitemShut  [1]{\csname bibitem#1\endcsname}%
\let\auto@bib@innerbib\@empty
\bibitem [{\citenamefont {Temme}\ \emph {et~al.}(2017)\citenamefont {Temme},
  \citenamefont {Bravyi},\ and\ \citenamefont {Gambetta}}]{TemBra2017}%
  \BibitemOpen
  \bibfield  {author} {\bibinfo {author} {\bibfnamefont {K.}~\bibnamefont
  {Temme}}, \bibinfo {author} {\bibfnamefont {S.}~\bibnamefont {Bravyi}},\ and\
  \bibinfo {author} {\bibfnamefont {J.~M.}\ \bibnamefont {Gambetta}},\ }\href
  {https://doi.org/10.1103/PhysRevLett.119.180509} {\bibfield  {journal}
  {\bibinfo  {journal} {Phys. Rev. Lett.}\ }\textbf {\bibinfo {volume} {119}},\
  \bibinfo {pages} {180509} (\bibinfo {year} {2017})}\BibitemShut {NoStop}%
\bibitem [{\citenamefont {Preskill}(2018)}]{Pre2018}%
  \BibitemOpen
  \bibfield  {author} {\bibinfo {author} {\bibfnamefont {J.}~\bibnamefont
  {Preskill}},\ }\href {https://doi.org/10.22331/q-2018-08-06-79} {\bibfield
  {journal} {\bibinfo  {journal} {{Quantum}}\ }\textbf {\bibinfo {volume}
  {2}},\ \bibinfo {pages} {79} (\bibinfo {year} {2018})}\BibitemShut {NoStop}%
\bibitem [{\citenamefont {Endo}\ \emph {et~al.}(2018)\citenamefont {Endo},
  \citenamefont {Benjamin},\ and\ \citenamefont {Li}}]{EndBen2018}%
  \BibitemOpen
  \bibfield  {author} {\bibinfo {author} {\bibfnamefont {S.}~\bibnamefont
  {Endo}}, \bibinfo {author} {\bibfnamefont {S.~C.}\ \bibnamefont {Benjamin}},\
  and\ \bibinfo {author} {\bibfnamefont {Y.}~\bibnamefont {Li}},\ }\href
  {https://doi.org/10.1103/PhysRevX.8.031027} {\bibfield  {journal} {\bibinfo
  {journal} {Phys. Rev. X}\ }\textbf {\bibinfo {volume} {8}},\ \bibinfo {pages}
  {031027} (\bibinfo {year} {2018})}\BibitemShut {NoStop}%
\bibitem [{\citenamefont {Kandala}\ \emph {et~al.}(2019)\citenamefont
  {Kandala}, \citenamefont {Temme}, \citenamefont {C{\'{o}}rcoles},
  \citenamefont {Mezzacapo}, \citenamefont {Chow},\ and\ \citenamefont
  {Gambetta}}]{KanTem2019}%
  \BibitemOpen
  \bibfield  {author} {\bibinfo {author} {\bibfnamefont {A.}~\bibnamefont
  {Kandala}}, \bibinfo {author} {\bibfnamefont {K.}~\bibnamefont {Temme}},
  \bibinfo {author} {\bibfnamefont {A.~D.}\ \bibnamefont {C{\'{o}}rcoles}},
  \bibinfo {author} {\bibfnamefont {A.}~\bibnamefont {Mezzacapo}}, \bibinfo
  {author} {\bibfnamefont {J.~M.}\ \bibnamefont {Chow}},\ and\ \bibinfo
  {author} {\bibfnamefont {J.~M.}\ \bibnamefont {Gambetta}},\ }\href
  {https://doi.org/10.1038/s41586-019-1040-7} {\bibfield  {journal} {\bibinfo
  {journal} {Nature}\ }\textbf {\bibinfo {volume} {567}},\ \bibinfo {pages}
  {491} (\bibinfo {year} {2019})}\BibitemShut {NoStop}%
\bibitem [{\citenamefont {Viola}\ \emph {et~al.}(1999)\citenamefont {Viola},
  \citenamefont {Knill},\ and\ \citenamefont {Lloyd}}]{Vio99}%
  \BibitemOpen
  \bibfield  {author} {\bibinfo {author} {\bibfnamefont {L.}~\bibnamefont
  {Viola}}, \bibinfo {author} {\bibfnamefont {E.}~\bibnamefont {Knill}},\ and\
  \bibinfo {author} {\bibfnamefont {S.}~\bibnamefont {Lloyd}},\ }\href
  {https://doi.org/10.1103/PhysRevLett.82.2417} {\bibfield  {journal} {\bibinfo
   {journal} {Phys. Rev. Lett.}\ }\textbf {\bibinfo {volume} {82}},\ \bibinfo
  {pages} {2417} (\bibinfo {year} {1999})}\BibitemShut {NoStop}%
\bibitem [{\citenamefont {Viola}\ and\ \citenamefont
  {Knill}(2003)}]{viola2003robust}%
  \BibitemOpen
  \bibfield  {author} {\bibinfo {author} {\bibfnamefont {L.}~\bibnamefont
  {Viola}}\ and\ \bibinfo {author} {\bibfnamefont {E.}~\bibnamefont {Knill}},\
  }\href {https://doi.org/10.1103/PhysRevLett.90.037901} {\bibfield  {journal}
  {\bibinfo  {journal} {Phys. Rev. Lett.}\ }\textbf {\bibinfo {volume} {90}},\
  \bibinfo {pages} {037901} (\bibinfo {year} {2003})}\BibitemShut {NoStop}%
\bibitem [{\citenamefont {Biercuk}\ \emph {et~al.}(2011)\citenamefont
  {Biercuk}, \citenamefont {Doherty},\ and\ \citenamefont
  {Uys}}]{biercuk2011dynamical}%
  \BibitemOpen
  \bibfield  {author} {\bibinfo {author} {\bibfnamefont {M.}~\bibnamefont
  {Biercuk}}, \bibinfo {author} {\bibfnamefont {A.}~\bibnamefont {Doherty}},\
  and\ \bibinfo {author} {\bibfnamefont {H.}~\bibnamefont {Uys}},\ }\href
  {https://doi.org/10.1088/0953-4075/44/15/154002} {\bibfield  {journal}
  {\bibinfo  {journal} {Journal of Physics B: Atomic, Molecular and Optical
  Physics}\ }\textbf {\bibinfo {volume} {44}},\ \bibinfo {pages} {154002}
  (\bibinfo {year} {2011})}\BibitemShut {NoStop}%
\bibitem [{\citenamefont {Ng}\ \emph {et~al.}(2011)\citenamefont {Ng},
  \citenamefont {Lidar},\ and\ \citenamefont {Preskill}}]{NgLid2011}%
  \BibitemOpen
  \bibfield  {author} {\bibinfo {author} {\bibfnamefont {H.~K.}\ \bibnamefont
  {Ng}}, \bibinfo {author} {\bibfnamefont {D.~A.}\ \bibnamefont {Lidar}},\ and\
  \bibinfo {author} {\bibfnamefont {J.}~\bibnamefont {Preskill}},\ }\href
  {https://doi.org/10.1103/PhysRevA.84.012305} {\bibfield  {journal} {\bibinfo
  {journal} {Phys. Rev. A}\ }\textbf {\bibinfo {volume} {84}},\ \bibinfo
  {pages} {012305} (\bibinfo {year} {2011})}\BibitemShut {NoStop}%
\bibitem [{\citenamefont {Souza}\ \emph {et~al.}(2011)\citenamefont {Souza},
  \citenamefont {\'Alvarez},\ and\ \citenamefont {Suter}}]{SouAlv2011}%
  \BibitemOpen
  \bibfield  {author} {\bibinfo {author} {\bibfnamefont {A.~M.}\ \bibnamefont
  {Souza}}, \bibinfo {author} {\bibfnamefont {G.~A.}\ \bibnamefont
  {\'Alvarez}},\ and\ \bibinfo {author} {\bibfnamefont {D.}~\bibnamefont
  {Suter}},\ }\href {https://doi.org/10.1103/PhysRevLett.106.240501} {\bibfield
   {journal} {\bibinfo  {journal} {Phys. Rev. Lett.}\ }\textbf {\bibinfo
  {volume} {106}},\ \bibinfo {pages} {240501} (\bibinfo {year}
  {2011})}\BibitemShut {NoStop}%
\bibitem [{\citenamefont {Medford}\ \emph {et~al.}(2012)\citenamefont
  {Medford}, \citenamefont {Cywi\ifmmode~\acute{n}\else \'{n}\fi{}ski},
  \citenamefont {Barthel}, \citenamefont {Marcus}, \citenamefont {Hanson},\
  and\ \citenamefont {Gossard}}]{MedCyw2012}%
  \BibitemOpen
  \bibfield  {author} {\bibinfo {author} {\bibfnamefont {J.}~\bibnamefont
  {Medford}}, \bibinfo {author} {\bibfnamefont {L.}~\bibnamefont
  {Cywi\ifmmode~\acute{n}\else \'{n}\fi{}ski}}, \bibinfo {author}
  {\bibfnamefont {C.}~\bibnamefont {Barthel}}, \bibinfo {author} {\bibfnamefont
  {C.~M.}\ \bibnamefont {Marcus}}, \bibinfo {author} {\bibfnamefont {M.~P.}\
  \bibnamefont {Hanson}},\ and\ \bibinfo {author} {\bibfnamefont {A.~C.}\
  \bibnamefont {Gossard}},\ }\href
  {https://doi.org/10.1103/PhysRevLett.108.086802} {\bibfield  {journal}
  {\bibinfo  {journal} {Phys. Rev. Lett.}\ }\textbf {\bibinfo {volume} {108}},\
  \bibinfo {pages} {086802} (\bibinfo {year} {2012})}\BibitemShut {NoStop}%
\bibitem [{\citenamefont {Paz-Silva}\ and\ \citenamefont
  {Lidar}(2013)}]{PazLid2013}%
  \BibitemOpen
  \bibfield  {author} {\bibinfo {author} {\bibfnamefont {G.~A.}\ \bibnamefont
  {Paz-Silva}}\ and\ \bibinfo {author} {\bibfnamefont {D.~A.}\ \bibnamefont
  {Lidar}},\ }\href {https://doi.org/10.1038/srep01530} {\bibfield  {journal}
  {\bibinfo  {journal} {Scientific Reports}\ }\textbf {\bibinfo {volume} {3}},\
  \bibinfo {pages} {1530} (\bibinfo {year} {2013})}\BibitemShut {NoStop}%
\bibitem [{\citenamefont {Zhang}\ \emph {et~al.}(2014)\citenamefont {Zhang},
  \citenamefont {Souza}, \citenamefont {Brandao},\ and\ \citenamefont
  {Suter}}]{ZhaSou2014}%
  \BibitemOpen
  \bibfield  {author} {\bibinfo {author} {\bibfnamefont {J.}~\bibnamefont
  {Zhang}}, \bibinfo {author} {\bibfnamefont {A.~M.}\ \bibnamefont {Souza}},
  \bibinfo {author} {\bibfnamefont {F.~D.}\ \bibnamefont {Brandao}},\ and\
  \bibinfo {author} {\bibfnamefont {D.}~\bibnamefont {Suter}},\ }\href
  {https://doi.org/10.1103/PhysRevLett.112.050502} {\bibfield  {journal}
  {\bibinfo  {journal} {Phys. Rev. Lett.}\ }\textbf {\bibinfo {volume} {112}},\
  \bibinfo {pages} {050502} (\bibinfo {year} {2014})}\BibitemShut {NoStop}%
\bibitem [{\citenamefont {Shor}(1995)}]{Shor1995}%
  \BibitemOpen
  \bibfield  {author} {\bibinfo {author} {\bibfnamefont {P.~W.}\ \bibnamefont
  {Shor}},\ }\href {https://doi.org/10.1103/PhysRevA.52.R2493} {\bibfield
  {journal} {\bibinfo  {journal} {Phys. Rev. A}\ }\textbf {\bibinfo {volume}
  {52}},\ \bibinfo {pages} {R2493} (\bibinfo {year} {1995})}\BibitemShut
  {NoStop}%
\bibitem [{\citenamefont {Steane}(1996)}]{Steane1996}%
  \BibitemOpen
  \bibfield  {author} {\bibinfo {author} {\bibfnamefont {A.~M.}\ \bibnamefont
  {Steane}},\ }\href {https://doi.org/10.1103/PhysRevLett.77.793} {\bibfield
  {journal} {\bibinfo  {journal} {Phys. Rev. Lett.}\ }\textbf {\bibinfo
  {volume} {77}},\ \bibinfo {pages} {793} (\bibinfo {year} {1996})}\BibitemShut
  {NoStop}%
\bibitem [{\citenamefont {Terhal}(2015)}]{Terhal2015}%
  \BibitemOpen
  \bibfield  {author} {\bibinfo {author} {\bibfnamefont {B.~M.}\ \bibnamefont
  {Terhal}},\ }\href {https://doi.org/10.1103/RevModPhys.87.307} {\bibfield
  {journal} {\bibinfo  {journal} {Rev. Mod. Phys.}\ }\textbf {\bibinfo {volume}
  {87}},\ \bibinfo {pages} {307} (\bibinfo {year} {2015})}\BibitemShut
  {NoStop}%
\bibitem [{\citenamefont {Gupta}\ \emph {et~al.}(2020)\citenamefont {Gupta},
  \citenamefont {Edmunds}, \citenamefont {Milne}, \citenamefont {Hempel},\ and\
  \citenamefont {Biercuk}}]{GupEdm2020}%
  \BibitemOpen
  \bibfield  {author} {\bibinfo {author} {\bibfnamefont {R.~S.}\ \bibnamefont
  {Gupta}}, \bibinfo {author} {\bibfnamefont {C.~L.}\ \bibnamefont {Edmunds}},
  \bibinfo {author} {\bibfnamefont {A.~R.}\ \bibnamefont {Milne}}, \bibinfo
  {author} {\bibfnamefont {C.}~\bibnamefont {Hempel}},\ and\ \bibinfo {author}
  {\bibfnamefont {M.~J.}\ \bibnamefont {Biercuk}},\ }\href
  {https://doi.org/10.1038/s41534-020-0286-0} {\bibfield  {journal} {\bibinfo
  {journal} {npj Quantum Information}\ }\textbf {\bibinfo {volume} {6}},\
  \bibinfo {pages} {53} (\bibinfo {year} {2020})}\BibitemShut {NoStop}%
\bibitem [{\citenamefont {Majumder}\ \emph {et~al.}(2020)\citenamefont
  {Majumder}, \citenamefont {{Andreta de Castro}},\ and\ \citenamefont
  {Brown}}]{MajAnd2020}%
  \BibitemOpen
  \bibfield  {author} {\bibinfo {author} {\bibfnamefont {S.}~\bibnamefont
  {Majumder}}, \bibinfo {author} {\bibfnamefont {L.}~\bibnamefont {{Andreta de
  Castro}}},\ and\ \bibinfo {author} {\bibfnamefont {K.~R.}\ \bibnamefont
  {Brown}},\ }\href {https://doi.org/10.1038/s41534-020-0251-y} {\bibfield
  {journal} {\bibinfo  {journal} {npj Quantum Information}\ }\textbf {\bibinfo
  {volume} {6}},\ \bibinfo {pages} {19} (\bibinfo {year} {2020})}\BibitemShut
  {NoStop}%
\bibitem [{\citenamefont {Singh}\ \emph {et~al.}(2022)\citenamefont {Singh},
  \citenamefont {Conor E.~Bradley}, \citenamefont {Ramesh}, \citenamefont
  {White},\ and\ \citenamefont {Bernien}}]{SinBra2022}%
  \BibitemOpen
  \bibfield  {author} {\bibinfo {author} {\bibfnamefont {K.}~\bibnamefont
  {Singh}}, \bibinfo {author} {\bibfnamefont {C.~E.}\ \bibnamefont {Bradley}}, \bibinfo {author} {\bibfnamefont {S.}~\bibnamefont {Anand}},\bibinfo {author} {\bibfnamefont {V.}~\bibnamefont {Ramesh}},
  \bibinfo {author} {\bibfnamefont {R.}~\bibnamefont {White}},\ and\ \bibinfo
  {author} {\bibfnamefont {H.}~\bibnamefont {Bernien}},\ }\href@noop {}
  {\bibfield  {journal} {\bibinfo  {journal} {arXiv:2208.11716 [quant-ph]}\ }
  (\bibinfo {year} {2022})}\BibitemShut {NoStop}%
\bibitem [{\citenamefont {Hanson}\ \emph {et~al.}(2007)\citenamefont {Hanson},
  \citenamefont {Kouwenhoven}, \citenamefont {Petta}, \citenamefont {Tarucha},\
  and\ \citenamefont {Vandersypen}}]{HanKou2007}%
  \BibitemOpen
  \bibfield  {author} {\bibinfo {author} {\bibfnamefont {R.}~\bibnamefont
  {Hanson}}, \bibinfo {author} {\bibfnamefont {L.~P.}\ \bibnamefont
  {Kouwenhoven}}, \bibinfo {author} {\bibfnamefont {J.~R.}\ \bibnamefont
  {Petta}}, \bibinfo {author} {\bibfnamefont {S.}~\bibnamefont {Tarucha}},\
  and\ \bibinfo {author} {\bibfnamefont {L.~M.~K.}\ \bibnamefont
  {Vandersypen}},\ }\href {https://doi.org/10.1103/RevModPhys.79.1217}
  {\bibfield  {journal} {\bibinfo  {journal} {Rev. Mod. Phys.}\ }\textbf
  {\bibinfo {volume} {79}},\ \bibinfo {pages} {1217} (\bibinfo {year}
  {2007})}\BibitemShut {NoStop}%
\bibitem [{\citenamefont {Pla}\ \emph {et~al.}(2012)\citenamefont {Pla},
  \citenamefont {Tan}, \citenamefont {Dehollain}, \citenamefont {Lim},
  \citenamefont {Morton}, \citenamefont {Jamieson}, \citenamefont {Dzurak},\
  and\ \citenamefont {Morello}}]{PlaTan2012}%
  \BibitemOpen
  \bibfield  {author} {\bibinfo {author} {\bibfnamefont {J.~J.}\ \bibnamefont
  {Pla}}, \bibinfo {author} {\bibfnamefont {K.~Y.}\ \bibnamefont {Tan}},
  \bibinfo {author} {\bibfnamefont {J.~P.}\ \bibnamefont {Dehollain}}, \bibinfo
  {author} {\bibfnamefont {W.~H.}\ \bibnamefont {Lim}}, \bibinfo {author}
  {\bibfnamefont {J.~J.~L.}\ \bibnamefont {Morton}}, \bibinfo {author}
  {\bibfnamefont {D.~N.}\ \bibnamefont {Jamieson}}, \bibinfo {author}
  {\bibfnamefont {A.~S.}\ \bibnamefont {Dzurak}},\ and\ \bibinfo {author}
  {\bibfnamefont {A.}~\bibnamefont {Morello}},\ }\href
  {https://doi.org/10.1038/nature11449} {\bibfield  {journal} {\bibinfo
  {journal} {Nature}\ }\textbf {\bibinfo {volume} {489}},\ \bibinfo {pages}
  {541} (\bibinfo {year} {2012})}\BibitemShut {NoStop}%
\bibitem [{\citenamefont {Morello}\ \emph {et~al.}(2020)\citenamefont
  {Morello}, \citenamefont {Pla}, \citenamefont {Bertet},\ and\ \citenamefont
  {Jamieson}}]{MorPla2020}%
  \BibitemOpen
  \bibfield  {author} {\bibinfo {author} {\bibfnamefont {A.}~\bibnamefont
  {Morello}}, \bibinfo {author} {\bibfnamefont {J.~J.}\ \bibnamefont {Pla}},
  \bibinfo {author} {\bibfnamefont {P.}~\bibnamefont {Bertet}},\ and\ \bibinfo
  {author} {\bibfnamefont {D.~N.}\ \bibnamefont {Jamieson}},\ }\href
  {https://doi.org/https://doi.org/10.1002/qute.202000005} {\bibfield
  {journal} {\bibinfo  {journal} {Advanced Quantum Technologies}\ }\textbf
  {\bibinfo {volume} {3}},\ \bibinfo {pages} {2000005} (\bibinfo {year}
  {2020})}\BibitemShut {NoStop}%
\bibitem [{\citenamefont {Morello}\ \emph {et~al.}(2010)\citenamefont
  {Morello}, \citenamefont {Pla}, \citenamefont {Zwanenburg}, \citenamefont
  {Chan}, \citenamefont {Tan}, \citenamefont {Huebl}, \citenamefont
  {M{\"o}tt{\"o}nen}, \citenamefont {Nugroho}, \citenamefont {Yang},
  \citenamefont {van Donkelaar}, \citenamefont {Alves}, \citenamefont
  {Jamieson}, \citenamefont {Escott}, \citenamefont {Hollenberg}, \citenamefont
  {Clark},\ and\ \citenamefont {Dzurak}}]{MorPla2010}%
  \BibitemOpen
  \bibfield  {author} {\bibinfo {author} {\bibfnamefont {A.}~\bibnamefont
  {Morello}}, \bibinfo {author} {\bibfnamefont {J.~J.}\ \bibnamefont {Pla}},
  \bibinfo {author} {\bibfnamefont {F.~A.}\ \bibnamefont {Zwanenburg}},
  \bibinfo {author} {\bibfnamefont {K.~W.}\ \bibnamefont {Chan}}, \bibinfo
  {author} {\bibfnamefont {K.~Y.}\ \bibnamefont {Tan}}, \bibinfo {author}
  {\bibfnamefont {H.}~\bibnamefont {Huebl}}, \bibinfo {author} {\bibfnamefont
  {M.}~\bibnamefont {M{\"o}tt{\"o}nen}}, \bibinfo {author} {\bibfnamefont
  {C.~D.}\ \bibnamefont {Nugroho}}, \bibinfo {author} {\bibfnamefont
  {C.}~\bibnamefont {Yang}}, \bibinfo {author} {\bibfnamefont {J.~A.}\
  \bibnamefont {van Donkelaar}}, \bibinfo {author} {\bibfnamefont {A.~D.~C.}\
  \bibnamefont {Alves}}, \bibinfo {author} {\bibfnamefont {D.~N.}\ \bibnamefont
  {Jamieson}}, \bibinfo {author} {\bibfnamefont {C.~C.}\ \bibnamefont
  {Escott}}, \bibinfo {author} {\bibfnamefont {L.~C.~L.}\ \bibnamefont
  {Hollenberg}}, \bibinfo {author} {\bibfnamefont {R.~G.}\ \bibnamefont
  {Clark}},\ and\ \bibinfo {author} {\bibfnamefont {A.~S.}\ \bibnamefont
  {Dzurak}},\ }\href {https://doi.org/10.1038/nature09392} {\bibfield
  {journal} {\bibinfo  {journal} {Nature}\ }\textbf {\bibinfo {volume} {467}},\
  \bibinfo {pages} {687} (\bibinfo {year} {2010})}\BibitemShut {NoStop}%
\bibitem [{\citenamefont {Keith}\ \emph {et~al.}(2019)\citenamefont {Keith},
  \citenamefont {House}, \citenamefont {Donnelly}, \citenamefont {Watson},
  \citenamefont {Weber},\ and\ \citenamefont {Simmons}}]{KeiHou2019}%
  \BibitemOpen
  \bibfield  {author} {\bibinfo {author} {\bibfnamefont {D.}~\bibnamefont
  {Keith}}, \bibinfo {author} {\bibfnamefont {M.~G.}\ \bibnamefont {House}},
  \bibinfo {author} {\bibfnamefont {M.~B.}\ \bibnamefont {Donnelly}}, \bibinfo
  {author} {\bibfnamefont {T.~F.}\ \bibnamefont {Watson}}, \bibinfo {author}
  {\bibfnamefont {B.}~\bibnamefont {Weber}},\ and\ \bibinfo {author}
  {\bibfnamefont {M.~Y.}\ \bibnamefont {Simmons}},\ }\href
  {https://doi.org/10.1103/PhysRevX.9.041003} {\bibfield  {journal} {\bibinfo
  {journal} {Phys. Rev. X}\ }\textbf {\bibinfo {volume} {9}},\ \bibinfo {pages}
  {041003} (\bibinfo {year} {2019})}\BibitemShut {NoStop}%
\bibitem [{\citenamefont {Blumoff}\ \emph {et~al.}(2022)\citenamefont
  {Blumoff}, \citenamefont {Pan}, \citenamefont {Keating}, \citenamefont
  {Andrews}, \citenamefont {Barnes}, \citenamefont {Brecht}, \citenamefont
  {Croke}, \citenamefont {Euliss}, \citenamefont {Fast}, \citenamefont
  {Jackson}, \citenamefont {Jones}, \citenamefont {Kerckhoff}, \citenamefont
  {Lanza}, \citenamefont {Raach}, \citenamefont {Thomas}, \citenamefont
  {Velunta}, \citenamefont {Weinstein}, \citenamefont {Ladd}, \citenamefont
  {Eng}, \citenamefont {Borselli}, \citenamefont {Hunter},\ and\ \citenamefont
  {Rakher}}]{BluPan2022}%
  \BibitemOpen
  \bibfield  {author} {\bibinfo {author} {\bibfnamefont {J.~Z.}\ \bibnamefont
  {Blumoff}}, \bibinfo {author} {\bibfnamefont {A.~S.}\ \bibnamefont {Pan}},
  \bibinfo {author} {\bibfnamefont {T.~E.}\ \bibnamefont {Keating}}, \bibinfo
  {author} {\bibfnamefont {R.~W.}\ \bibnamefont {Andrews}}, \bibinfo {author}
  {\bibfnamefont {D.~W.}\ \bibnamefont {Barnes}}, \bibinfo {author}
  {\bibfnamefont {T.~L.}\ \bibnamefont {Brecht}}, \bibinfo {author}
  {\bibfnamefont {E.~T.}\ \bibnamefont {Croke}}, \bibinfo {author}
  {\bibfnamefont {L.~E.}\ \bibnamefont {Euliss}}, \bibinfo {author}
  {\bibfnamefont {J.~A.}\ \bibnamefont {Fast}}, \bibinfo {author}
  {\bibfnamefont {C.~A.~C.}\ \bibnamefont {Jackson}}, \bibinfo {author}
  {\bibfnamefont {A.~M.}\ \bibnamefont {Jones}}, \bibinfo {author}
  {\bibfnamefont {J.}~\bibnamefont {Kerckhoff}}, \bibinfo {author}
  {\bibfnamefont {R.~K.}\ \bibnamefont {Lanza}}, \bibinfo {author}
  {\bibfnamefont {K.}~\bibnamefont {Raach}}, \bibinfo {author} {\bibfnamefont
  {B.~J.}\ \bibnamefont {Thomas}}, \bibinfo {author} {\bibfnamefont
  {R.}~\bibnamefont {Velunta}}, \bibinfo {author} {\bibfnamefont {A.~J.}\
  \bibnamefont {Weinstein}}, \bibinfo {author} {\bibfnamefont {T.~D.}\
  \bibnamefont {Ladd}}, \bibinfo {author} {\bibfnamefont {K.}~\bibnamefont
  {Eng}}, \bibinfo {author} {\bibfnamefont {M.~G.}\ \bibnamefont {Borselli}},
  \bibinfo {author} {\bibfnamefont {A.~T.}\ \bibnamefont {Hunter}},\ and\
  \bibinfo {author} {\bibfnamefont {M.~T.}\ \bibnamefont {Rakher}},\ }\href
  {https://doi.org/10.1103/PRXQuantum.3.010352} {\bibfield  {journal} {\bibinfo
   {journal} {PRX Quantum}\ }\textbf {\bibinfo {volume} {3}},\ \bibinfo {pages}
  {010352} (\bibinfo {year} {2022})}\BibitemShut {NoStop}%
\bibitem [{\citenamefont {Culcer}\ and\ \citenamefont
  {Zimmerman}(2013)}]{CulZim2013}%
  \BibitemOpen
  \bibfield  {author} {\bibinfo {author} {\bibfnamefont {D.}~\bibnamefont
  {Culcer}}\ and\ \bibinfo {author} {\bibfnamefont {N.~M.}\ \bibnamefont
  {Zimmerman}},\ }\href@noop {} {\bibfield  {journal} {\bibinfo  {journal}
  {Applied Physics Letters}\ }\textbf {\bibinfo {volume} {102}},\ \bibinfo
  {pages} {232108} (\bibinfo {year} {2013})}\BibitemShut {NoStop}%
\bibitem [{\citenamefont {Bermeister}\ \emph {et~al.}(2014)\citenamefont
  {Bermeister}, \citenamefont {Keith},\ and\ \citenamefont
  {Culcer}}]{BerKei2014}%
  \BibitemOpen
  \bibfield  {author} {\bibinfo {author} {\bibfnamefont {A.}~\bibnamefont
  {Bermeister}}, \bibinfo {author} {\bibfnamefont {D.}~\bibnamefont {Keith}},\
  and\ \bibinfo {author} {\bibfnamefont {D.}~\bibnamefont {Culcer}},\
  }\href@noop {} {\bibfield  {journal} {\bibinfo  {journal} {Applied Physics
  Letters}\ }\textbf {\bibinfo {volume} {105}},\ \bibinfo {pages} {192102}
  (\bibinfo {year} {2014})}\BibitemShut {NoStop}%
\bibitem [{\citenamefont {Culcer}\ \emph {et~al.}(2009)\citenamefont {Culcer},
  \citenamefont {Hu},\ and\ \citenamefont {Das~Sarma}}]{CulHu2009}%
  \BibitemOpen
  \bibfield  {author} {\bibinfo {author} {\bibfnamefont {D.}~\bibnamefont
  {Culcer}}, \bibinfo {author} {\bibfnamefont {X.}~\bibnamefont {Hu}},\ and\
  \bibinfo {author} {\bibfnamefont {S.}~\bibnamefont {Das~Sarma}},\ }\href
  {https://doi.org/10.1063/1.3194778} {\bibfield  {journal} {\bibinfo
  {journal} {Applied Physics Letters}\ }\textbf {\bibinfo {volume} {95}},\
  \bibinfo {pages} {073102} (\bibinfo {year} {2009})}\BibitemShut {NoStop}%
\bibitem [{\citenamefont {Song}\ \emph {et~al.}(2022)\citenamefont {Song},
  \citenamefont {Chantasri}, \citenamefont {Tonekaboni},\ and\ \citenamefont
  {Wiseman}}]{PRL}%
  \BibitemOpen
  \bibfield  {author} {\bibinfo {author} {\bibfnamefont {H.}~\bibnamefont
  {Song}}, \bibinfo {author} {\bibfnamefont {A.}~\bibnamefont {Chantasri}},
  \bibinfo {author} {\bibfnamefont {B.}~\bibnamefont {Tonekaboni}},\ and\
  \bibinfo {author} {\bibfnamefont {H.~M.}\ \bibnamefont {Wiseman}},\
  }\href@noop {} {\bibfield  {journal} {\bibinfo  {journal} {arXiv:2205.12567
  [quant-ph]}\ } (\bibinfo {year} {2022})}\BibitemShut {NoStop}%
\bibitem [{\citenamefont {Zorin}\ \emph {et~al.}(1996)\citenamefont {Zorin},
  \citenamefont {Ahlers}, \citenamefont {Niemeyer}, \citenamefont {Weimann},
  \citenamefont {Wolf}, \citenamefont {Krupenin},\ and\ \citenamefont
  {Lotkhov}}]{ZorAhl1996}%
  \BibitemOpen
  \bibfield  {author} {\bibinfo {author} {\bibfnamefont {A.~B.}\ \bibnamefont
  {Zorin}}, \bibinfo {author} {\bibfnamefont {F.-J.}\ \bibnamefont {Ahlers}},
  \bibinfo {author} {\bibfnamefont {J.}~\bibnamefont {Niemeyer}}, \bibinfo
  {author} {\bibfnamefont {T.}~\bibnamefont {Weimann}}, \bibinfo {author}
  {\bibfnamefont {H.}~\bibnamefont {Wolf}}, \bibinfo {author} {\bibfnamefont
  {V.~A.}\ \bibnamefont {Krupenin}},\ and\ \bibinfo {author} {\bibfnamefont
  {S.~V.}\ \bibnamefont {Lotkhov}},\ }\href
  {https://doi.org/10.1103/PhysRevB.53.13682} {\bibfield  {journal} {\bibinfo
  {journal} {Phys. Rev. B}\ }\textbf {\bibinfo {volume} {53}},\ \bibinfo
  {pages} {13682} (\bibinfo {year} {1996})}\BibitemShut {NoStop}%
\bibitem [{\citenamefont {Paladino}\ \emph {et~al.}(2002)\citenamefont
  {Paladino}, \citenamefont {Faoro}, \citenamefont {Falci},\ and\ \citenamefont
  {Fazio}}]{PalFao2002}%
  \BibitemOpen
  \bibfield  {author} {\bibinfo {author} {\bibfnamefont {E.}~\bibnamefont
  {Paladino}}, \bibinfo {author} {\bibfnamefont {L.}~\bibnamefont {Faoro}},
  \bibinfo {author} {\bibfnamefont {G.}~\bibnamefont {Falci}},\ and\ \bibinfo
  {author} {\bibfnamefont {R.}~\bibnamefont {Fazio}},\ }\href
  {https://doi.org/10.1103/PhysRevLett.88.228304} {\bibfield  {journal}
  {\bibinfo  {journal} {Phys. Rev. Lett.}\ }\textbf {\bibinfo {volume} {88}},\
  \bibinfo {pages} {228304} (\bibinfo {year} {2002})}\BibitemShut {NoStop}%
\bibitem [{\citenamefont {Fujisawa}\ and\ \citenamefont
  {Hirayama}(2000)}]{fujisawa2000charge}%
  \BibitemOpen
  \bibfield  {author} {\bibinfo {author} {\bibfnamefont {T.}~\bibnamefont
  {Fujisawa}}\ and\ \bibinfo {author} {\bibfnamefont {Y.}~\bibnamefont
  {Hirayama}},\ }\href {https://doi.org/10.1063/1.127038} {\bibfield  {journal}
  {\bibinfo  {journal} {Applied Physics Letters}\ }\textbf {\bibinfo {volume}
  {77}},\ \bibinfo {pages} {543} (\bibinfo {year} {2000})}\BibitemShut
  {NoStop}%
\bibitem [{\citenamefont {Paladino}\ \emph {et~al.}(2014)\citenamefont
  {Paladino}, \citenamefont {Galperin}, \citenamefont {Falci},\ and\
  \citenamefont {Altshuler}}]{PalGal2014}%
  \BibitemOpen
  \bibfield  {author} {\bibinfo {author} {\bibfnamefont {E.}~\bibnamefont
  {Paladino}}, \bibinfo {author} {\bibfnamefont {Y.~M.}\ \bibnamefont
  {Galperin}}, \bibinfo {author} {\bibfnamefont {G.}~\bibnamefont {Falci}},\
  and\ \bibinfo {author} {\bibfnamefont {B.~L.}\ \bibnamefont {Altshuler}},\
  }\href {https://doi.org/10.1103/RevModPhys.86.361} {\bibfield  {journal}
  {\bibinfo  {journal} {Rev. Mod. Phys.}\ }\textbf {\bibinfo {volume} {86}},\
  \bibinfo {pages} {361} (\bibinfo {year} {2014})}\BibitemShut {NoStop}%
\bibitem [{\citenamefont {Itakura}\ and\ \citenamefont
  {Tokura}(2003)}]{ItaTok2003}%
  \BibitemOpen
  \bibfield  {author} {\bibinfo {author} {\bibfnamefont {T.}~\bibnamefont
  {Itakura}}\ and\ \bibinfo {author} {\bibfnamefont {Y.}~\bibnamefont
  {Tokura}},\ }\href {https://doi.org/10.1103/PhysRevB.67.195320} {\bibfield
  {journal} {\bibinfo  {journal} {Phys. Rev. B}\ }\textbf {\bibinfo {volume}
  {67}},\ \bibinfo {pages} {195320} (\bibinfo {year} {2003})}\BibitemShut
  {NoStop}%
\bibitem [{\citenamefont {Bergli}\ \emph {et~al.}(2006)\citenamefont {Bergli},
  \citenamefont {Galperin},\ and\ \citenamefont {Altshuler}}]{BerGal2006}%
  \BibitemOpen
  \bibfield  {author} {\bibinfo {author} {\bibfnamefont {J.}~\bibnamefont
  {Bergli}}, \bibinfo {author} {\bibfnamefont {Y.~M.}\ \bibnamefont
  {Galperin}},\ and\ \bibinfo {author} {\bibfnamefont {B.~L.}\ \bibnamefont
  {Altshuler}},\ }\href {https://doi.org/10.1103/PhysRevB.74.024509} {\bibfield
   {journal} {\bibinfo  {journal} {Phys. Rev. B}\ }\textbf {\bibinfo {volume}
  {74}},\ \bibinfo {pages} {024509} (\bibinfo {year} {2006})}\BibitemShut
  {NoStop}%
\bibitem [{\citenamefont {Bergli}\ \emph {et~al.}(2009)\citenamefont {Bergli},
  \citenamefont {Galperin},\ and\ \citenamefont {Altshuler}}]{BerGal2009}%
  \BibitemOpen
  \bibfield  {author} {\bibinfo {author} {\bibfnamefont {J.}~\bibnamefont
  {Bergli}}, \bibinfo {author} {\bibfnamefont {Y.~M.}\ \bibnamefont
  {Galperin}},\ and\ \bibinfo {author} {\bibfnamefont {B.~L.}\ \bibnamefont
  {Altshuler}},\ }\href {https://doi.org/10.1088/1367-2630/11/2/025002}
  {\bibfield  {journal} {\bibinfo  {journal} {New Journal of Physics}\ }\textbf
  {\bibinfo {volume} {11}},\ \bibinfo {pages} {025002} (\bibinfo {year}
  {2009})}\BibitemShut {NoStop}%
\bibitem [{\citenamefont {Muhonen}\ \emph {et~al.}(2014)\citenamefont
  {Muhonen}, \citenamefont {Dehollain}, \citenamefont {Laucht}, \citenamefont
  {Hudson}, \citenamefont {Kalra}, \citenamefont {Sekiguchi}, \citenamefont
  {Itoh}, \citenamefont {Jamieson}, \citenamefont {McCallum}, \citenamefont
  {Dzurak},\ and\ \citenamefont {Morello}}]{MuhDeh2014}%
  \BibitemOpen
  \bibfield  {author} {\bibinfo {author} {\bibfnamefont {J.~T.}\ \bibnamefont
  {Muhonen}}, \bibinfo {author} {\bibfnamefont {J.~P.}\ \bibnamefont
  {Dehollain}}, \bibinfo {author} {\bibfnamefont {A.}~\bibnamefont {Laucht}},
  \bibinfo {author} {\bibfnamefont {F.~E.}\ \bibnamefont {Hudson}}, \bibinfo
  {author} {\bibfnamefont {R.}~\bibnamefont {Kalra}}, \bibinfo {author}
  {\bibfnamefont {T.}~\bibnamefont {Sekiguchi}}, \bibinfo {author}
  {\bibfnamefont {K.~M.}\ \bibnamefont {Itoh}}, \bibinfo {author}
  {\bibfnamefont {D.~N.}\ \bibnamefont {Jamieson}}, \bibinfo {author}
  {\bibfnamefont {J.~C.}\ \bibnamefont {McCallum}}, \bibinfo {author}
  {\bibfnamefont {A.~S.}\ \bibnamefont {Dzurak}},\ and\ \bibinfo {author}
  {\bibfnamefont {A.}~\bibnamefont {Morello}},\ }\href
  {https://doi.org/10.1038/nnano.2014.211} {\bibfield  {journal} {\bibinfo
  {journal} {Nature Nanotechnology}\ }\textbf {\bibinfo {volume} {9}},\
  \bibinfo {pages} {986} (\bibinfo {year} {2014})}\BibitemShut {NoStop}%
\bibitem [{\citenamefont {Gardiner}(1985)}]{Gar85}%
  \BibitemOpen
  \bibfield  {author} {\bibinfo {author} {\bibfnamefont {C.~W.}\ \bibnamefont
  {Gardiner}},\ }\href@noop {} {\emph {\bibinfo {title} {Handbook of Stochastic
  Methods}}}\ (\bibinfo  {publisher} {Spring\-er},\ \bibinfo {address}
  {Berlin},\ \bibinfo {year} {1985})\BibitemShut {NoStop}%
\bibitem [{\citenamefont {Jacobs}(2010)}]{jacobs2010stochastic}%
  \BibitemOpen
  \bibfield  {author} {\bibinfo {author} {\bibfnamefont {K.}~\bibnamefont
  {Jacobs}},\ }\href@noop {} {\emph {\bibinfo {title} {Stochastic processes for
  physicists: understanding noisy systems}}}\ (\bibinfo  {publisher} {Cambridge
  University Press},\ \bibinfo {year} {2010})\BibitemShut {NoStop}%
\bibitem [{\citenamefont {Christensen}\ \emph {et~al.}(2019)\citenamefont
  {Christensen}, \citenamefont {Wilen}, \citenamefont {Opremcak}, \citenamefont
  {Nelson}, \citenamefont {Schlenker}, \citenamefont {Zimonick}, \citenamefont
  {Faoro}, \citenamefont {Ioffe}, \citenamefont {Rosen}, \citenamefont {DuBois}
  \emph {et~al.}}]{christensen2019anomalous}%
  \BibitemOpen
  \bibfield  {author} {\bibinfo {author} {\bibfnamefont {B.~G.}~\bibnamefont
  {Christensen}}, \bibinfo {author} {\bibfnamefont {C.~D.}~\bibnamefont {Wilen}},
  \bibinfo {author} {\bibfnamefont {A.}~\bibnamefont {Opremcak}}, \bibinfo
  {author} {\bibfnamefont {J.}~\bibnamefont {Nelson}}, \bibinfo {author}
  {\bibfnamefont {F.}~\bibnamefont {Schlenker}}, \bibinfo {author}
  {\bibfnamefont {C.~H.}~\bibnamefont {Zimonick}}, \bibinfo {author}
  {\bibfnamefont {L.}~\bibnamefont {Faoro}}, \bibinfo {author} {\bibfnamefont
  {L.~B.}~\bibnamefont {Ioffe}}, \bibinfo {author} {\bibfnamefont
  {Y.~J.}~\bibnamefont {Rosen}}, \bibinfo {author} {\bibfnamefont
  {J.~L.}~\bibnamefont {DuBois}}, \emph {et~al.},\ }\href
  {https://doi.org/10.1103/PhysRevB.100.140503} {\bibfield  {journal} {\bibinfo
   {journal} {Physical Review B}\ }\textbf {\bibinfo {volume} {100}},\ \bibinfo
  {pages} {140503} (\bibinfo {year} {2019})}\BibitemShut {NoStop}%
\bibitem [{\citenamefont {Freeman}\ \emph {et~al.}(2016)\citenamefont
  {Freeman}, \citenamefont {Schoenfield},\ and\ \citenamefont
  {Jiang}}]{freeman2016comparison}%
  \BibitemOpen
  \bibfield  {author} {\bibinfo {author} {\bibfnamefont {B.~M.}\ \bibnamefont
  {Freeman}}, \bibinfo {author} {\bibfnamefont {J.~S.}\ \bibnamefont
  {Schoenfield}},\ and\ \bibinfo {author} {\bibfnamefont {H.}~\bibnamefont
  {Jiang}},\ }\href {https://doi.org/10.1063/1.4954700} {\bibfield  {journal}
  {\bibinfo  {journal} {Applied Physics Letters}\ }\textbf {\bibinfo {volume}
  {108}},\ \bibinfo {pages} {253108} (\bibinfo {year} {2016})}\BibitemShut
  {NoStop}%
\bibitem [{\citenamefont {Kim}\ \emph {et~al.}(2015)\citenamefont {Kim},
  \citenamefont {Mamin}, \citenamefont {Sherwood}, \citenamefont {Ohno},
  \citenamefont {Awschalom},\ and\ \citenamefont {Rugar}}]{kim2015decoherence}%
  \BibitemOpen
  \bibfield  {author} {\bibinfo {author} {\bibfnamefont {M.}~\bibnamefont
  {Kim}}, \bibinfo {author} {\bibfnamefont {H.~J.}~\bibnamefont {Mamin}}, \bibinfo
  {author} {\bibfnamefont {M.~H.}~\bibnamefont {Sherwood}}, \bibinfo {author}
  {\bibfnamefont {K.}~\bibnamefont {Ohno}}, \bibinfo {author} {\bibfnamefont
  {D.~D.}~\bibnamefont {Awschalom}},\ and\ \bibinfo {author} {\bibfnamefont
  {D.}~\bibnamefont {Rugar}},\ }\href
  {https://doi.org/10.1103/PhysRevLett.115.087602} {\bibfield  {journal}
  {\bibinfo  {journal} {Physical review letters}\ }\textbf {\bibinfo {volume}
  {115}},\ \bibinfo {pages} {087602} (\bibinfo {year} {2015})}\BibitemShut
  {NoStop}%
\bibitem [{\citenamefont {Brownnutt}\ \emph {et~al.}(2015)\citenamefont
  {Brownnutt}, \citenamefont {Kumph}, \citenamefont {Rabl},\ and\ \citenamefont
  {Blatt}}]{brownnutt2015ion}%
  \BibitemOpen
  \bibfield  {author} {\bibinfo {author} {\bibfnamefont {M.}~\bibnamefont
  {Brownnutt}}, \bibinfo {author} {\bibfnamefont {M.}~\bibnamefont {Kumph}},
  \bibinfo {author} {\bibfnamefont {P.}~\bibnamefont {Rabl}},\ and\ \bibinfo
  {author} {\bibfnamefont {R.}~\bibnamefont {Blatt}},\ }\href
  {https://doi.org/10.1103/RevModPhys.87.1419} {\bibfield  {journal} {\bibinfo
  {journal} {Reviews of Modern Physics}\ }\textbf {\bibinfo {volume} {87}},\
  \bibinfo {pages} {1419} (\bibinfo {year} {2015})}\BibitemShut {NoStop}%
\bibitem [{\citenamefont {Galperin}\ \emph {et~al.}(2006)\citenamefont
  {Galperin}, \citenamefont {Altshuler}, \citenamefont {Bergli},\ and\
  \citenamefont {Shantsev}}]{GalAlt2006}%
  \BibitemOpen
  \bibfield  {author} {\bibinfo {author} {\bibfnamefont {Y.~M.}\ \bibnamefont
  {Galperin}}, \bibinfo {author} {\bibfnamefont {B.~L.}\ \bibnamefont
  {Altshuler}}, \bibinfo {author} {\bibfnamefont {J.}~\bibnamefont {Bergli}},\
  and\ \bibinfo {author} {\bibfnamefont {D.~V.}\ \bibnamefont {Shantsev}},\
  }\href {https://doi.org/10.1103/PhysRevLett.96.097009} {\bibfield  {journal}
  {\bibinfo  {journal} {Phys. Rev. Lett.}\ }\textbf {\bibinfo {volume} {96}},\
  \bibinfo {pages} {097009} (\bibinfo {year} {2006})}\BibitemShut {NoStop}%
\bibitem [{\citenamefont {Kuhlmann}\ \emph {et~al.}(2013)\citenamefont
  {Kuhlmann}, \citenamefont {Houel}, \citenamefont {Ludwig}, \citenamefont
  {Greuter}, \citenamefont {Reuter}, \citenamefont {Wieck}, \citenamefont
  {Poggio},\ and\ \citenamefont {Warburton}}]{kuhlmann2013charge}%
  \BibitemOpen
  \bibfield  {author} {\bibinfo {author} {\bibfnamefont {A.~V.}\ \bibnamefont
  {Kuhlmann}}, \bibinfo {author} {\bibfnamefont {J.}~\bibnamefont {Houel}},
  \bibinfo {author} {\bibfnamefont {A.}~\bibnamefont {Ludwig}}, \bibinfo
  {author} {\bibfnamefont {L.}~\bibnamefont {Greuter}}, \bibinfo {author}
  {\bibfnamefont {D.}~\bibnamefont {Reuter}}, \bibinfo {author} {\bibfnamefont
  {A.~D.}\ \bibnamefont {Wieck}}, \bibinfo {author} {\bibfnamefont
  {M.}~\bibnamefont {Poggio}},\ and\ \bibinfo {author} {\bibfnamefont {R.~J.}\
  \bibnamefont {Warburton}},\ }\href {https://doi.org/10.1038/nphys2688}
  {\bibfield  {journal} {\bibinfo  {journal} {Nature Physics}\ }\textbf
  {\bibinfo {volume} {9}},\ \bibinfo {pages} {570} (\bibinfo {year}
  {2013})}\BibitemShut {NoStop}%
\bibitem [{\citenamefont {Daniotti}\ \emph {et~al.}(2018)\citenamefont
  {Daniotti}, \citenamefont {Benedetti},\ and\ \citenamefont
  {Paris}}]{DanBen2018}%
  \BibitemOpen
  \bibfield  {author} {\bibinfo {author} {\bibfnamefont {S.}~\bibnamefont
  {Daniotti}}, \bibinfo {author} {\bibfnamefont {C.}~\bibnamefont
  {Benedetti}},\ and\ \bibinfo {author} {\bibfnamefont {M.~G.~A.}\ \bibnamefont
  {Paris}},\ }\href {https://doi.org/10.1140/epjd/e2018-90450-x} {\bibfield
  {journal} {\bibinfo  {journal} {The European Physical Journal D}\ }\textbf
  {\bibinfo {volume} {72}},\ \bibinfo {pages} {208} (\bibinfo {year}
  {2018})}\BibitemShut {NoStop}%
\bibitem [{\citenamefont {Fisher}(1995)}]{fisher1995statistical}%
  \BibitemOpen
  \bibfield  {author} {\bibinfo {author} {\bibfnamefont {N.~I.}\ \bibnamefont
  {Fisher}},\ }\href@noop {} {\emph {\bibinfo {title} {Statistical analysis of
  circular data}}}\ (\bibinfo  {publisher} {Cambridge University Press},\
  \bibinfo {year} {1995})\BibitemShut {NoStop}%
\bibitem [{\citenamefont {Joseph}\ and\ \citenamefont {Tou}(1961)}]{JosTou61}%
  \BibitemOpen
  \bibfield  {author} {\bibinfo {author} {\bibfnamefont {D.~P.}\ \bibnamefont
  {Joseph}}\ and\ \bibinfo {author} {\bibfnamefont {T.~J.}\ \bibnamefont
  {Tou}},\ }\href {https://doi.org/10.1109/TAI.1961.6371743} {\bibfield
  {journal} {\bibinfo  {journal} {Transactions of the American Institute of
  Electrical Engineers, Part II: Applications and Industry}\ }\textbf {\bibinfo
  {volume} {80}},\ \bibinfo {pages} {193} (\bibinfo {year} {1961})}\BibitemShut
  {NoStop}%
\bibitem [{\citenamefont {{\AA}str{\" o}m}(1970)}]{Astrom1970}%
  \BibitemOpen
  \bibfield  {author} {\bibinfo {author} {\bibfnamefont {K.~J.}\ \bibnamefont
  {{\AA}str{\" o}m}},\ }\href@noop {} {\emph {\bibinfo {title} {Introduction to
  Stochastic Control Theory}}}\ (\bibinfo  {publisher} {Academic Press},\
  \bibinfo {address} {New York and London},\ \bibinfo {year}
  {1970})\BibitemShut {NoStop}%
\bibitem [{\citenamefont {Cormen}\ \emph {et~al.}(2009)\citenamefont {Cormen},
  \citenamefont {Leiserson}, \citenamefont {Rivest},\ and\ \citenamefont
  {Stein}}]{cormen2009AlgorithmGreedy}%
  \BibitemOpen
  \bibfield  {author} {\bibinfo {author} {\bibfnamefont {T.~H.}\ \bibnamefont
  {Cormen}}, \bibinfo {author} {\bibfnamefont {C.~E.}\ \bibnamefont
  {Leiserson}}, \bibinfo {author} {\bibfnamefont {R.~L.}\ \bibnamefont
  {Rivest}},\ and\ \bibinfo {author} {\bibfnamefont {C.}~\bibnamefont
  {Stein}},\ }\href@noop {} {\emph {\bibinfo {title} {Introduction to
  algorithms}}}\ (\bibinfo  {publisher} {MIT press},\ \bibinfo {year} {2009})\
  Chap.~\bibinfo {chapter} {16}\BibitemShut {NoStop}%
\bibitem [{\citenamefont {Berry}\ and\ \citenamefont
  {Wiseman}(2000)}]{BerWis00}%
  \BibitemOpen
  \bibfield  {author} {\bibinfo {author} {\bibfnamefont {D.~W.}\ \bibnamefont
  {Berry}}\ and\ \bibinfo {author} {\bibfnamefont {H.~M.}\ \bibnamefont
  {Wiseman}},\ }\href {https://doi.org/10.1103/PhysRevLett.85.5098} {\bibfield
  {journal} {\bibinfo  {journal} {Phys. Rev. Lett.}\ }\textbf {\bibinfo
  {volume} {85}},\ \bibinfo {pages} {5098} (\bibinfo {year}
  {2000})}\BibitemShut {NoStop}%
\bibitem [{\citenamefont {Berry}\ \emph {et~al.}(2001)\citenamefont {Berry},
  \citenamefont {Wiseman},\ and\ \citenamefont {Breslin}}]{BerWisBre01}%
  \BibitemOpen
  \bibfield  {author} {\bibinfo {author} {\bibfnamefont {D.~W.}\ \bibnamefont
  {Berry}}, \bibinfo {author} {\bibfnamefont {H.~M.}\ \bibnamefont {Wiseman}},\
  and\ \bibinfo {author} {\bibfnamefont {J.~K.}\ \bibnamefont {Breslin}},\
  }\href {https://doi.org/10.1103/PhysRevA.63.053804} {\bibfield  {journal}
  {\bibinfo  {journal} {Phys. Rev. A}\ }\textbf {\bibinfo {volume} {63}},\
  \bibinfo {pages} {053804} (\bibinfo {year} {2001})}\BibitemShut {NoStop}%
\bibitem [{\citenamefont {Helstrom}(1976)}]{Helstrom}%
  \BibitemOpen
  \bibfield  {author} {\bibinfo {author} {\bibfnamefont {C.~W.}\ \bibnamefont
  {Helstrom}},\ }\href@noop {} {\emph {\bibinfo {title} {Quantum Detection and
  Estimation Theory}}},\ \bibinfo {series} {Mathematics in Science and
  Engineering}, Vol.\ \bibinfo {volume} {123}\ (\bibinfo  {publisher} {Academic
  Press},\ \bibinfo {address} {New York},\ \bibinfo {year} {1976})\BibitemShut
  {NoStop}%
\bibitem [{\citenamefont {Wiseman}\ and\ \citenamefont
  {Milburn}(2010)}]{WisMil10}%
  \BibitemOpen
  \bibfield  {author} {\bibinfo {author} {\bibfnamefont {H.~M.}\ \bibnamefont
  {Wiseman}}\ and\ \bibinfo {author} {\bibfnamefont {G.~J.}\ \bibnamefont
  {Milburn}},\ }\href@noop {} {\emph {\bibinfo {title} {Quantum Measurement and
  Control}}}\ (\bibinfo  {publisher} {Cambridge University Press},\ \bibinfo
  {address} {Cambridge, England},\ \bibinfo {year} {2010})\BibitemShut
  {NoStop}%
\bibitem [{\citenamefont {Draper}\ and\ \citenamefont
  {Smith}(1998)}]{DraSmi1998}%
  \BibitemOpen
  \bibfield  {author} {\bibinfo {author} {\bibfnamefont {N.~R.}\ \bibnamefont
  {Draper}}\ and\ \bibinfo {author} {\bibfnamefont {H.}~\bibnamefont {Smith}},\
  }\href@noop {} {\emph {\bibinfo {title} {Applied regression analysis}}},\
  Vol.\ \bibinfo {volume} {326}\ (\bibinfo  {publisher} {John Wiley \& Sons},\
  \bibinfo {year} {1998})\BibitemShut {NoStop}%
\bibitem [{\citenamefont {Paz}(1971)}]{Paz1971}%
  \BibitemOpen
  \bibfield  {author} {\bibinfo {author} {\bibfnamefont {A.}~\bibnamefont
  {Paz}},\ }\href@noop {} {\emph {\bibinfo {title} {Introduction to
  Probabilistic Automata}}}\ (\bibinfo  {publisher} {Academic Press},\ \bibinfo
  {year} {1971})\BibitemShut {NoStop}%
\bibitem [{\citenamefont {Shallit}\ and\ \citenamefont
  {Wang}(2001)}]{shawan2001}%
  \BibitemOpen
  \bibfield  {author} {\bibinfo {author} {\bibfnamefont {J.}~\bibnamefont
  {Shallit}}\ and\ \bibinfo {author} {\bibfnamefont {M.-W.}\ \bibnamefont
  {Wang}},\ }\href {https://doi.org/10.25596/jalc-2001-537} {\bibfield
  {journal} {\bibinfo  {journal} {Journal of Automata, Languages and
  Combinatorics}\ }\textbf {\bibinfo {volume} {6}},\ \bibinfo {pages} {537}
  (\bibinfo {year} {2001})}\BibitemShut {NoStop}%
\bibitem [{\citenamefont {Paz-Silva}\ \emph {et~al.}(2017)\citenamefont
  {Paz-Silva}, \citenamefont {Norris},\ and\ \citenamefont
  {Viola}}]{paz2017multiqubit}%
  \BibitemOpen
  \bibfield  {author} {\bibinfo {author} {\bibfnamefont {G.~A.}\ \bibnamefont
  {Paz-Silva}}, \bibinfo {author} {\bibfnamefont {L.~M.}\ \bibnamefont
  {Norris}},\ and\ \bibinfo {author} {\bibfnamefont {L.}~\bibnamefont
  {Viola}},\ }\href {https://doi.org/10.1103/PhysRevA.95.022121} {\bibfield
  {journal} {\bibinfo  {journal} {Phys. Rev. A}\ }\textbf {\bibinfo {volume}
  {95}},\ \bibinfo {pages} {022121} (\bibinfo {year} {2017})}\BibitemShut
  {NoStop}%
\bibitem [{\citenamefont {von L{\"u}pke}\ \emph {et~al.}(2020)\citenamefont
  {von L{\"u}pke}, \citenamefont {Beaudoin}, \citenamefont {Norris},
  \citenamefont {Sung}, \citenamefont {Winik}, \citenamefont {Qiu},
  \citenamefont {Kjaergaard}, \citenamefont {Kim}, \citenamefont {Yoder},
  \citenamefont {Gustavsson} \emph {et~al.}}]{von2020two}%
  \BibitemOpen
  \bibfield  {author} {\bibinfo {author} {\bibfnamefont {U.}~\bibnamefont {von
  L{\"u}pke}}, \bibinfo {author} {\bibfnamefont {F.}~\bibnamefont {Beaudoin}},
  \bibinfo {author} {\bibfnamefont {L.~M.}\ \bibnamefont {Norris}}, \bibinfo
  {author} {\bibfnamefont {Y.}~\bibnamefont {Sung}}, \bibinfo {author}
  {\bibfnamefont {R.}~\bibnamefont {Winik}}, \bibinfo {author} {\bibfnamefont
  {J.~Y.}\ \bibnamefont {Qiu}}, \bibinfo {author} {\bibfnamefont
  {M.}~\bibnamefont {Kjaergaard}}, \bibinfo {author} {\bibfnamefont
  {D.}~\bibnamefont {Kim}}, \bibinfo {author} {\bibfnamefont {J.}~\bibnamefont
  {Yoder}}, \bibinfo {author} {\bibfnamefont {S.}~\bibnamefont {Gustavsson}},
  \emph {et~al.},\ }\href {https://doi.org/10.1103/PRXQuantum.1.010305}
  {\bibfield  {journal} {\bibinfo  {journal} {PRX Quantum}\ }\textbf {\bibinfo
  {volume} {1}},\ \bibinfo {pages} {010305} (\bibinfo {year}
  {2020})}\BibitemShut {NoStop}%
\bibitem [{\citenamefont {Chalermpusitarak}\ \emph {et~al.}(2021)\citenamefont
  {Chalermpusitarak}, \citenamefont {Tonekaboni}, \citenamefont {Wang},
  \citenamefont {Norris}, \citenamefont {Viola},\ and\ \citenamefont
  {Paz-Silva}}]{chalermpusitarak2021frame}%
  \BibitemOpen
  \bibfield  {author} {\bibinfo {author} {\bibfnamefont {T.}~\bibnamefont
  {Chalermpusitarak}}, \bibinfo {author} {\bibfnamefont {B.}~\bibnamefont
  {Tonekaboni}}, \bibinfo {author} {\bibfnamefont {Y.}~\bibnamefont {Wang}},
  \bibinfo {author} {\bibfnamefont {L.~M.}\ \bibnamefont {Norris}}, \bibinfo
  {author} {\bibfnamefont {L.}~\bibnamefont {Viola}},\ and\ \bibinfo {author}
  {\bibfnamefont {G.~A.}\ \bibnamefont {Paz-Silva}},\ }\href
  {https://doi.org/10.1103/PRXQuantum.2.030315} {\bibfield  {journal} {\bibinfo
   {journal} {PRX Quantum}\ }\textbf {\bibinfo {volume} {2}},\ \bibinfo {pages}
  {030315} (\bibinfo {year} {2021})}\BibitemShut {NoStop}%
\bibitem [{\citenamefont {Youssry}\ \emph {et~al.}(2021)\citenamefont
  {Youssry}, \citenamefont {Paz-Silva},\ and\ \citenamefont
  {Ferrie}}]{youssry2021noise}%
  \BibitemOpen
  \bibfield  {author} {\bibinfo {author} {\bibfnamefont {A.}~\bibnamefont
  {Youssry}}, \bibinfo {author} {\bibfnamefont {G.~A.}\ \bibnamefont
  {Paz-Silva}},\ and\ \bibinfo {author} {\bibfnamefont {C.}~\bibnamefont
  {Ferrie}},\ }\bibfield  {journal} {\bibinfo  {journal} {arXiv preprint}\
  }\href {https://doi.org/10.48550/arXiv.2103.13018}
  {10.48550/arXiv.2103.13018} (\bibinfo {year} {2021})\BibitemShut {NoStop}%
\bibitem [{\citenamefont {Youssry}\ \emph {et~al.}(2020)\citenamefont
  {Youssry}, \citenamefont {Paz-Silva},\ and\ \citenamefont
  {Ferrie}}]{youssry2020characterization}%
  \BibitemOpen
  \bibfield  {author} {\bibinfo {author} {\bibfnamefont {A.}~\bibnamefont
  {Youssry}}, \bibinfo {author} {\bibfnamefont {G.~A.}\ \bibnamefont
  {Paz-Silva}},\ and\ \bibinfo {author} {\bibfnamefont {C.}~\bibnamefont
  {Ferrie}},\ }\href {https://doi.org/10.1038/s41534-020-00332-8} {\bibfield
  {journal} {\bibinfo  {journal} {npj Quantum Information}\ }\textbf {\bibinfo
  {volume} {6}},\ \bibinfo {pages} {1} (\bibinfo {year} {2020})}\BibitemShut
  {NoStop}%
\bibitem [{\citenamefont {Korotkov}(1999)}]{KorPRB99}%
  \BibitemOpen
  \bibfield  {author} {\bibinfo {author} {\bibfnamefont {A.~N.}\ \bibnamefont
  {Korotkov}},\ }\href {https://doi.org/10.1103/PhysRevB.60.5737} {\bibfield
  {journal} {\bibinfo  {journal} {Phys. Rev. B}\ }\textbf {\bibinfo {volume}
  {60}},\ \bibinfo {pages} {5737} (\bibinfo {year} {1999})}\BibitemShut
  {NoStop}%
\bibitem [{\citenamefont {Goan}\ \emph {et~al.}(2001)\citenamefont {Goan},
  \citenamefont {Milburn}, \citenamefont {Wiseman},\ and\ \citenamefont
  {Sun}}]{GoaMilWisSunPRB01}%
  \BibitemOpen
  \bibfield  {author} {\bibinfo {author} {\bibfnamefont {H.-S.}\ \bibnamefont
  {Goan}}, \bibinfo {author} {\bibfnamefont {G.~J.}\ \bibnamefont {Milburn}},
  \bibinfo {author} {\bibfnamefont {H.~M.}\ \bibnamefont {Wiseman}},\ and\
  \bibinfo {author} {\bibfnamefont {H.-B.}\ \bibnamefont {Sun}},\ }\href
  {https://doi.org/10.1103/PhysRevB.63.125326} {\bibfield  {journal} {\bibinfo
  {journal} {Phys. Rev. B}\ }\textbf {\bibinfo {volume} {63}},\ \bibinfo
  {pages} {125326} (\bibinfo {year} {2001})}\BibitemShut {NoStop}%
\end{thebibliography}
%

\end{document}